\documentclass[twocolumn]{aastex631}

\usepackage{xcolor}
\usepackage{xspace}
\usepackage{mwe}
\usepackage{amsmath}
\usepackage{changepage}

\newcommand\cod{\mbox{$\rm CO(3-2)$}}
\newcommand\coa{\mbox{$\rm CO(6-5)$}}
\newcommand\cob{\mbox{$\rm CO(7-6)$}}
\newcommand\coc{\mbox{$\rm CO(10-9)$}}
\newcommand\coe{\mbox{$\rm CO(1-0)$}}
\newcommand\ci{\mbox{[C{\scriptsize I}]}}
\newcommand\cii{\mbox{[C{\scriptsize II}]}}
\newcommand\sii{\mbox{[Si{\scriptsize II}]}}
\newcommand\oi{\mbox{[O{\scriptsize I}]}}
\newcommand\oiii{\mbox{[O{\scriptsize III}]}}
\newcommand\cia{\mbox{[C{\scriptsize I}]$\rm (1-0)$}}
\newcommand\cib{\mbox{[C{\scriptsize I}]$\rm (2-1)$}}

\newcommand\watera{\mbox{p-$\rm H_{2} O (2_{1,1}-2_{0,2})$}}
\newcommand\waterb{\mbox{p-$\rm H_{2} O (3_{1,2}-2_{2,1})$}}
\newcommand\water{\mbox{$\rm H_{2}O$}}
\newcommand\lwater{\mbox{$\rm L_{H_{2}O}$}}

\newcommand\lfir{\mbox{$\rm L_{FIR}$}}
\newcommand\lir{\mbox{$\rm L_{IR}$}}

\DeclareUnicodeCharacter{2212}{\textendash}

\def\Illinoisa{Department of Astronomy, University of Illinois, 1002 West Green St., Urbana, IL 61801, USA}
\def\Illinoisb{Department of Physics, University of Illinois at Urbana-Champaign, 1110 W Green St Loomis Laboratory, Urbana, IL 61801, USA}
\def\Illinoisc{Center for AstroPhysical Surveys, National Center for Supercomputing Applications, Urbana, IL, 61801, USA}

\def\Arizona{Steward Observatory, University of Arizona, 933 North Cherry Avenue, Tucson, AZ 85721, USA}
\def\MPIfR{Max-Planck-Institut f\"{u}r Radioastronomie, Auf dem H\"{u}gel 69 D-53121 Bonn, Germany}
\def\UFlorida{Department of Astronomy, University of Florida, Gainesville, FL 32611, USA}
\def\ESO{European Southern Observatory, Alonso de C{\'o}rdova 3107, Vitacura, Casilla 19001, Santiago de Chile, Chile}
\def\Arizona{Steward Observatory, University of Arizona, 933 North Cherry Avenue, Tucson, AZ 85721, USA}
\def\Diego{N\'ucleo de Astronom\'ia, Facultad de Ingenier\'ia, Universidad Diego Portales, Av. Ej\'ercito 441, Santiago, Chile}
\def\Flatiron{Center for Computational Astrophysics, Flatiron Institute, 162 Fifth Avenue, New York, NY 10010, USA}
\def\canadaa{Department of Physics and Astronomy, University of British Columbia, 6225 Agricultural Rd., Vancouver, V6T 1Z1, Canada}
\def\canadab{Department of Physics and Atmospheric Science, Dalhousie University, Halifax, Nova Scotia, Canada}
\def\dawn{Cosmic Dawn Center (DAWN), DTU-Space, Technical University of Denmark, Elektrovej 327, DK-2800 Kgs. Lyngby, Denmark}

\shortauthors{Jarugula, et al.}

\graphicspath{{./}{figures/}}

\begin{document}

\title{Molecular Line Observations in Two Dusty Star-Forming Galaxies at $z$ = 6.9}


\author[0000-0002-5386-7076]{Sreevani~Jarugula}\footnote{\href{mailto:jarugul2@illinois.edu}{jarugul2@illinois.edu}}
\affiliation{\Illinoisa}
\author[0000-0001-7192-3871]{Joaquin~D.~Vieira}
\affiliation{\Illinoisa}
\affiliation{\Illinoisb}
\affiliation{\Illinoisc}
\author[0000-0003-4678-3939]{Axel~Wei{\ss}}
\affiliation{\MPIfR}
\author[0000-0003-3256-5615]{Justin S. Spilker}
\affiliation{Department of Astronomy, University of Texas at Austin, 2515 Speedway, Stop C1400, Austin, TX 78712, USA}
\affiliation{NHFP Hubble Fellow}
\author[0000-0002-6290-3198]{Manuel~Aravena}
\affiliation{\Diego}
\author[0000-0002-0517-9842]{Melanie~Archipley}
\affil{\Illinoisa}
\affil{\Illinoisc}
\author[0000-0002-3915-2015]{Matthieu~B{\'e}thermin}
\affil{Aix Marseille Univ., CNRS, CNES, LAM, Marseille, France}
\author{Scott ~C. Chapman}
\affil{Eureka Scientific, Inc. 2452 Delmer Street Suite 100, Oakland, CA 94602-3017}

\author[0000-0002-5823-0349]{Chenxing~Dong}
\affiliation{\UFlorida}
\author{Thomas~R.~Greve}
\affiliation{\dawn}
\affil{Department of Physics and Astronomy, University College London, Gower Street, London WC1E 6BT, UK}
\author{Kevin~Harrington}
\affiliation{\ESO}
\author[0000-0003-4073-3236]{Christopher~C.~Hayward}
\affiliation{\Flatiron}
\author[0000-0002-8669-5733]{Yashar~Hezaveh}
\affil{D\'{e}partement de Physique, Universit\'{e} de Montr\'{e}al, Montreal, Quebec, H3T 1J4, Canada}
\affiliation{\Flatiron}
\author{Ryley~Hill}
\affiliation{\canadaa}
\author[0000-0002-4208-3532]{Katrina~C.~Litke}
\affiliation{\Arizona}
\author[0000-0001-6919-1237]{Matthew~A. Malkan}
\affil{Department of Physics and Astronomy, University of California, Los Angeles, CA 90095-1547, USA
}
\author[0000-0002-2367-1080]{Daniel~P. Marrone}
\affiliation{\Arizona}
\author[0000-0002-7064-4309]{Desika~Narayanan}
\affil{Department of Astronomy, University of Florida, 211 Bryant Space Sciences Center, Gainesville, FL 32611, USA}
\affil{University of Florida Informatics Institute, 432 Newell Drive, CISE Bldg E251, Gainesville, FL 32611, USA}
\affiliation{\dawn}
\author[0000-0001-7946-557X]{Kedar~A~Phadke}
\affiliation{\Illinoisa}
\author[0000-0001-7477-1586]{Cassie~Reuter}
\affil{\Illinoisa}
\author{Kaja~M. Rotermund}
\affiliation{\canadab}


\begin{abstract}

\noindent SPT0311-58 is the most massive infrared luminous system discovered so far during the Epoch of Reionization (EoR). In this paper, we present a detailed analysis of the molecular interstellar medium at \mbox{z = 6.9}, through high resolution observations of the \coa, \cob, \coc, \cib, and \watera\ lines and dust continuum emission with the Atacama Large Millimeter/submillimeter Array (ALMA). The system consists of a pair of intensely star-forming gravitationally lensed galaxies (labelled West and East). The intrinsic far-infrared luminosity is \mbox{(16 $\pm$ 4)$\times\rm 10^{12} \ \rm L_{\odot}$} in West and (27 $\pm$ 4)$\times\rm 10^{11} \ \rm L_{\odot}$ in East. We model the dust, CO, and \ci\ using non-local thermodynamic equilibrium radiative transfer models and estimate the intrinsic gas mass to be \mbox{(5.4 $\pm$ 3.4)$\times\rm 10^{11} \ \rm M_{\odot}$} in West and (3.1 $\pm$ 2.7)$\times\rm 10^{10} \ \rm M_{\odot}$ in East. We find that the CO spectral line energy distribution in West and East are typical of high-redshift sub-millimeter galaxies (SMGs). The CO-to-$\rm H_{2}$ conversion factor ($\rm \alpha_{CO}$) and the gas depletion time scales estimated from the model are consistent with the high-redshift SMGs in the literature within the uncertainties. We find no evidence of evolution of depletion time with redshift in SMGs at \mbox{z $>$ 3}. This is the most detailed study of molecular gas content of a galaxy in the EoR to-date, with the most distant detection of \water\ in a galaxy without any evidence for active galactic nuclei in the literature.

\end{abstract}



\section{Introduction} \label{sec:intro}
Dusty star-forming galaxies (DSFGs) are dust enshrouded, intensely star-forming galaxies with tens to thousands of stars forming per year \citep[e.g.,][]{casey14}. These starburst galaxies are bright at sub-millimeter wavelengths as the ultraviolet (UV) photons from the young stars are absorbed and re-radiated by dust into the far-infrared (FIR) wavelengths giving rise to total infrared luminosities (\lir) greater than $\rm 10^{12}\ L_{\odot}$. Observations of these galaxies have the advantage of the negative-K correction \citep{blain93} at long wavelengths, enabling galaxy selection which is independent of redshift. DSFGs contribute significantly to the cosmic star formation history at high redshift \citep{casey14} and are thought to be the progenitors of present day massive ellipticals \citep[e.g.,][]{simpson14}. However, the theoretical understanding of the DSFG population has been challenging \citep[e.g.,][]{narayanan15,hayward21}, especially during the Epoch of Reionization (EoR), between z$\sim$15$-$6 \citep{madau97}. 

To understand the star formation in DSFGs, it is important to study the molecular gas content of the interstellar medium (ISM) \citep[e.g.,][]{carilli13}. However, cold molecular hydrogen ($\rm H_{2}$), which is the most abundant molecule and the fuel for star formation, is difficult to observe directly because it lacks a permanent dipole moment. The most commonly used tracer of $\rm H_{2}$ is carbon monoxide ($\rm ^{12}C^{16}O$, hereafter CO), which is the second most abundant molecule in the universe and also an important coolant. The observations of multiple CO rotational transitions from local galaxies \citep[e.g.,][]{greve14,rosenberg15, lu17}, high-redshift ultraluminous infrared galaxies (ULIRGs), and quasars \citep[e.g.,][]{weiss07,yang17} and simulations \citep{kamenetzky18} have shown that the CO spectral line energy distribution (SLED) can probe the physical conditions of the galaxies, such as density and temperature. The low-J CO transitions are emitted from the low density ($\rm n_{H_{2}} \lesssim10^{3}\ cm^{-3}$) diffuse ISM \citep[e.g.,][]{harris10,ivison11}. The \mbox{mid-J} CO transitions, such as \coa\ and \cob\ are excited in denser regions of the galaxy ($\rm n_{H_{2}}\sim10^{4}\ cm^{-3}$) where star formation mainly occurs \citep{lu15,lu17}. The high-J CO transitions \mbox{(J $>$ 10)} originate in the compact high density regions and may be further enhanced in the presence of active galactic nuclei (AGN) \citep[e.g.,][]{weiss07,bradford09,lu17}. The molecular gas mass is traditionally calculated from the CO luminosity by using a CO-to-$\rm H_{2}$ gas conversion factor, $\rm \alpha_{CO}$ \citep[e.g.,][]{bolatto13}. However, there is a large uncertainty in the estimation of gas mass because of factors such as the dependence of carbon abundance on optical depth, metallicity, and the destruction of CO due to UV radiation \citep[e.g.,][]{narayanan11b,bolatto13}. Observations have shown that \cia\ can be an independent tracer of molecular gas \citep[e.g.,][]{papadopoulos04,weiss05,walter11,bothwell17}, given its simpler excitation level structure.

Another abundant molecule in the universe after $\rm H_{2}$ and CO is water, \water\ \citep{neufeld95}. The complex level populations of \water\ are tightly coupled to the infrared radiation field. This is due to infrared pumping, where the molecular transitions in \water\ are mainly excited by the FIR photons. The different transitions of \water\ have been observed to be correlated with \lir\ and hence, star formation over more than three orders of magnitude, both in the local and high-redshift (U)LIRGs \citep[e.g.,][]{yang13,omont13,yang16,liu17,jarugula19}. \water\ is observed to be more tightly correlated to \lir\ than $\rm CO(6-5)$ or \cii\ \citep{jarugula19}.

SPT0311-58 consists of a pair of gravitationally lensed galaxies discovered in the South Pole Telescope (SPT) Survey \citep{vieira13,everett20}. The spatially unresolved observations of low and \mbox{mid-J} CO transitions from the Atacama Large Millimeter/submillimeter Array (ALMA), Australia  Telescope  Compact  Array  (ATCA), and Atacama Pathfinder Experiment (APEX) are presented in \citet{strandet17}. These observations confirm the redshift of the source at z = 6.9. High resolution dust, \cii, and \oiii\ ALMA observations of \mbox{SPT0311-58} are discussed in \citet{marrone18} and a detailed lens modeling shows a pair of galaxies separated by a projected distance of $\rm \sim$8 kpc ($\sim$1.5$\arcsec$), likely in a state of merging. In this paper, we combine the previous observations with high resolution molecular line imaging with ALMA, including the high-J CO transition, \coc\ and \watera. We use realistic non-Local Thermodynamic Equilibrium (non-LTE) models and estimate the physical properties of the ISM in both the galaxies. 

Throughout the paper, we refer to SPT0311-58 West as W and SPT0311-58 East as E. In Section \ref{sec:observations}, we present the observations used in this analysis and the data reduction. In Section \ref{sec:data_analysis}, we detail the data analysis procedure to estimate the continuum and line luminosities. The lens modeling is also discussed in this section. The results are presented in Section \ref{sec:results}. In the first half of Section \ref{sec:results}, we present the results from observations and in the second half, we discuss the radiative transfer models and results. In Section \ref{sec:discussion}, we discuss the results and in Section \ref{sec:conlusion}, we conclude with a summary. We use flat $\Lambda$CDM cosmology where $h$ = 0.677, \mbox{$\Omega_{m}$ = 0.307}, and \mbox{$\Omega_{\Lambda}$ = 0.693} \citep{planck16cosmo}. We estimate the \lir\ as flux integrated from 8$-$1000 $\rm \mu m$ and total far-infrared luminosity (\lfir) from 42.5$-$122.5 $\rm \mu m$ in the rest frame \citep{helou85}.

\section{Observations} \label{sec:observations}
\begin{figure}[h!t]
\centering
\includegraphics[trim={3.0cm 1.0cm 0.0cm 2.5cm},clip,width=0.6\textwidth]{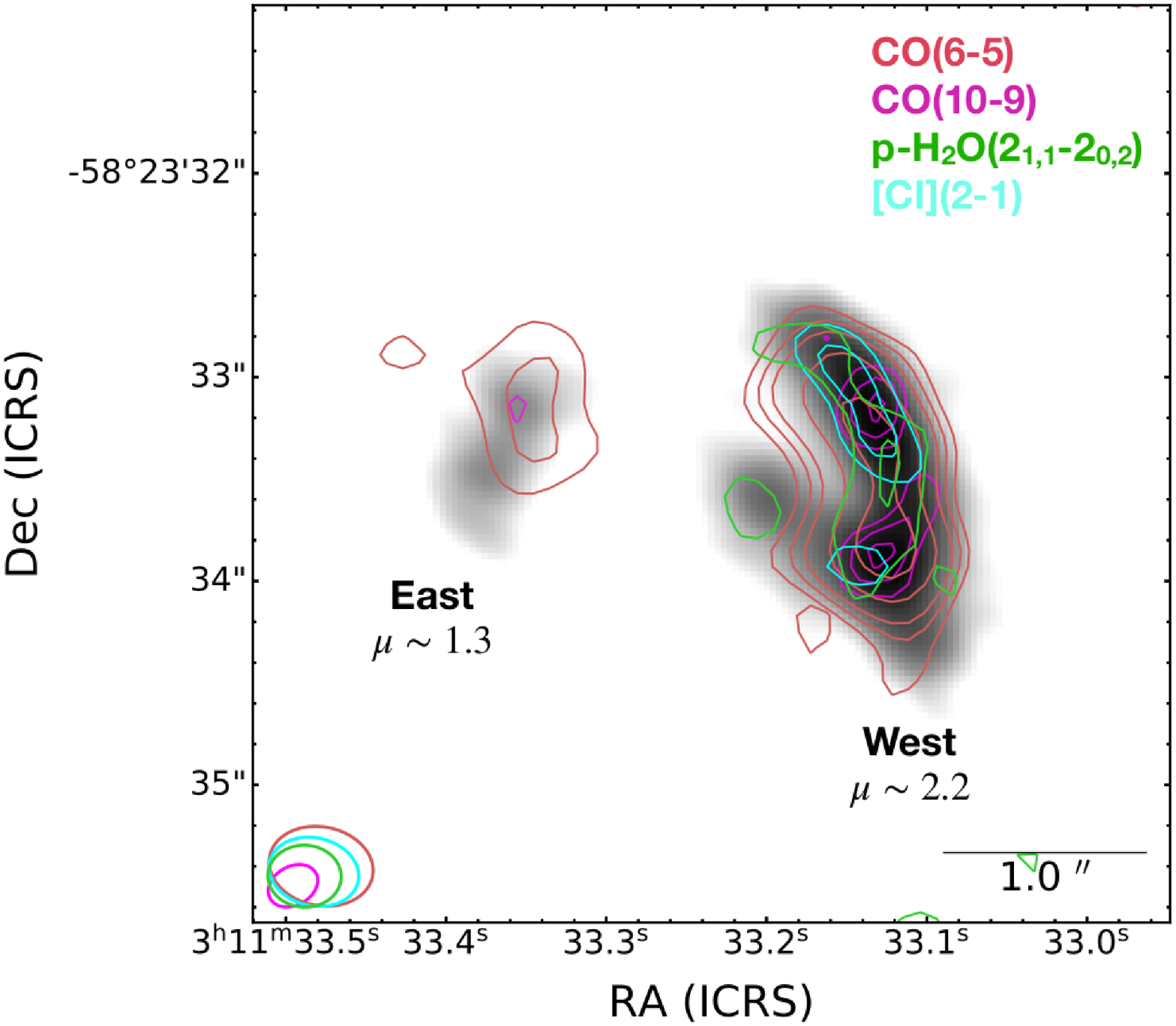}
\caption{SPT0311-58 West and East. The background grey scale image is ALMA 423 GHz Band 8 continuum with magnification of $\sim$1.3 in East and $\sim$2.2 in West \citep{marrone18}. The moment 0 contours of \coa\ (red), \coc\ (magenta), \watera\ (green), and \cib\ (cyan) from this analysis are overlaid on the continuum image. The CO contours are at [3,4,5,7,9] $\times \sigma$ and \ci\ and \water\ contours at [2,3,4] $\times \sigma$. The synthesized beams are shown in the bottom left and the 1.0$\arcsec$ scale bar, which corresponds to $\sim$5.4 kpc at z = 6.9, is at the bottom right.}
\label{fig:figure1}
\end{figure}

We obtained observations of different emission lines and the continuum using ALMA Band 3 and 4 over three tunings (Project ID: 2017.1.01168.S, PI: Vieira). Table \ref{almaobstab} presents the details of the observations. 
We detect \coa, \cob, and \coc\ in both the components, and \cib\ and \watera\ in W.
In Figure \ref{fig:figure1}, we show the high resolution ALMA Band 8 continuum image from \citet{marrone18} and \coa, \coc, \cia, and \watera\ observations from this analysis (see also Figure \ref{spectrumfig}). 

\subsection{Data Reduction and Imaging} \label{sec:imaging}
The data reduction and imaging were performed using the Common Astronomy Software Application package \texttt{CASA} \citep{mcmullin07}. We use the calibrated data products from the ALMA data reduction pipeline (\texttt{CASA}  version 5.1.1). The continuum images are produced by combining the lower and upper side bands (LSB, USB) and excluding the channels containing the line emission. Briggs weighting and a robust parameter of 0.5 are used, which provide a good compromise between resolution and noise. This gives a synthesized beam of $\sim$0.5\arcsec\ at \mbox{95 GHz} and  $\sim$0.3\arcsec\ at 140 GHz. To generate the spectral cubes, we use the same weighting as with the continuum maps and a velocity averaging of \mbox{100 $\rm km\ s^{-1}$}, after subtracting the continuum using \texttt{CASA} task \texttt{uvcontsub} with a polynomial fit of order 1. 

 To generate the velocity integrated intensity maps (moment 0), we produce single channel cubes with a width of 1000 $\rm km\ s^{-1}$ in W and 500 $\rm km\ s^{-1}$ in E, which is \mbox{$\sim$2 $\times$} full width at half maximum (FWHM) of the lines. We consider a velocity separation of $\sim$750 $\rm km\ s^{-1}$ between W and E based on the high resolution \cii\ observations \citep{marrone18}. The synthesized beam of the \mbox{moment 0} maps for \watera\ is $\sim$0.3\arcsec, \coa\ and \cob\ is $\sim$0.5\arcsec, and \coc\ is $\sim$0.2\arcsec.

The signal-to-noise ratio of the lines and the flux densities of the continuum are given in Table \ref{almaobstab} and \ref{continuum_prop}, respectively. The continuum images with moment 0 contours overlaid on top are shown in Figure \ref{spectrumfig}.

\begin{table*}
\centering
\caption{ALMA observations}
\label{almaobstab}
\begin{tabular}{c c c c c c c c c}
\hline\hline
Tuning & Time on source & $\nu_{\rm central}$  &   Line &  $\nu_{\rm rest}^{\rm line}$ & $\rm FWHM_{cont}$ & $\sigma_{\rm cont}$& SNR$_{\rm line}$ \\
  & & LSB, USB  & &  &  &  &\\
 & [minutes]  & [GHz]    & & [GHz] &  [ $\arcsec$ ] & [$\mu$Jy/beam] & W, E\\\\

 \hline
1 & 47 &95, 107   &    \watera & 752.03  & 0.31 $\times$ 0.22 & 16.19 &  4, - \\
  
\hline
 2& 43 &   88, 90    &    \coa & 691.50  & 0.47 $\times$ 0.36 & 8.83 & 10, 4 \\
  &  & 100, 102    &    \cob & 806.65  &  &  & 8, 2\\
  &  &  &     \cib & 809.34  &  &  & 5, - \\

\hline
 3&46  & 134, 146   &    \coc & 1151.98  & 0.27 $\times$ 0.20 & 8.27 & 6, 2\\
 &  &  &       \waterb & 1153.13 & &  & - \\

\hline\hline
\multicolumn{8}{p{\textwidth}}{\hspace{+0.2in}{NOTE. - $\nu_{\rm central}$ corresponds to the central frequency in the lower and upper side band (LSB, USB) of the observations. $\nu_{\rm rest}^{\rm line}$ is the rest frequency of the lines. $\rm FWHM_{cont}$ is the resolution of the continuum image. $\sigma_{\rm cont}$ is the RMS noise in the continuum image by combining LSB and USB. SNR$_{\rm line}$ is the signal-to-noise ratio of the lines in W and E, which is calculated by using the maximum flux density of the scaled \cii\ spectrum and the standard deviation of the flux density in the line free channels (Section \ref{sec:spectrum}). SNR$_{\rm line}$ $<$ 2 are not shown.}}
\end{tabular}
\end{table*}

\begin{table*}
\centering
\caption{ALMA Band 3 and Band 4 continuum properties}
\label{continuum_prop}
\begin{tabular}{c*{8}{>{}p{6.3em}}}
\hline\hline
Tuning & Source & $\nu_{\rm obs}$ & S$_{\rm obs}$ & $\mu$ & S$_{\rm int}$ &   L$_{\rm  FIR \ int}$ &    L$_{\rm  IR \ int}$ \\
 & & [GHz] &  [mJy] & & [mJy] & [10$^{12}$ $\rm L_{\odot}$] & [10$^{12}$ $\rm L_{\odot}$]\\
\hline
1 \& 2 & W & 95 & 1.26 $\pm$ 0.04 & 2.09 $\pm$ 0.06 & 0.60 $\pm$ 0.09 & 16 $\pm$ 4 & 26 $\pm$ 12\\
3 & W & 140  & 4.12 $\pm$ 0.01 & 2.09 $\pm$ 0.10 & 1.97 $\pm$ 0.31 & &\\
1 \& 2 & E & 95  & 0.04 $\pm$ 0.02 & 1.3  & 0.03 $\pm$ 0.02  & 3.0 $\pm$ 0.4 & 3.5 $\pm$ 0.7\\
3 & E & 140  & 0.24 $\pm$ 0.01 & 1.32 $\pm$ 0.06 & 0.18 $\pm$ 0.03 & &\\
\hline\hline
\multicolumn{8}{p{\textwidth}}{\hspace{+0.2in}{NOTE. -  $\nu_{\rm obs}$ corresponds to the observed frequency of the continuum. $\rm S_{obs}$ is the observed flux density from the continuum image. $\rm \mu$ is the flux weighted magnification of the continuum obtained from lens modeling. Note that, due to low SNR at 95 GHz in E, we take magnification from \citet{marrone18}. $\rm S_{int}$ is intrinsic flux density ($\rm S_{obs}/ \mu$) with 15\% additional uncertainty to account for uncertainties in the absolute flux calibration and lens modeling, added in quadrature.  $\rm L_{FIR \ int}$ is the intrinsic $\rm L_{FIR}$ calculated from the SED for each source (flux integrated from 42.5$-$122.5 $\rm \mu m$ in the rest frame). $\rm L_{IR \ int}$ is the intrinsic $\rm L_{IR}$ (flux integrated from 8$-$1000 $\rm \mu m$ in the rest frame).}}
\end{tabular}
\end{table*}

\subsection{Ancillary Data}
In addition to the molecular line observations and the continuum flux density at 95 GHz (B3) and 140 GHz (B4), we include ancillary data on SPT0311-58 from the literature in this analysis \citep{strandet16,strandet17,marrone18}. The source has been observed in previous ALMA cycles (2015.1.00504.S, PI: Strandet and 2016.1.01293.S, PI: Marrone) in B3, B6, B7, and B8 corresponding to 95, 240, 340, and 420 GHz, respectively. For the continuum flux density in B3, we use the current observations which are at a higher resolution of $\sim$0.5\arcsec\ compared to previous observations at $\sim$3.5\arcsec. We also include \textit{Herschel}/SPIRE photometry at 250, 350, and 500 $\mu$m (project ID: DDT$\_$mstrande$\_$1, \citealt{strandet16}). The intrinsic continuum flux densities (corrected for magnification) in W and E are obtained by performing lens modeling described in detail in \citet{marrone18}.

We include \cod\ data observed with ATCA to constrain the CO spectral line energy distribution (SLED). The data reduction and line properties are present in \citet{strandet17}. Since \cod\ is spatially unresolved ($\sim$5 $-$ 6\arcsec) into W and E, we distribute the flux density of \cod\ by scaling to \coa\ in the two galaxies.

\section{Data Analysis} \label{sec:data_analysis}

\subsection{Spectral Line Decomposition} \label{sec:spectrum}
\begin{figure*}[h!t]
\hspace*{0.6cm}
\includegraphics[trim={0cm 0 0.0cm 0.0},clip,width=0.212\textwidth]{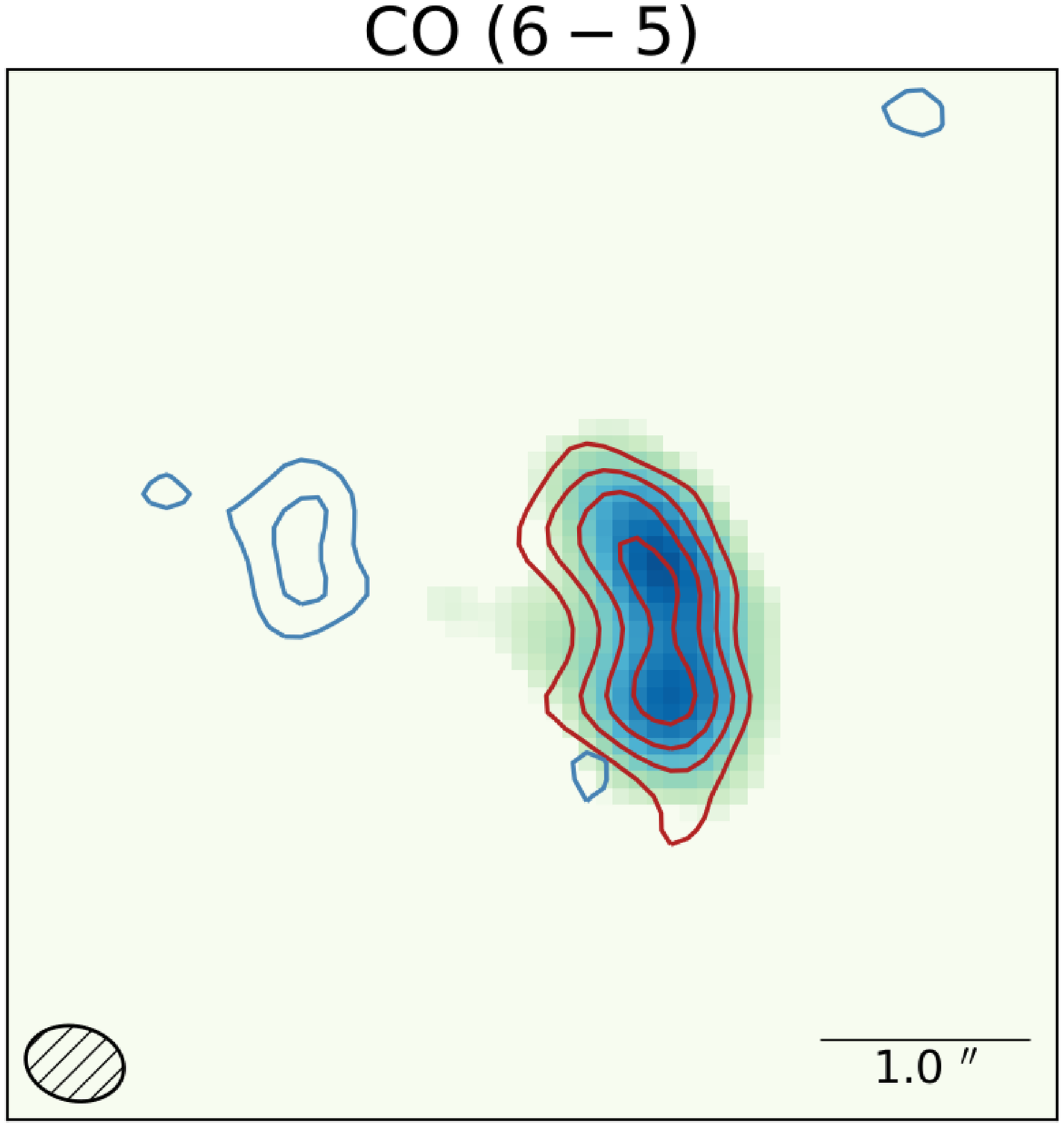}
\hspace*{0.22cm} \includegraphics[trim={0cm 0 0.0cm 0.0},clip,width=0.212\textwidth]{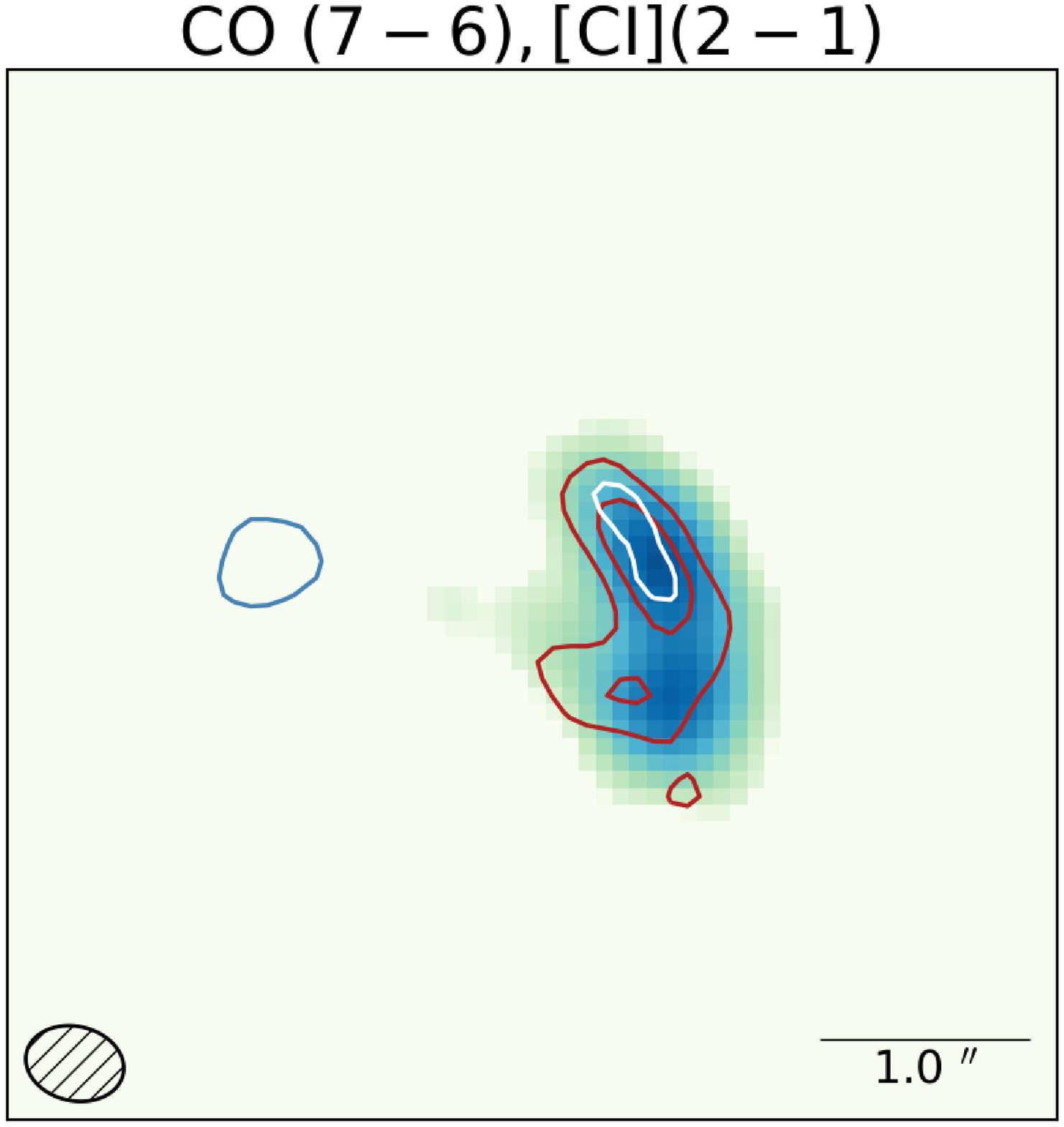}
\hspace*{0.38cm} \includegraphics[trim={0cm 0 0.0cm 0.0},clip,width=0.212\textwidth]{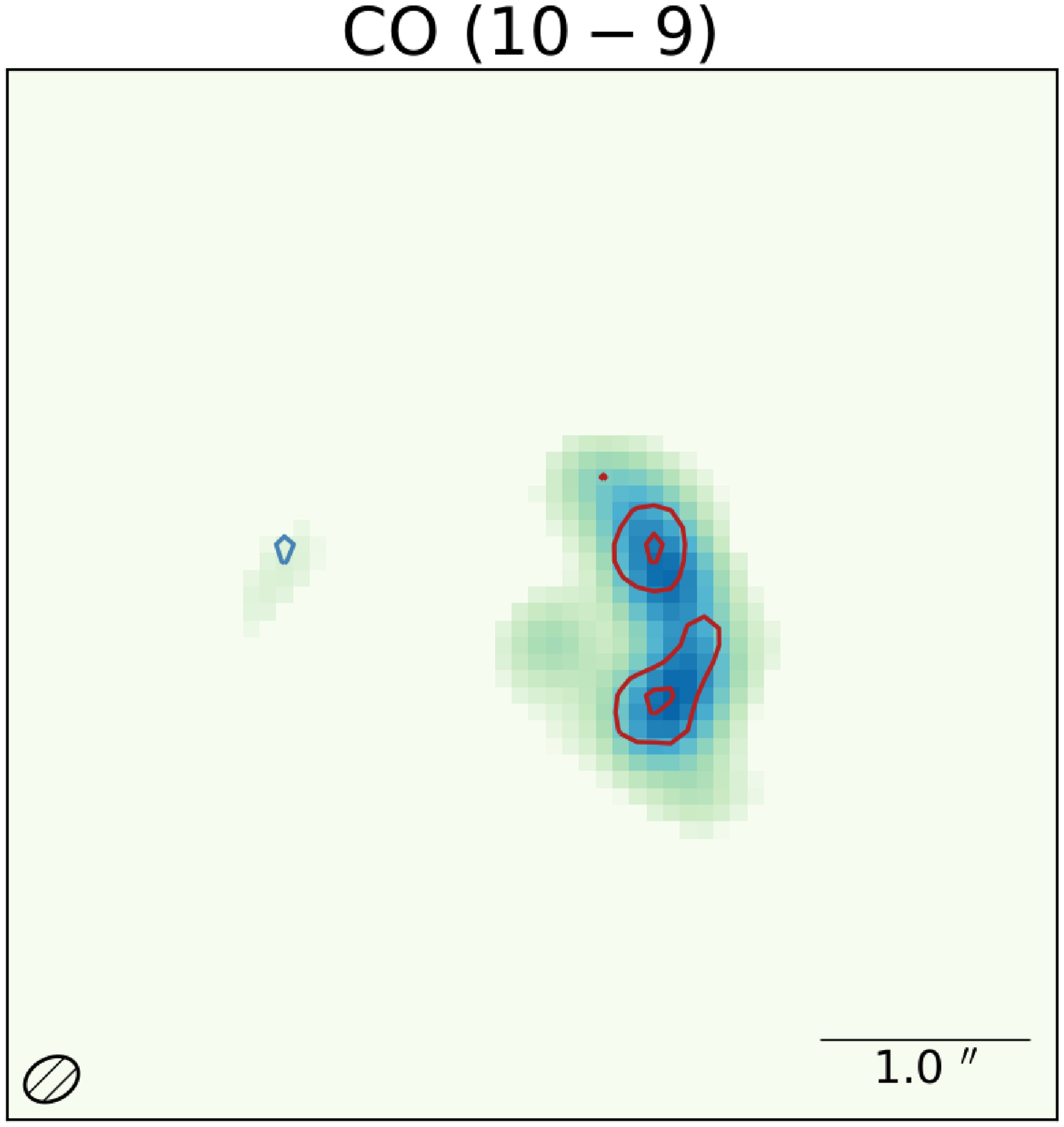}
\hspace*{0.42cm} \includegraphics[trim={0cm 0 0.0cm 0.0},clip,width=0.212\textwidth]{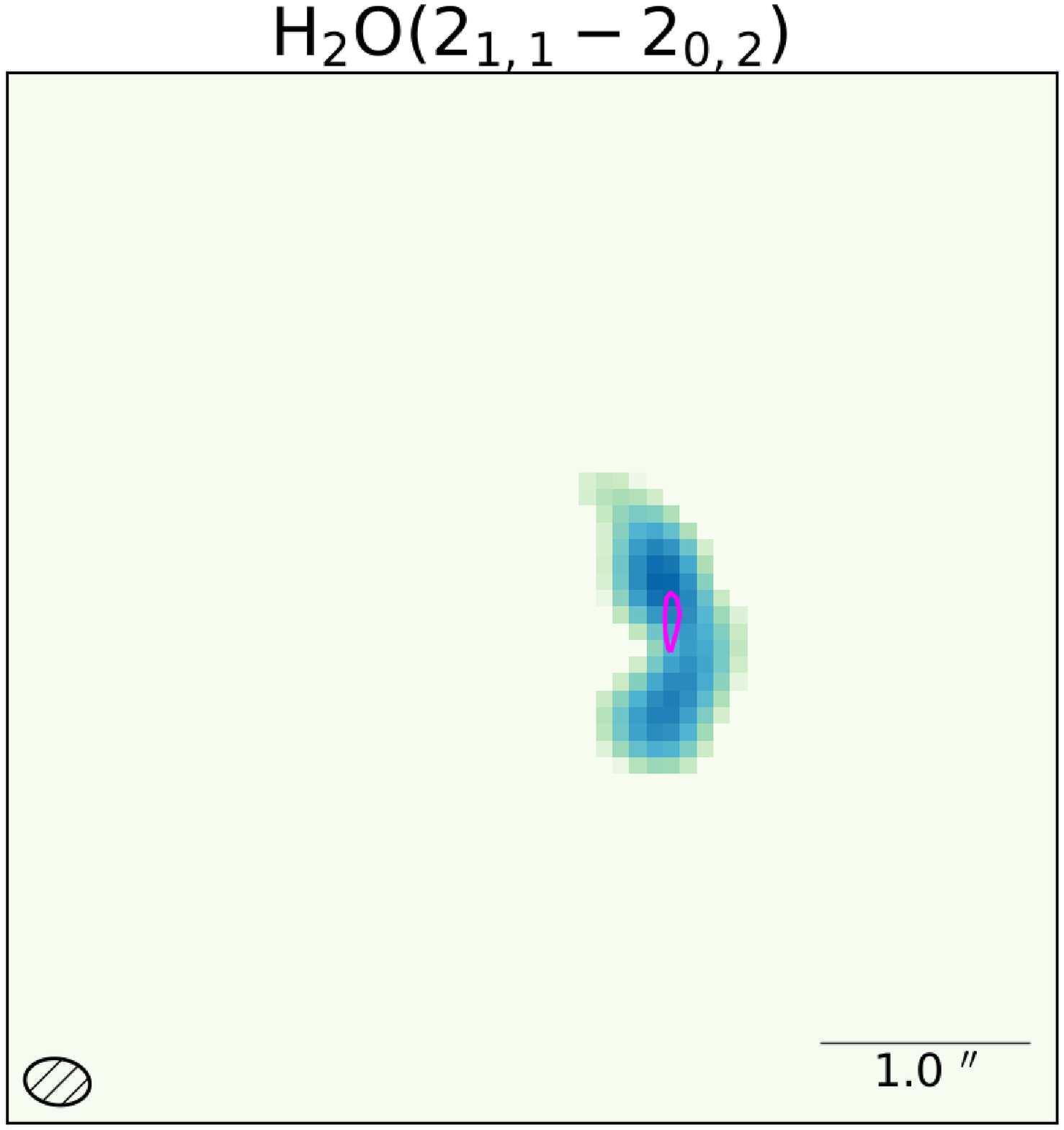}\\
\hspace*{0.0cm}
	\includegraphics[trim={0.0cm 0 0.0cm 2.0},clip,width=0.247\textwidth]{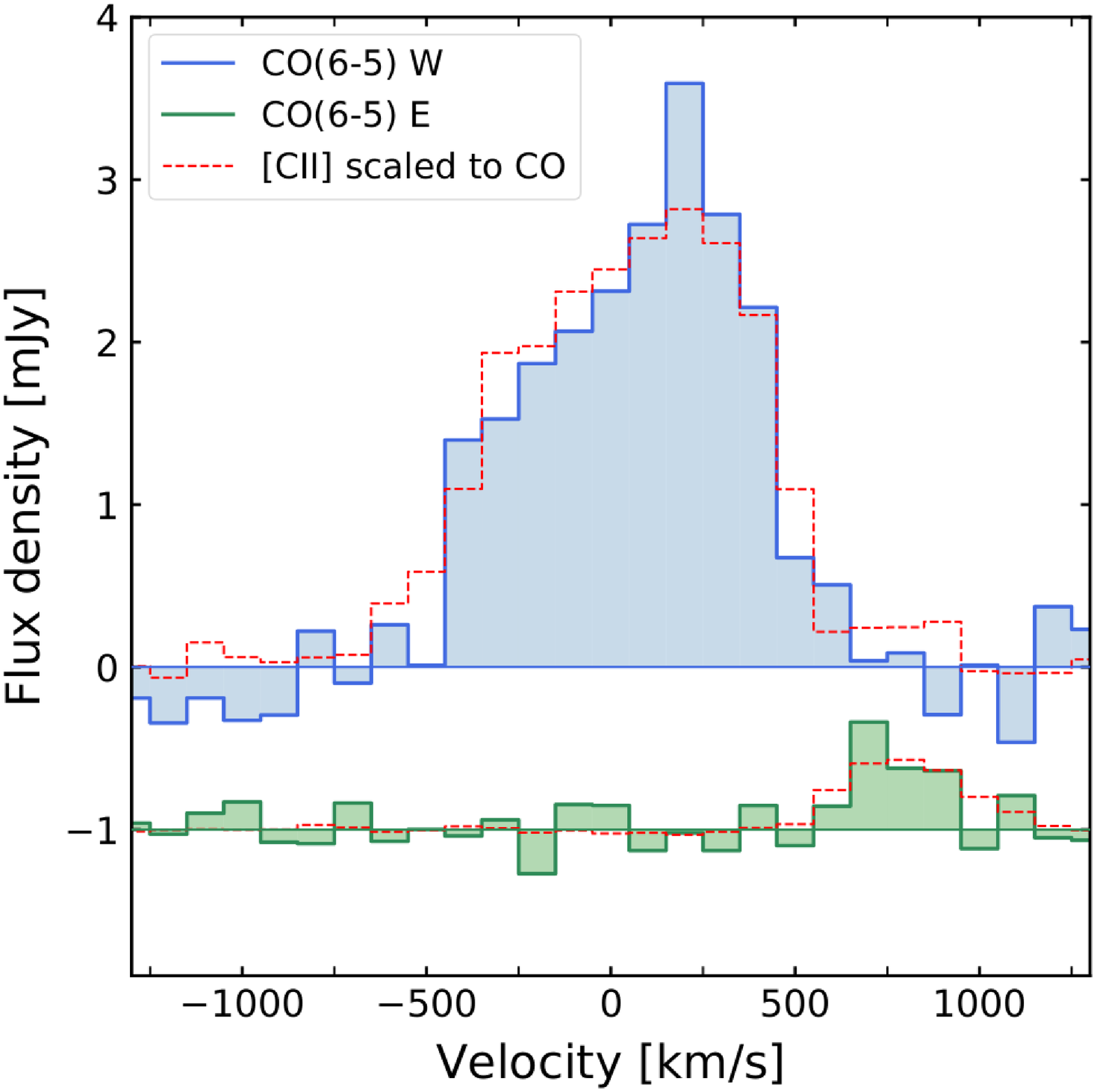}
	\hspace{0.0 cm}
	\includegraphics[trim={1.5cm 0 0.3cm 2.0},clip,width=0.23\textwidth]{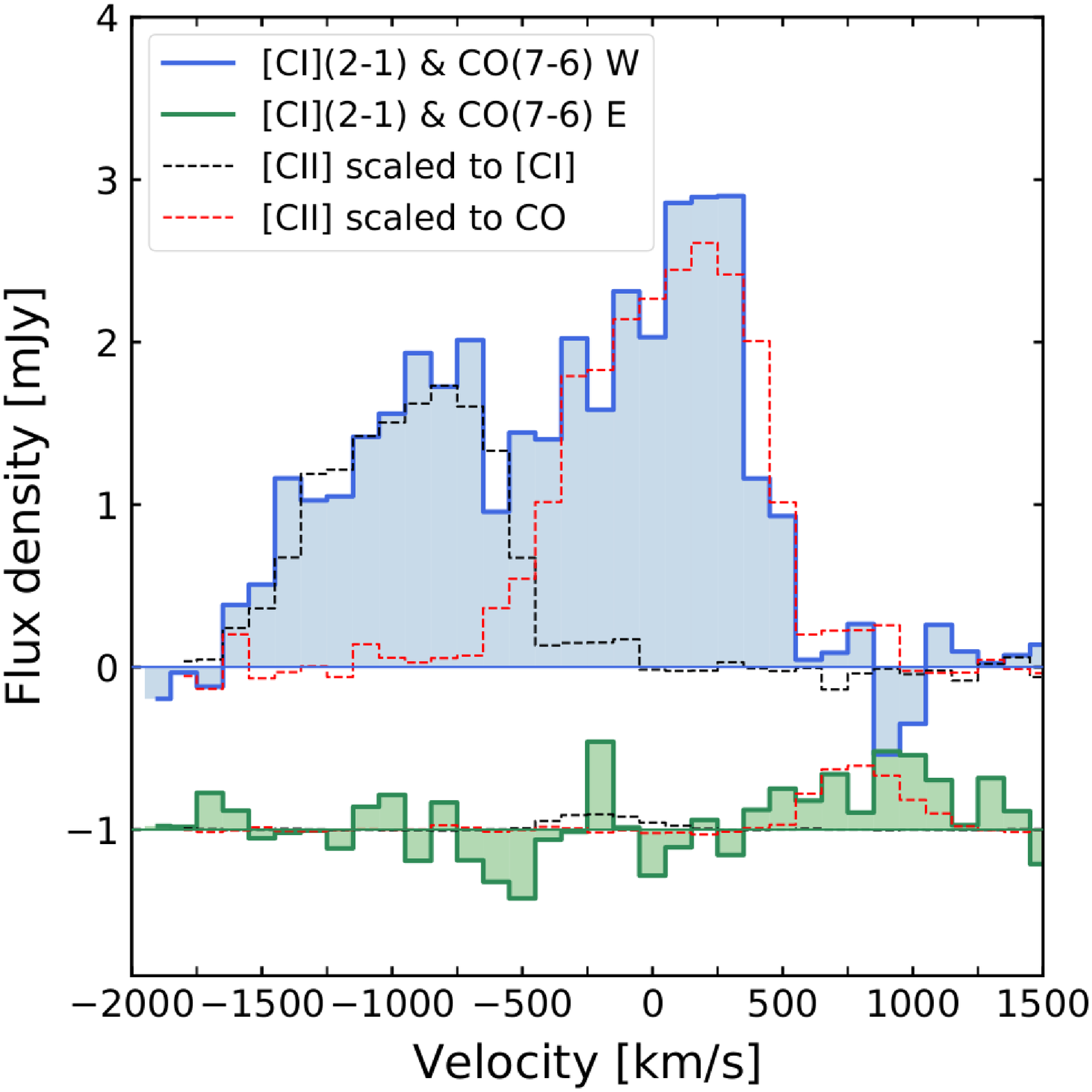}
	\hspace{0.0 cm}
	\includegraphics[trim={1.5cm 0 0.3cm 2.0},clip,width=0.23\textwidth]{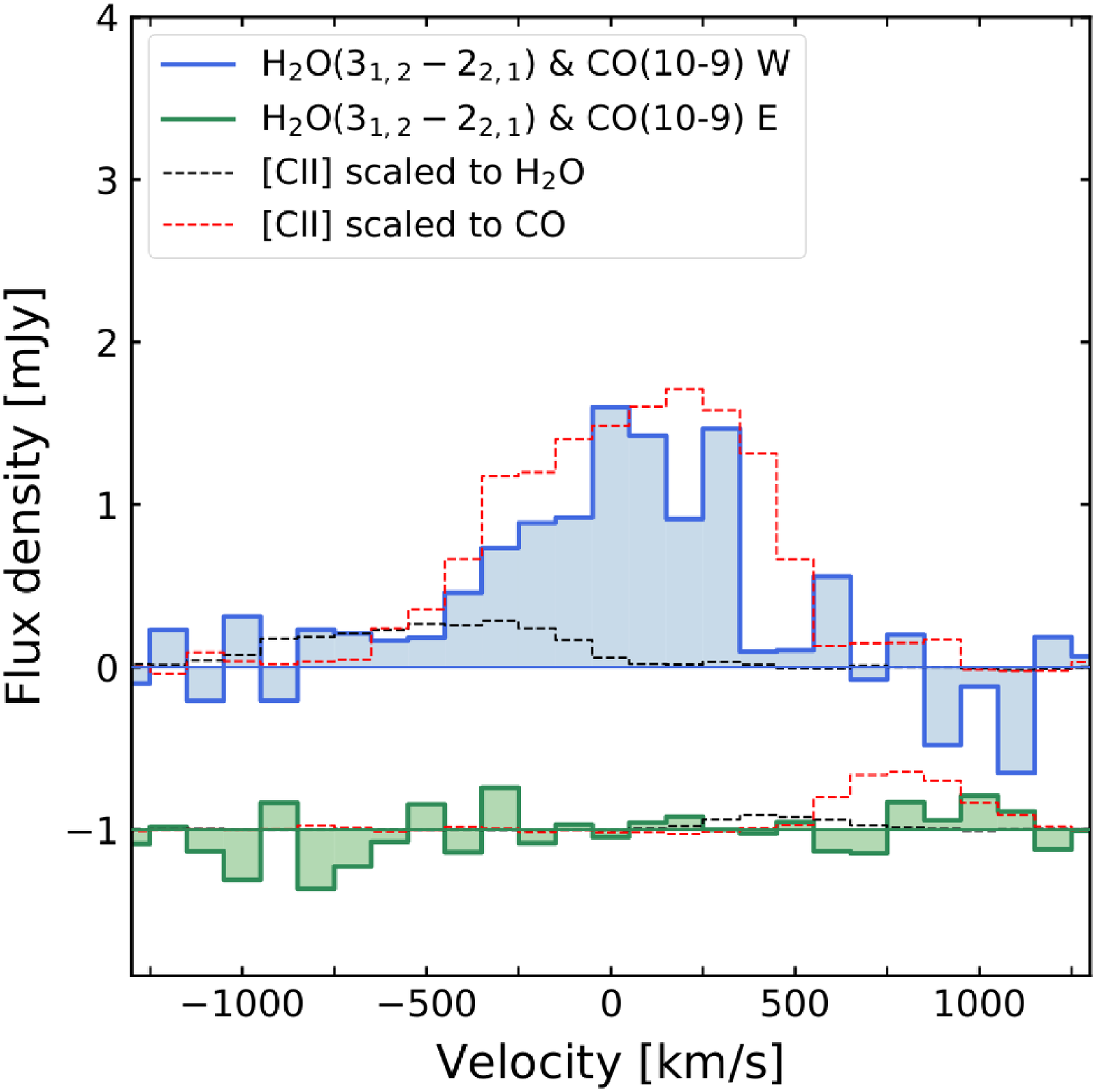}
	\hspace{0.0 cm}
	\includegraphics[trim={1.5cm 0 0.0cm 2.0},clip,width=0.233\textwidth]{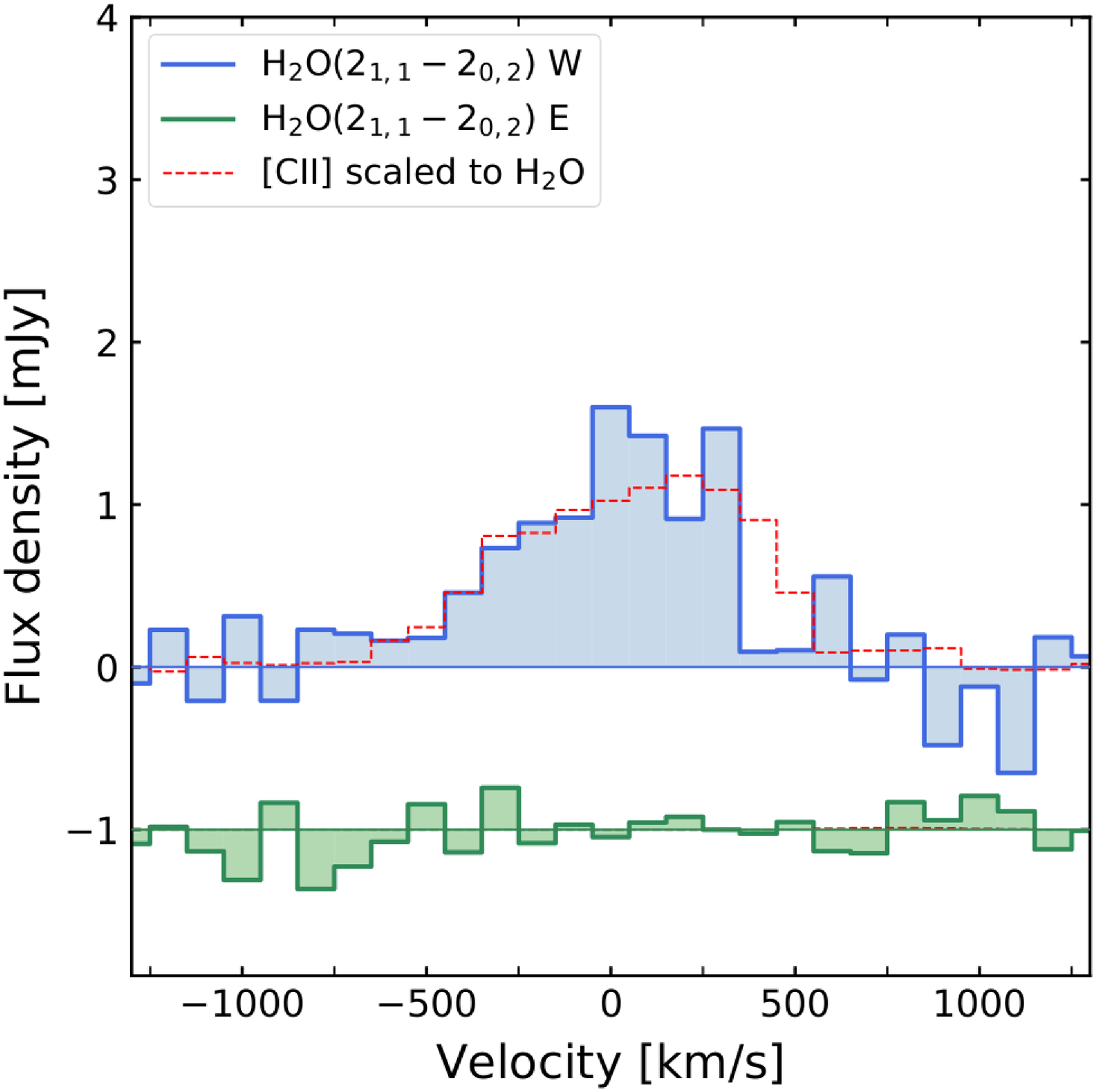}
	\hspace{0.0 cm}
\caption{\textbf{Top:} The continuum is shown as the background in square root scale with minimum pixel value as 3$\rm \sigma_{cont}$, where $\rm \sigma_{cont}$ is the RMS of the continuum map. The moment 0 contours of the lines are overlaid at [3,5,7,9] $\times\ \sigma$ in W and at [3,4] $\times\ \sigma$ in E, where $\sigma$ is the RMS noise in the moment 0 map. Red and blue contours correspond to \coa, \cob, and \coc\ in the W and E sources, respectively. The \ci\ 3$\sigma$ contour (in the W source of second panel) is shown in white and the \watera\ 3$\sigma$ contour  (in the W source of last panel) is shown in magenta. The synthesized beam of the continuum image is shown in the lower left and the 1.0$^{\prime\prime}$ scale bar is at the lower right.  \textbf{Bottom:} The spectrum of each line is shown in W (blue) and E (green), shifted down for clarity. The scaled \cii\ spectrum is shown in red for CO and \watera\ lines and in black for \ci\ and \waterb. The scaling procedure is described in Section \ref{sec:spectrum}.}
\label{spectrumfig}
\end{figure*}

The spectra in W and E are extracted from the \mbox{100 $\rm km\ s^{-1}$} spectral cubes using an aperture of 2.5$\arcsec$ and 1.2$\arcsec$, respectively. 
The continuum subtracted spectrum is shown in Figure \ref{spectrumfig}. Since the spectral lines are non-Gaussian and blended in the case of \cob\ and \cib, and \coc\ and \waterb, we use \cii\ from the ALMA 240 GHz observations \citep{marrone18} as a template to derive the spectral properties. \cii\ is observed at high spatial and spectral resolution with a signal-to-noise ratio of $\sim$56 in W and $\sim$23 in E. We re-imaged the \cii\ data using the same weighting and velocity averaging as described in Section \ref{sec:imaging}. The \cii\ spectrum obtained has a FWHM of 779 $\pm$ 25 $\rm km\ s^{-1}$ in W and 371 $\pm$ 12 $\rm km\ s^{-1}$ in E. This \cii\ spectrum is scaled to the observed spectrum of the lines using a least squares fit. We adopt the standard deviation of the flux densities in line-free channels as the error in each velocity bin. The velocity integrated observed line flux, I$_{\rm obs}$ (\mbox{Jy $\rm km\ s^{-1}$}), is obtained by adding the flux density under the scaled \cii\ spectrum from line center $-$ 3$\sigma_{\nu}$  to line center + 3$\sigma_{\nu}$ $\rm km\ s^{-1}$ where $\sigma_{\nu}$ is FWHM/2.35. We estimate the line properties by repeating the scaling 3000 times with random Gaussian noise added to the flux density and taking the median value of all the fits. The uncertainty on the values is taken as the standard deviation of all the fits. In the case of blended lines, we shift the \cii\ spectrum to the centers of the two lines and perform a joint fit.

Using the velocity integrated line flux densities, we calculate the line luminosities using the equation from \citet{solomon97} as given below:
\begin{align}
\begin{split}
\rm L_{line}  &= \rm (1.04 \times 10^{-3}) \ I_{obs} \ \nu_{rest}\ D_{L}^{2}\ (1+z)^{-1} \\
\rm L^{\prime}_{line} &=  \rm (3.25 \times 10^{7}) \ I_{obs} \ D_{L}^{2} \ (1+z)^{-3} \ \nu_{obs}^{-2}
\end{split}
\end{align}
where $\rm L_{line}$ is the total observed line luminosity in units of $\rm L_{\odot}$, $\rm L^{\prime}_{line}$ is the luminosity in units of $\rm K\ km\ s^{-1}\ pc^{-2}$, $\rm \nu_{rest}$ and $\rm \nu_{obs}$ are the rest and observed frequencies of the line in GHz, and $\rm D_{L}$ is the luminosity distance to the source at a redshift z, in Mpc. The line properties are given in Table \ref{line_prop}.

\begin{table*}
\centering
\caption{Line properties}
\label{line_prop}
\begin{tabular}{c*{4}{>{}p{10em}}}
\hline\hline
Line & Source & I$_{\rm obs}$ & L$_{\rm obs}$ & $\mu$\\
  & & [Jy $\rm km\ s^{-1}$] &  [$\times$ 10$^{8}$ L$_{\odot}$]&  \\ 
\hline
\coa    & W & 2.33 $\pm$ 0.09 & 10.22 $\pm$ 0.39 & 2.08 $\pm$ 0.21 \\
        & E & 0.18 $\pm$ 0.03 & 0.79 $\pm$ 0.13  & 1.3 \\
\cob    & W & 2.15 $\pm$ 0.12 & 11.00 $\pm$ 0.61 & 2.13 $\pm$ 0.14 \\
        & E & 0.17 $\pm$ 0.04 & 0.87 $\pm$ 0.20  &  1.3 \\
\coc    & W & 1.41 $\pm$ 0.10 & 10.30  $\pm$ 0.73& 2.16 $\pm$ 0.11 \\
        & E & 0.14 $\pm$ 0.04 & 1.02 $\pm$ 0.29  & 1.3 \\
\cib     & W & 1.42 $\pm$ 0.12 & 7.28 $\pm$ 0.61  &  2.13 $\pm$ 0.14 \\
        & E & 0.04 $\pm$ 0.04 & 0.21 $\pm$ 0.21  & 1.3 \\
\waterb & W & 0.22 $\pm$ 0.10 & 1.63 $\pm$ 0.74  &  2.2 \\
        & E & 0.03 $\pm$ 0.03 & 0.23 $\pm$ 0.26  &  1.3 \\
\watera & W & 0.97 $\pm$ 0.09 & 4.63 $\pm$ 0.43  &  2.2  \\
        & E & 0.01 $\pm$ 0.04 & 0.05 $\pm$ 0.19 &  1.3 \\
\hline\hline
\multicolumn{5}{p{\textwidth}}{\hspace{+0.2in}{NOTE. - I$_{\rm obs}$ is the observed (not corrected for magnification) integrated flux density in Jy $\rm km\ s^{-1}$ (Section \ref{sec:spectrum}). L$_{\rm obs}$ is the observed line luminosity in L$_{\odot}$. $\mu$ is the magnification of the lines. In W, the CO magnification is obtained from lens modeling using velocity integrated measurement sets (Section \ref{sec:lensmodel}). For the \water\ lines, we adopt $\mu$ = 2.2 in W. For all the lines in E, we adopt $\mu$ = 1.3 \citep{marrone18}. \waterb\ is not detected in W and E. 
\cib\ and \watera\ are not detected in E. The integrated flux densities in these lines are obtained through template fitting procedure described in \mbox{Section \ref{sec:spectrum}}.
}}
\end{tabular}
\end{table*}

\subsection{Lens Modeling} \label{sec:lensmodel}
SPT0311-58 is a gravitationally lensed system. To infer the intrinsic properties of the source, a lensing reconstruction has to be performed. Lens modeling using a pixellated lensing reconstruction technique \citep{hezaveh16} on high resolution (0.3$\arcsec$) rest-frame \mbox{160 $\mu$m}, 110 $\mu$m, 90 $\mu$m continuum and \cii\ gives a magnification of $\rm \mu_{E}$=1.3, $\rm \mu_{W}$=2.2, and  $\rm \mu_{tot}$=2.0 \citep{marrone18}. 
In this analysis, we perform lens modeling on the 140 and 101 GHz ($\sim$272 and \mbox{375 $\mu$m} rest-frame, respectively) continuum, and \coa, \cob, and \coc\ molecular line transitions in W. 
Since the signal-to-noise is not sufficiently high for pixellated modeling, we use a parametric lens modeling  code, \texttt{visilens} \citep{spilker16}. In this code, the background source is parameterized by one or more S$\rm \acute{e}$rsic \citep{sersic68} profiles with seven free parameters: source position relative to the lens ($\rm x_{S},  y_{S}$), flux density (S), half light radius major axis ($\rm a_{S}$), S$\rm \acute{e}$rsic index ($\rm n_{S}$), axis ratio ($\rm b_{S}/a_{S}$), and position angle ($\rm \phi_{S}$). The source profile can be constrained to be circular with $\rm n_{S}$ = 0.5 (Gaussian profile). 
The lensing source is represented by one or more Singular Isothermal Ellipsoids (SIEs), which has five free parameters: the center of the lens relative to phase center ($\rm x_{L}$,  $\rm y_{L}$), the lens mass ($\rm M_{L}$) which determines the strength of lensing, the ellipticity of the lens ($\rm e_{L}$), and the position angle of the lens major axis ($\rm \phi_{L}$) in degrees east of north. This model uses a Markov Chain Monte Carlo (MCMC) algorithm (\texttt{emcee} package; \citealt{foremanmackey13}) to sample the parameter space. The analysis is performed in the visibility plane instead of the image plane to account for the correlated noise in the interferometric images.

For lens modeling in this analysis, we average the data 
to decrease the number of visibilities. The performance of different models is measured by Deviance Information Criterion (DIC; \citealt{spiegelhalter02}) to determine the number of sources to include in the model. As the best-fit parameters, we take the median value of the 1000 MCMC chains with 1$\sigma$ uncertainty. For all the lensing reconstruction in this paper, we vary the lens parameters only within the best-fit values obtained from the high resolution 140 GHz measurement set using a single lens. The lens parameters obtained from the \mbox{140 GHz} data agree with \citet{marrone18} model.
For the lines, we fit single channel models with the width of 1000 $\rm km\ s^{-1}$, which is equivalent to the $\sim$2 $\times$ FWHM to estimate the overall magnification. 
The best-fit parameters for continuum and lines in W are given in the Appendix (Tables \ref{tab:lens_parameters}, \ref{tab:cont_parameters}, and \ref{tab:co_parameters}). Since we do not have a good signal-to-noise for lensing reconstruction in E, we adopt a magnification of 1.3 for all the molecular lines. For the same reason, we adopt $\mu = 2.2$ from \citet{marrone18} in \watera\ in W. The magnification values used in this analysis for continuum and molecular lines are given in Tables \ref{continuum_prop} and \ref{line_prop}, respectively.

\subsection{SED Fitting and Dust Mass}\label{sec:sedfitting}
To estimate \lir\ and \lfir\ in the two sources, we fit a dust SED to the magnification corrected photometry given in Appendix \ref{app:photometry} (Table \ref{tab:photometry}). A 15\% absolute calibration error is added in quadrature to the statistical error to account for the uncertainty from the absolute flux calibration and lens modeling. Here, we fit a simplistic SED model to the dust, and in Section \ref{sec:lvgmodeling} we perform a joint fit of dust, CO, and \ci, based on 
radiative transfer modeling (See Figures \ref{westsed} and \ref{eastsed}). In this section, we assume a uniform dust temperature ($\rm T_{dust}$) in each source and fit a modified blackbody function to the photometry greater than 50 $\mu m$ rest frame and the equations are adopted from \citep{weiss07}:
\begin{equation}
\label{eqn:blackbody}
    \rm S_{\nu} = \frac{\pi r^{2}}{{D_{A}^{2}}} \ \frac{1}{(1+z)^3} \ [B_{\nu}(T_{dust}) - B_{\nu}(T_{CMB})](1-e^{-\tau_{\nu}})
\end{equation}
where B$_{\nu}(\rm T_{dust})$ is the Planck function at dust \mbox{temperature}, $\rm D_{A}$ is the angular diameter distance and r is the effective radius of the source. The optical depth $\tau_{\nu}$ is given by 
\begin{equation}
\label{eqn:optical_depth}
\rm \tau_{\nu} = \frac{\kappa(\nu) \ M_{dust}}{\pi r^{2}}
\end{equation}
where $\rm M_{dust}$ is the dust mass. The frequency dependent dust absorption coefficient is adopted from \citep{draine03} 
\begin{equation}
\rm \kappa_{\nu} = 0.038(\nu/372.7 \ GHz)^{\beta_{T_{d}}} \ [m^{2} \ kg^{-1}] 
\end{equation}
Here, $\beta_{T_{d}}$ is the spectral index which determines the slope of the Rayleigh$-$Jeans tail of the blackbody. We fix $\beta_{T_{d}}$ to 2.0 \citep{greve12}.

In the case of SPT0311-58, we have an estimate of the effective radius of the sources from the 95 GHz and \mbox{140 GHz} lens models. We use MCMC algorithm to sample the dust mass and the dust temperature by giving an upper limit on the effective radius as \mbox{5000 pc} in W and \mbox{1300 pc} in E. 

From the SED modeling in SPT0311-58, we get \mbox{$\rm T_{dust} = 69 \pm 20$ K} in W and \mbox{$49 \pm 9$ K} in E, and an intrinsic \mbox{$\rm M_{dust} = (1.3 \pm 0.5)\times10^{9}\ M_{\odot}$} in W and \mbox{$\rm (1.8 \pm 0.6)\times10^{8}\ M_{\odot}$} in E. We estimate intrinsic \lir\ = (26 $\pm$ 12)$\times \rm 10^{12}\ \rm L_{\odot}$ and \lfir\ = (16 $\pm$ 4)$\times \rm 10^{12}\ \rm L_{\odot}$ in W, and \lir\ = (35 $\pm$ 7)$\times \rm 10^{11}\ \rm L_{\odot}$ and \lfir\ = \mbox{(27 $\pm$ 4)$\times \rm 10^{11}\ \rm L_{\odot}$} in E. These values agree with the luminosities inferred from \texttt{CIGALE} SED fits from \citet{marrone18} to within the uncertainties. 

For comparison with SPT0311-58, we have also fitted SEDs to the SPT-SMG sample and the other literature sources at low and high redshift. In the case of the literature sources, we do not constrain the radius in the SED fit. Instead, we adopt the \mbox{$\rm \lambda_{0}-T_{dust}$} relationship from \citet{spilker16}, where $\lambda_{0}$ is the wavelength at which the optical depth is unity. We then estimate the radius using $\rm \sqrt{\kappa(\nu_{0}) M_{dust}}$, where $\kappa(\nu_{0})$ is the dust absorption coefficient at \mbox{$\nu_{0}=c/\lambda_{0}$}. The MCMC sampling of the dust mass and dust temperature is the same as that followed for SPT0311-58.

\section{Results} \label{sec:results}

\subsection{$L_{H_{2}O} - L_{FIR}$ Correlation}\label{sec:lh2olfir}
Water is the third most abundant molecule in the universe \citep{neufeld95} and its line intensity can be as bright as mid-J CO transitions in high-redshift ULIRGs \citep[e.g.,][]{yang13,omont13}. Multiple \water\ transitions from low-redshift galaxies are detected by \textit{Herschel Space Observatory} while ground based facilities, such as CSO, NOEMA, and ALMA have detected redshifted \water\ transitions from high-redshift galaxies. From these observations, it has been shown that \water\ traces \lir, both in low-redshift galaxies and high-redshift (U)LIRGs \citep{yang13,omont13,yang16,jarugula19}. This can be explained by the excitation mechanism of \water\ molecules where infrared pumping plays a major role \citep[e.g.,][]{gonzalezalfonso10,gonzalezalfonso12}: the higher transitions, such as \watera, p-$\rm H_{2}O (2_{2,0}-1_{1,1})$, and p-$\rm H_{2}O (2_{2,0}-2_{1,1})$ are pumped by the \mbox{101 $\mu$m} far-infrared photons from the base 1$\rm _{1,1}$ level. These lines are mainly found in the warm star-forming regions of galaxies.

\begin{figure}[h]
\hspace{-0.7cm}
\includegraphics[trim={0.0cm 0 0.0cm 0.0},clip,width=0.55\textwidth]{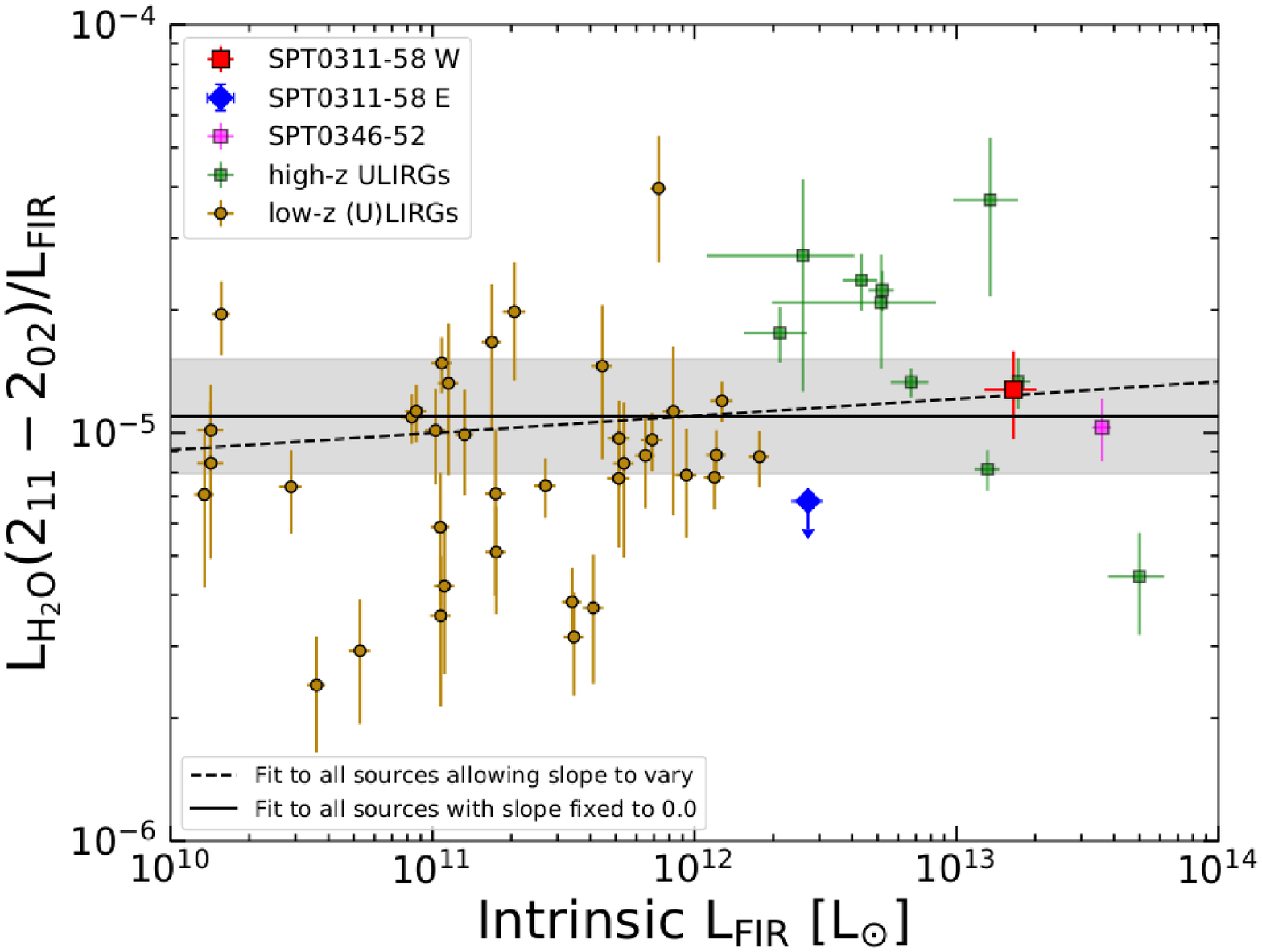}
\caption{
 SPT0311-58 W is shown in red and a 1$\sigma$ upper limit in E is shown in blue. The \water\ observations in local (U)LIRGs are from \citet{yang13} and the high-redshift ULIRGs are from \citet{omont13} and \citet{yang16}. We also include SPT0346-52 observations from \citet{apostolovski19} shown in magenta. The dashed black line shows the slightly super-linear correlation. The fit by fixing the slope to 0.0 is shown as the thick black line and the error as the grey shaded region.
}
\label{lh2olfir}
\end{figure}

We detect \watera\ in W with a signal-to-noise $\sim$4 above the continuum. This is the most distant detection of \water\ in the literature in a galaxy without an AGN. \water\ has been detected in the host galaxy of a quasar candidate at redshift 7 \citep{koptelova19}. 
Since the \watera\ is not significantly detected in E, we give a flux density based on the template fit (see Section \ref{sec:spectrum} for template fitting procedure). We compare the \water\ luminosity (\lwater) in the two galaxies of SPT0311-58 with the local and other high-redshift galaxies as shown in Figure \ref{lh2olfir}. For all the galaxies in the plot, we derive \lfir\ by fitting a modified blackbody function to the photometry. In case of the local galaxies, the continuum flux density values are from \citet{sanders03} and \lwater\ from \citet{yang13}. The photometry and magnification in the high-redshift galaxies are from \citet{weiss07,riechers09,riechers13, bussmann13, cooray14} and \lwater\ from \citet{omont13,yang16}. We additionally include another SPT source, SPT0346-52 from \citet{apostolovski19}, in the plot. The \watera\ luminosity in SPT0311-58 W and E is consistent within the scatter of the other high-redshift ULIRGs.

To get the correlation between \lwater\ and \lfir, we perform two MCMC linear regressions to all the sources in log-log space as shown in Figure \ref{lh2olfir}. We assume that there is no differential magnification between \water\ and continuum emission in the gravitationally lensed sources. In Figure \ref{lh2olfir}, one of the fits fixes the slope of log$_{\rm 10}$(\lwater/\lfir) versus log$_{\rm 10}$(\lfir) to zero (thick black line) and the other allows the slope to vary (dashed line). From the second fit, we see that the relationship is slightly super-linear, which is discussed in the literature \citep[e.g.,][]{omont13,yang16}. One explanation could be the increase in the optical depth at 100 $\mu$m with increasing \lfir, which further enhances the \watera\ emission. A similar super-linear correlation is observed in another \water\ transition,  \mbox{p-$\rm H_{2}O (2_{0,2}-1_{1,1})$}, which is also excited by 100 $\mu$m photons \citep[e.g.,][]{yang16,jarugula19}. The fit with the slope fixed to zero gives:
\begin{equation} \label{eqn:lh2o_lfir}
    \rm \frac{L_{H_{2}O}}{L_{FIR}} = (1.10 \pm 0.4) \times 10^{-5}
\end{equation}

The star formation rate (SFR) is traditionally calculated from \lfir, which is a good tracer of star formation under the assumption that young stars are dust obscured. This assumption holds true in (U)LIRGs. However, to estimate \lfir, the peak of the SED at $\rm \lambda_{rest}=100\ \mu m$ has to be well-sampled, which is observationally expensive. The long-wavelength spectral lines, such as \water, which is bright and well-correlated with \lfir, can be used as an alternative tracer of star formation. We use the $\rm L_{H_{2}O} - L_{FIR}$ correlation to estimate SFR in W and E. The SFR scaling relation from \lir\ is taken from \citet{kennicutt12}:
\begin{equation}\label{eqn:sfr}
    \rm SFR\ [M_{\odot}/yr] = 1.47 \times 10^{-10}\ L_{IR}\ [L_{\odot}]
\end{equation}
We calibrate this SFR with \lwater\ using \lir/\lfir\ values from the SED fit and Equation \ref{eqn:lh2o_lfir}. We use the following equation to estimate SFR from \watera:
\begin{equation}\label{eqn:sfrwater}
    \rm SFR\ [M_{\odot}/yr] = 2.07 \times 10^{-5}\ \lwater\ [L_{\odot}]
\end{equation}
We estimate an intrinsic SFR of (4356 $\pm$ 2143) $\rm M_{\odot}$/yr in W and a 1$\sigma$ upper limit of 385 $\rm M_{\odot}$/yr in E. 

\subsection{Gas Mass from \ci} \label{sec:cigasmass}
The molecular gas mass is traditionally estimated from \coe\ emission line luminosity by assuming a CO-to-$\rm H_{2}$ conversion factor ($\rm \alpha_{CO}$)\citep[e.g.,][]{ bolatto13}. However, $\rm \alpha_{CO}$ can vary significantly depending on the physical environments of the galaxies, such as gas density, temperature, and starburst phase driven by mergers \citep[e.g.,][]{maloney88,narayanan11, bolatto13}. Moreover, it is challenging to observe \coe\ from high redshift galaxies where the CMB temperature is high and can dominate the CO signal. The dependence of CO brightness temperature on the CMB temperature is discussed in the literature \citep[e.g.,][]{dacunha13,tunnard16}.

An alternative molecular gas tracer is \cia\ \mbox{(492 GHz)}, whose luminosity is observed to be linearly correlated with \coe\ luminosity across a wide range of environments both in local and high-redshift galaxies \citep[e.g.,][]{papadopoulos04,alaghbandzadeh13, jiao17,valentino20}. In our current analysis, we have observations of \cib\ (809 GHz), which is brighter than \cia\ and which comes for free with \cob. We estimate gas mass from \cib\ using the equation from \citet{weiss03}\footnote{Note that the coefficient in \citet{weiss03} should be \mbox{4.556 $\times 10^{-4}$}. This correction is mentioned in the footnote in \citet{weiss05}.}. To consider the effect of the background CMB temperature at redshift 6.9, which is $\sim$21.5 K, we include a factor $\rm B_{\nu}(T_{ex})/(B_{\nu}(T_{ex}) - B_{\nu}(T_{CMB}))$ to the equation.

\begin{multline}\label{eqn:gasmassci21}
   \rm  M_{CI} = 4.566 \times 10^{-4} \ Q(T_{ex}) \ \frac{1}{5} e^{62.5/T_{ex}} \\ \rm \frac{B_{\nu}(T_{ex})}{(B_{\nu}(T_{ex}) - B_{\nu}(T_{CMB}))} \ L^{\prime}_{CI(2-1)} \ [M_{\odot}]
\end{multline}
Here, $\rm Q(T_{ex})$ is the partition function, which is given by:
\begin{equation}\label{eqn:qex}
    \rm Q(T_{ex}) = 1 + 3e^{\textendash T_{1}/T_{ex}} + 5e^{\textendash T_{2}/T_{ex}}
\end{equation}
where $\rm T_{1}$ = 23.6 K and $\rm T_{2}$ = 62.5 K are the excitation energy transitions of the two transitions of \ci. The excitation temperature, $\rm T_{ex}$, can be estimated from the \ci\ line ratios \citep{stutzki97}.
However, since we only have observations of \cib, we adopt a typical value of $\rm T_{ex}$ = 30 K \citep{walter11}. 

The gas mass from \cib\ is converted into the total $\rm H_{2}$ gas mass by assuming $\rm [CI]/[H_{2}]$ abundance of \mbox{(8.4 $\pm$ 3.5) $\times 10^{-5}$} \citep{walter11}. To include the contribution from Helium, we multiply the $\rm H_{2}$ gas mass by 1.36. We estimate an intrinsic gas mass of \mbox{(24.9 $\pm$ 10.7)$\times 10^{10} \ \rm M_{\odot}$} in W and \mbox{(11.4 $\pm$ 11.4)$\times 10^{9} \ \rm M_{\odot}$} in E. 
Note that \cib\ is not detected in E and we present the gas mass derived from the flux density obtained by template fitting (see Section \ref{sec:spectrum} for template fitting procedure).


\subsection{CO SLED and AGN Fraction}\label{sec:co109agn}
CO emission lines have traditionally been used as tracers of molecular gas in the local and high-redshift galaxies. \coe\ has a critical density of $\sim$10$^{3}$ cm$^{-3}$ and traces the bulk of the molecular gas in the ISM. However, star formation occurs in dense molecular clouds and this warm dense molecular gas is better traced by mid-J CO lines, such as \coa\ and \cob\ with a critical density of $\sim$10$^{4}$ cm$^{-3}$. A linear correlation has been observed between mid-J CO luminosity (4 $\lesssim$ J$_{\rm up}$ $\lesssim$ 8) and \lir\ in local and high-redshift (U)LIRGs \citep[e.g.,][]{greve14,rosenberg15, lu17,yang17}. The high-J CO transitions (J$_{\rm up}$ $\gtrsim$ 9) require high gas densities ($\ge$ 10$^{5}$ cm$^{-3}$) and high temperatures. Such favourable conditions for high-J CO excitations can arise in the presence of AGN \citep{vanderwerf10}, warm photo-dissociation regions (PDRs), or shocks \citep{mashian15}.

\begin{figure*}
\includegraphics[trim={0.0cm 0 0.0cm 0.0cm},clip,width=1.0\textwidth]{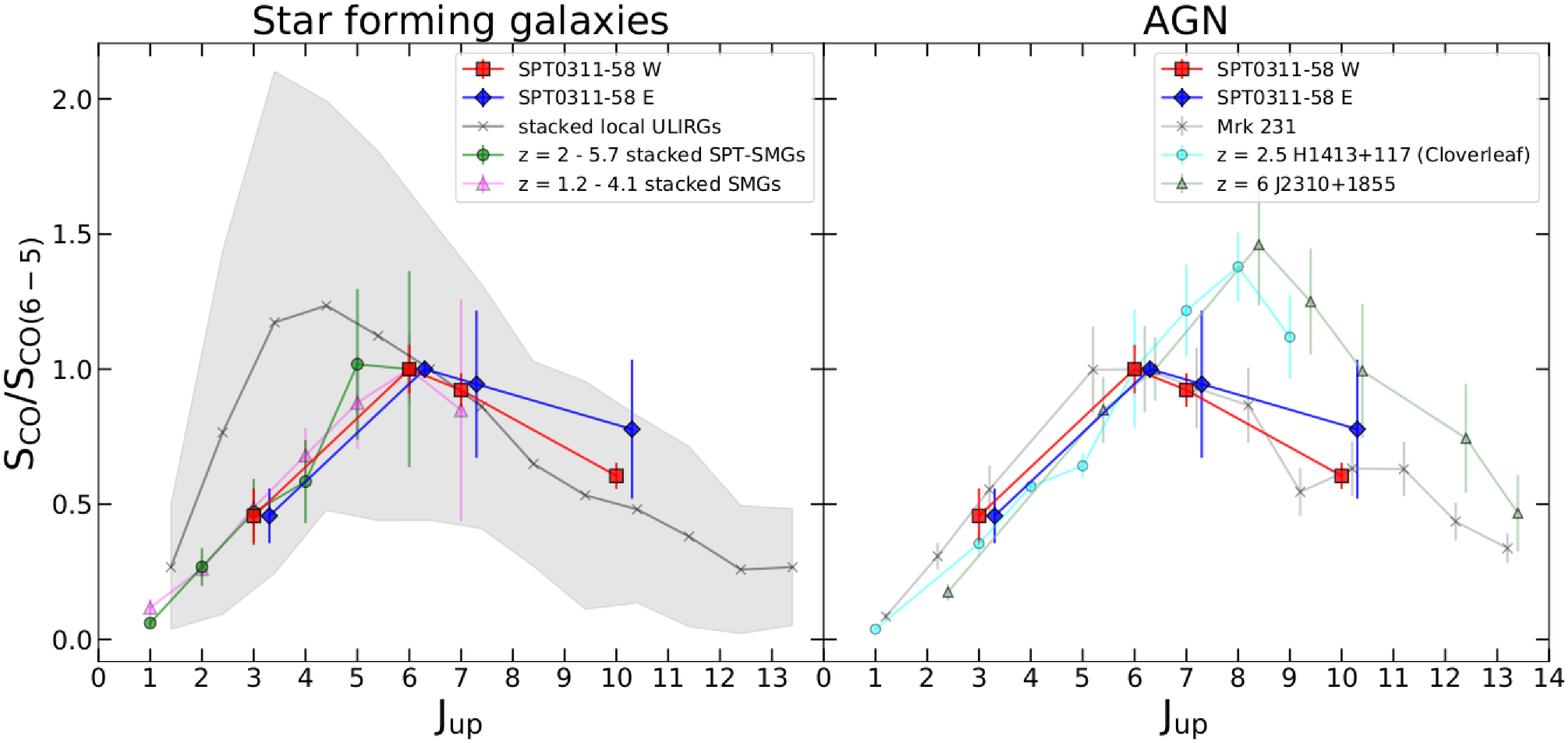}
\caption{The CO SLED normalized by \coa. In the left panel, we compare the two galaxies, W (red) and E (blue), with star-forming galaxies. The local galaxies shown in grey are from \citet{rosenberg15}. The high-redshift SMGs are from \citet{spilker14} (z = 2 $-$ 5.7) and \citet{bothwell13} (z = 1.2 $-$ 4.1 ). In the right panel, we compare W and E with representative local and high-redshift AGN. Mrk 231 observations are from \citet{vanderwerf10}. The high-redshift quasars are Cloverleaf \citep{bradford09, uzgil16} and J2310+1855 \citep{li20}. We observe that the SED in W and E peak at transitions similar to high-redshift star-forming galaxies.
}
\label{fig:COSLED_compare}
\end{figure*}

The CO spectral line energy distribution (SLED) shape provides information about the physical conditions of the molecular gas. In Figure \ref{fig:COSLED_compare}, we compare the SLED shape of the two galaxies of SPT0311-58 with the local and high-redshift starburst galaxies in the left panel, and known AGN in the right panel. All the flux densities are normalized to \coa, which is detected at high signal-to-noise in SPT0311-58 W and E. The local ULIRGs sample includes 29 galaxies taken from \citet{rosenberg15}, excluding \mbox{NGC 6240} which is an outlier due to galaxy wide shocks \citep{meijerink13}. We also compare SPT0311-58 with the average of 22 SPT-SMGs from \mbox{z = 2 $-$ 5.7} \citep{spilker14} and 32 SMGs from \mbox{z = 1.2 $-$ 4.1} \citep{bothwell13}.
The local starburst galaxies peak at J$_{\rm up}\sim$ 4 while the high-redshift sample peaks at mid-J CO (J$_{\rm up}$ $\sim$ 6) transitions. In both the local and high-redshift galaxies, there is a drastic decrease in the high-J CO emission at J$_{\rm up}$ $\gtrsim$ 8. In the right panel of Figure \ref{fig:COSLED_compare}, we see that Mrk231 peaks at J$_{\rm up}$ $\sim$ 5 \citep{weiss05}, higher than local starburst galaxies, and the CO emission can be explained by an XDR model detailed in \citet{vanderwerf10}. The high-redshift quasars, Cloverleaf at \mbox{z = 2.56} \citep{bradford09} and J2310+1855 at \mbox{z = 6.0} \citep{li20}, peak at \mbox{J$_{\rm up}$ $\sim$ 9} and the CO emission is luminous even after J\mbox{$_{\rm up}$ $\gtrsim$ 10}. Both the galaxies in SPT0311-58 peak at \mbox{J$_{\rm up}$ $\sim$ 6}, similar to other high-redshift starburst galaxies.


\begin{figure}[h!]
\hspace*{-0.8cm}
\includegraphics[trim={0.0cm 0 0.0cm 0.0cm},clip,width=0.55\textwidth]{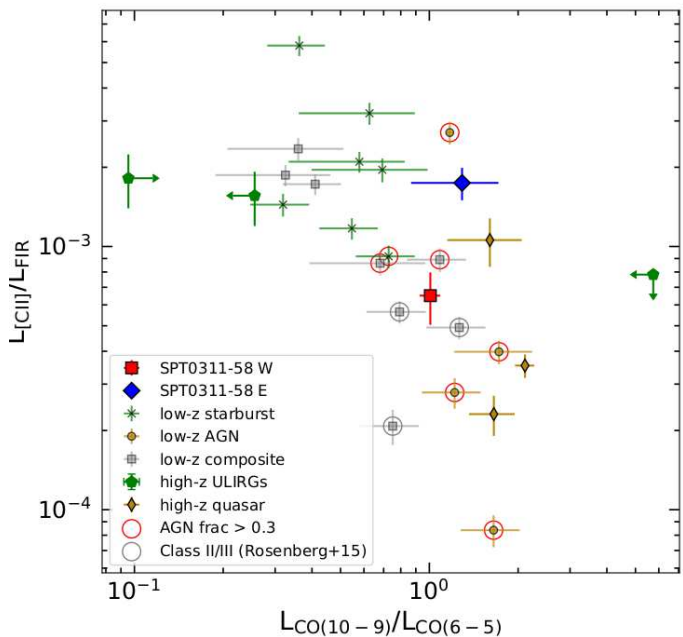}
\caption{The bottom-right corner of the plot with enhanced higher-J CO emission and lower \cii\ emission, might correspond to heating from X-rays and/or mechanical heating from processes such as AGN activity, stellar winds, or supernovae. We compare the two galaxies, W (red) and E (blue), with starburst, composite, and AGN at local and high redshift. Local galaxies with AGN fraction $>$ 0.3 are indicated by red circles and other Class II or Class III objects (high ratio of high-J CO to mid-J CO emission) are shown by grey circles \citep{rosenberg15}. The local galaxy observations are detailed in \citet{diazsantos14,rosenberg15,mashian15} while the high-redshift sample is from \citet{danielson11,frayer11,yang17,zhang18,andreani18,li19,yang19b,shao19,wang19,rybak20}. The position of \mbox{SPT0311-58} E and W on the plot indicates that there might be additional heating mechanisms beyond photoelectric heating. }
\label{fig:AGNfraction}
\end{figure}

We use the ratio of high-J CO to mid-J CO transitions (\coc/\coa) to explore the presence of heating mechanisms in addition to photoelectric heating in SPT0311-58. The ratio of high-J to mid-J CO has been used in literature to characterize starburst and AGN activity \citep[e.g.,][]{rosenberg15,mashian15}, which defines the drop-off slope of the CO SLED. In Figure \ref{fig:AGNfraction}, we explore the correlation between $\rm L_{[CII]}/L_{FIR}$, $\rm L_{CO(10-9)}/L_{CO(6-5)}$, and the presence of AGN. We compare the two galaxies of SPT0311-58 with low-redshift starburst, AGN, and composite galaxies and high-redshift SMGs, and quasars. The low-redshift \cii\ sample is from \citet{diazsantos14} and the CO observations are from \citet{rosenberg15,mashian15}. The AGN fraction in the local sample is detailed in \citet{rosenberg15}. The CO and \cii\ observations in the high-redshift galaxies are from \citet{danielson11,frayer11,yang17,zhang18,andreani18,li19,yang19b,shao19,wang19,rybak20}. 

\cii\ is a dominant cooling line in the ISM and has been widely studied. It has been observed that $\rm L_{[CII]}/L_{FIR}$ decreases with increasing \lfir\ \citep[e.g.,][]{malhotra01,brauher08,diazsantos13,gullberg15,litke19}. Several mechanisms have been proposed to explain this \cii\ deficit such as positively charged dust grains in the presence of ionizing UV photons, which reduce the photoelectric heating (UV heating) efficiency, saturation of \cii\ emission in PDR regions, and self absorption \citep[e.g.,][]{narayanan14,munoz16,narayanan17}. If the \cii\ deficit is due to the reduced photoelectric heating, this effect will also result in reduced $\rm L_{CO(10-9)}/L_{CO(6-5)}$ since the main collision partner, $\rm H_{2}$, is heated by electrons deep within the molecular clouds in a pure PDR context. However, in Figure \ref{fig:AGNfraction}, we observe that $\rm L_{CO(10-9)}/L_{CO(6-5)}$ is increasing with decreasing $\rm L_{[CII]}/L_{FIR}$. This indicates that other heating mechanisms (e.g., mechanical processes), in addition to the photoelectric effect, result in the observed trend or that the \cii\ deficit is probably not due to the effect of reduced photoelectric heating efficiency. The presence of AGN can also decrease \cii\ emission due to possible destruction of dust grains \citep{smith17}. There could also be an apparent deficit in \cii/FIR due to contribution to the total infrared emission from AGN. However, it has been observed that this effect is not significant \citep{diazsantos13}.
As discussed in \citet{rosenberg15}, heating mechanisms in addition to photoelectric heating are required to explain the high ratio of high-J to mid-J CO emission.  

In Figure \ref{fig:AGNfraction}, reduced \cii\ emission and increased \coc\ might indicate X-ray heating from AGN and/or mechanical heating from AGN or other sources, such as stellar winds, mergers, or supernovae explosions in addition to photoelectric heating. We observe that, in Figure \ref{fig:AGNfraction}, the known AGN and other Class II or III composite galaxies (galaxies with high or low AGN fraction and a high ratio of high-J CO to mid-J CO emission; \citealt{rosenberg15}) are mostly separated from starburst galaxies.  Both the galaxies of SPT0311-58 have a higher $\rm L_{CO(10-9)}/L_{CO(6-5)}$ comparable to AGN and Class II or III galaxies. W has a lower $\rm L_{[CII]}/L_{FIR}$ ratio possibly due to increased star formation density or higher metallicity than E \citep{marrone18}. We conclude that there is a possibility of other heating mechanisms in addition to photoelectric heating in SPT0311-58, but we cannot infer the presence of AGN. We further discuss mechanical heating from shocks and stellar feedback in detail in Section \ref{sec:energybudget}.


\subsection{Radiative Transfer Modeling} \label{sec:lvgmodeling}
To model the dust emission, observed CO, and \ci\ flux densities in SPT0311-58 W and E, we use two non-local thermodynamic equilibrium (non-LTE) large velocity gradient (LVG) radiative transfer modeling methods: \textit{N-component} and \textit{Turbulence}. The first model is described in detail in \citet{weiss07} and an updated version in \citet{strandet17}. The second modeling method, based on the equations from \citet{weiss07}, is summarized in \citet{harrington20}.

In the radiative transfer models, we model the excitation of the dust continuum, CO and \ci\ flux densities simultaneously. 

Dust continuum: We model the dust continuum using a modified blackbody function shown in Equation \ref{eqn:blackbody}. The dust optical depth (Equation \ref{eqn:optical_depth}) depends on the dust mass and the dust absorption coefficient. In the radiative transfer models, we adopt $\rm \kappa_{\nu}/m^{2}\ kg^{-1}\ = 0.04 \times (\nu/250\ GHz)^{\beta_{T_{d}}}$. The dust mass is obtained from the gas mass by using the gas-to-dust mass ratio which is a free parameter in the models.

Line emission:  For the CO and \ci\ line emission modeling, the infrared radiation is considered as background radiation in addition to the CMB radiation at z=6.9. The line flux density is modeled by:
\begin{equation}
    \rm S_{CO/[CI]} = T_{b} \ 2k \ \nu_{obs}^{2} \ \Omega / \ [ \ c^{2} \ (1+z) ]
\end{equation}
where c is the speed of light, $\rm \nu_{obs}$ is the observed frequency of the CO or \ci\ lines and $\rm \Omega$ is the source solid angle given by $\rm \pi r^{2}/D_{A}^2$. The brightness temperature of the line, $\rm T_{b}$, is calculated using the excitation temperature of the line, background temperature, dust temperature, and the dust opacity. These are further dependent on the gas number density, kinetic temperature, the gas phase abundance per velocity gradient and the gas-to-dust mass ratio. All these parameters are discussed in detail in this section.

In this section, we also highlight the differences between the two models and present the best-fit and derived parameters obtained from the radiative transfer modeling.\\

\subsubsection{N-component and Turbulence Models}\label{2comp-turb}
Both the \textit{N-component} and \textit{Turbulence} codes model the dust, CO, and \ci\ flux densities simultaneously. The \textit{N-component} model is the more basic of these two codes and can describe the ISM using N components. However, due to the increased degeneracies with increased number of free parameters, we use 1 and 2 components in this analysis. This model will henceforth be referred to as \textit{1-component} and \textit{2-component}, which has 10 and 20 input parameters, respectively (see Table \ref{ncomp_parameters} in the Appendix). The \textit{Turbulence} model realistically models the ISM with 12 input parameters (see Table \ref{turb_parameters}).

One of the main differences between the \textit{Turbulence} and \textit{N-component} models is that, in the \textit{Turbulence} code, the ISM is modelled with the dependence of the source solid angle ($\Omega$) on the gas volume density (n$_{\rm H_{2}}$, normalized to the mean density). This is given by:
\begin{equation}
    \rm \Omega = \Omega_{s} \ x  \ dp(n^{\prime}, dv_{turb})
\end{equation}
where $\rm \Omega_{s}$ is the solid angle of the total emission region. $\rm dp(n^{\prime}, dv_{turb})$ is the log-normal probability distribution function (PDF) in supersonically isothermal turbulent gas as given in \citet{krumholz05}, which depends on the normalized gas density ($\rm n^{\prime}$) and the turbulence line width ($\rm dv_{turb}$). The normalized gas density is given by $\rm n^{\prime} = n_{H_{2}} / <n_{H_{2}}>$ where $\rm <n_{H_{2}}>$ is the mean gas density. The PDF is given by:
\begin{equation}
    \rm dp(n^{\prime}, dv_{turb}) = \frac{1}{\sqrt{(2 \pi \sigma^{2})}} \ x \ \frac{1}{n^{\prime}} e^{ [ - log(n^{\prime})  / 2 \sigma^{2} ]}
\end{equation}
where the spread of the PDF, sigma is defined as $\rm \sigma = [log ( 1 + \frac{3}{4}  Mach^{2}) ] ^{1/2}$ and 
\mbox{$\rm Mach = dv_{turb} / [\sqrt{(k \ T_{K}) / 2 \ m_{H}}]$} where $\rm m_{H}$ is the proton mass and $\rm T_{K}$ is the kinetic temperature.
For each value of $\rm < n_{H_{2}}>$, we sample 50 gas densities within $\rm 10 - 10^{10} \ cm^{-3}$. The final dust and gas SED, which describes the observations, is the sum of all the 50 SEDs.

In the \textit{Turbulence} model, the kinetic temperature (T$_{\rm K}$) is coupled to the molecular gas density as \mbox{T$_{\rm K}$ $\propto$ (n$_{\rm H_{2}})^{\rm \beta_{T_{K}}}$} where $\rm \beta_{T_{K}}$ is a negative power-law index. Chemical modeling and simulations have shown that at gas densities $\lesssim$ 10$^{5}$ cm$^{-3}$, the temperature slightly decreases with increasing densities. This correlation is because photoelectric and cosmic ray heating has a linear density dependence, but cooling due to CO and \cii\  has a super linear relationship \citep{larson05,meijerink07,krumholz14}, which leads to an overall cooling per unit mass in this regime. In addition to T$_{\rm K}$, the \ci\ abundance relative to H$_{2}$ is also coupled to the gas density as a power-law with a negative index, $\rm \beta_{[CI]}$. At gas densities $\rm > 10^{4} \ cm^{-3}$, the \ci\ line ratio, \cib/\cia, is shown to increase in the PDR models \citep[e.g.,][]{meijerink07}. However, to reproduce the subthermally excited \ci\ gas, i.e., lower \cib/\cia, as typically observed, it is therefore required to reduce the \ci\ abundance in the dense gas. Since, we do not have observations of both the \ci\ lines, this is of no importance in the current analysis.
The \textit{N-component} code, on the other hand, considers $\Omega$ and n$_{\rm H_{2}}$ to be independent and there is no explicit dependence of T$_{\rm K}$ and [CI] abundance on the gas density.

To constrain the \textit{2-component} model, which has more free parameters than \textit{1-component} and \textit{Turbulence}, we consider the maximum value of CO(13$-$12) from  \citet{rosenberg15} scaled to \coc\ in W and E as upper limits. We also include these upper limits in the \textit{1-component} and \textit{Turbulence} models, although the results do not change significantly when the limits are not included. 

The models are optimized to fit the observed flux densities of dust and gas using the Monte Carlo Bees Algorithm \citep{pham09}. To briefly summarize, the algorithm explores the parameter space within the given ranges by estimating $\chi^{2}$ for a few different models referred to as ``bees''. Extra bees are assigned to regions of best $\chi^{2}$ while the rest of the space continues to be explored by the other bees. We run the code 30 times to avoid artificially narrow PDFs for the solutions and hence have 30 final solutions. The best solution is the one with the best $\chi^{2}$.
We evaluate $\sim$10$^{7}$ models and the best-fit \textit{Turbulence} model parameter values shown in Table \ref{turb_parameters} are obtained by taking the mean and standard deviation from all the model runs. The best-fit \textit{1-component} and \textit{2-component} parameters are shown in Table \ref{ncomp_parameters} in the Appendix. 


\subsubsection{Model Parameters}
The free parameters and the corresponding ranges given as inputs in the \textit{Turbulence} model are shown in Table \ref{turb_parameters}.  We use the same ranges for all the parameters in the \textit{N-component} model, except $\rm \beta_{T_{K}}$, and $\rm \beta_{[CI]}$, since there is no coupling between T$_{\rm K}$ and [CI] abundance with the gas density in this model. In this analysis, we run two models. The first one is exploring a range of gas-to-dust mass ratio (GDMR) and CO abundance in SPT0311-58 W and E, corresponding to metallicity less than solar metallicity. The second model is by assuming solar metallicity and constraining the GDMR and CO abundance accordingly.
In starburst galaxies at high redshift, it is possible for the ISM to be enriched to solar metallicity \citep{novak19,debreuck19}. The parameters which are given as inputs to the models are discussed below:\\


\textit{Gas volume density (log(n$_{\rm H_{2}}$))}: We consider a wide range of gas densities, from $\rm 10^{1} - 10^{10} cm^{-3}$ in the \textit{Turbulence} model to sample the density PDF. The mean density of the PDF is sampled from $\rm 10^{1} - 10^{7} cm^{-3}$ and we use the same range in \textit{N-component} model.\\
\textit{Gas kinetic temperature (T$_{\rm K}$) and Dust temperature (T$_{\rm dust}$)}: 
The gas kinetic temperature is coupled to the gas volume density in the \textit{Turbulence} model as mentioned in Section \ref{2comp-turb}. We explore T$_{\rm K}$ with the CMB temperature at redshift 6.9 ($\sim$21 K) as the lower limit. The dust temperature is weakly coupled to the kinetic temperature through a free parameter, T$_{\rm K}$/T$_{\rm dust}$, which allows for additional heating mechanisms. We limit this parameter to 0.5 $-$ 6 in the models. \\
\textit{Kinetic temperature power law index ($\rm \beta_{T_{K}}$)}: As discussed in Section \ref{2comp-turb}, the kinetic temperature is expected to decrease with an increase in density below a certain threshold value. From theory and observations of nearby low mass and low density galaxies, \citet{larson05} gives the equation of state for this relationship where $\rm \beta_{T_{K}}$ = -0.27. In this analysis of high-redshift galaxies, we explore a wide range between -0.5 and \mbox{-0.05}. \\
\textit{Dust emissivity ($\rm \beta_{T_d}$)}:  We explore the dust spectral emissivity index within a range of 1.5 $-$ 2.0, which is consistent with the observations from high-redshift dusty star-forming galaxies \citep[e.g.,][]{conley11,casey14}.\\
\textit{Effective radius (R$_{\rm eff}$)}: This radius defines the source solid angle of the emitting region, which normalizes the density PDF in the \textit{Turbulence} model. 
In W and E, we give an upper limit of 5 kpc. This is larger than the size we obtain from lens modeling because low-J CO is more diffuse than mid and high-J CO, giving rise to a large radius. CO gas sizes are also found to be larger than the infrared emission \citep[e.g.,][]{spilker15,dong19,apostolovski19}.\\
\textit{Virial parameter ($\rm \kappa_{vir}$)}: The velocity gradient, which determines the escape velocity of the gas is coupled to the gas volume density through the virial parameter \citep{goldsmith01} as
\begin{equation}\label{eqn:vdvr}
   \rm  dv/dr = 3.1 \ \kappa_{vir} \sqrt{\frac{n_{H_{2}}}{10^{4}}}
\end{equation}
We explore a range of virial parameter from 1 $-$ 3 where $\rm \kappa_{vir}$ = 1 corresponds to virialized gas and $>$ 1 corresponds to unbound motions \citep{greve09}.\\
\textit{Turbulence line width (dv$_{\rm turb}$)}: The turbulence line width along the line of sight, in addition to the other free parameters, determine the gas mass including the contribution from Helium \citep{weiss07}, which is given by:
\begin{equation}\label{gasmasseqn}
    \rm M_{gas} = 1.36 \ \Omega \ \frac{n_{H_{2}}}{dv/dr} \ dv_{turb}
\end{equation}
dv$_{\rm turb}$/(dv/dr) is the equivalent path-length of the molecules. The velocity profile of CO (from the lens modeling in this analysis) and \cii\ \citep{marrone18} show a velocity gradient across SPT0311-58 W, which could indicate either rotation of the galaxy or a complicated merger pattern. The observed line width ($\sim$1000 km s$^{-1}$ in all the CO and \water\ lines in W) is a combination of both galaxy rotation and random motions due to turbulence. In this analysis, we explore a range from \mbox{5 $-$ 200 km s$^{-1}$} for the turbulence velocity line width. \\
\textit{Gas-to-Dust mass ratio (GDMR)}: This parameter is used to calculate the dust mass from the gas mass given in Equation \ref{gasmasseqn}. Observations and models suggest that GDMR has a dependence on metallicity where it increases with decreasing metallicity  \citep[e.g.,][]{sandstrom13,leroy11,li19}. We run two models: the first one assumes metallicity less than or similar to solar metallicity and explores the GDMR in the range of 90 $-$ 1100. In the second model, we use GDMR in the range of 120 $-$ 150 assuming the galaxies have enriched Milky Way metallicity \citep[e.g.,][]{draine07,elia17}. In starburst systems, dense regions can build up metals relatively early and approach solar metallicity \citep{cen99,novak19, debreuck19}. We fix the GDMR assuming solar metallicity to better constrain the models. 
To estimate the dust mass in the LVG models from the gas mass and GDMR, we adopt 
$\rm \kappa_{\nu}/m^{2}\ kg^{-1}\ = 0.04 \times (\nu/250\ GHz)^{\beta_{T_{d}}}$.\\
\textit{CO abundance ($\rm [CO/H_{2}$])}: We explore a range of 1.0$\times$10$^{-7}$ $-$ 2.0$\times$10$^{-4}$ for the CO abundance assuming metallicity less than or similar to solar metallicity. Under the assumption of solar metallicity conditions, we use the abundance in the Milky Way and nearby giant molecular clouds (GMCs) which is in the range of 1.0$\times$10$^{-4}$ $-$ 2.0$\times$10$^{-4}$ \citep[e.g.,][]{blake87,kulesa02}. \\
\textit{\ci\ abundance ($\rm [CI/H_{2}]$)}: \ci\ abundance is explored in the range of 1.0$\times$10$^{-7}$ $-$ 1.0$\times$10$^{-4}$. This range includes the \ci\ abundance values from GMCs \citep[e.g.,][]{simon15,fuente19}.\\
\textit{\ci\ abundance power-law index ($\rm \beta_{CI}$)}: As mentioned in Section \ref{2comp-turb}, the \ci\ abundance decreases with increasing gas density. This is modelled as a power-law where $\rm \beta_{CI}$ is explored in a wide range of -5 $-$ 0.

Throughout the paper, we refer to the \textit{Turbulence}, \textit{2-component}, and \textit{1-component} models as the models run by assuming that SPT0311-58 has metallicity which is less than the solar metallicity. In the case of models run with enriched solar metallicity assumption, we specify the assumption explicitly.

\begin{table*}
\caption{Turbulence model parameters}
\label{turb_parameters}
\begin{adjustwidth}{-2.0cm}{}
\begin{tabular}{c @{\hskip 0.05in}c@{\hskip 0.05in} c @{\hskip 0.05in}c@{\hskip 0.1in} c@{\hskip 0.1in} c@{\hskip 0.1in} c}
\hline\hline

Parameter &     unit &              range  &      West  &  West  &   East & East\\
 & & & & [Z$_{\odot}$] & & [Z$_{\odot}$] \\
\hline
Model input\\
\hline
log(n$_{\rm H_{2}}$) & cm$^{-3}$ & 1 $-$ 7  &  5.3 $\pm$ 1.3  & 5.1 $\pm$ 0.9  & 4.7 $\pm$ 2.4 & 4.4 $\pm$ 2.2 \\
T$_{\rm K}$ & K & 21  $-$ 600               &  116 $\pm$ 42 & 90 $\pm$ 24 & 166 $\pm$ 65 & 136 $\pm$ 57\\
T$_{\rm K}$/T$_{\rm dust}$ & - & 0.5 $-$ 6.0 &  2.0 $\pm$ 0.9 & 1.6 $\pm$ 0.4   & 3.4 $\pm$ 1.3 & 2.7 $\pm$ 1.1 \\
$\rm \beta_{T_{K}}$ & - & -0.5 $-$ -0.05     &  -0.1 $\pm$ 0.0 & -0.1 $\pm$ 0.0  & -0.1 $\pm$ 0.0 & -0.1 $\pm$ 0.0\\
$\rm \beta_{T_{dust}}$ & - & 1.5 $-$ 2.0        &  1.8 $\pm$ 0.1 &  1.9 $\pm$ 0.1  & 2.0 $\pm$ 0.0 & 2.0 $\pm$ 0.0\\
R$_{\rm eff}$ & pc & 0.1 $-$ 5000           &  3450 $\pm$ 750 & 3342 $\pm$ 628   & 2548 $\pm$ 954 & 2055 $\pm$ 870\\
$\rm \kappa_{vir}$ & $\rm km\ s^{-1}$ pc$^{-1}$ cm$^{3/2}$ & 1 $-$ 3 & 1.4 $\pm$ 0.4 & 1.3 $\pm$ 0.4 & 1.7 $\pm$ 0.6 & 1.5 $\pm$ 0.6\\
dv$_{\rm turb}$ & $\rm km\ s^{-1}$ & 5 $-$ 200   &  134 $\pm$ 46  & 127 $\pm$ 43  & 55 $\pm$ 61 & 71 $\pm$ 61\\
GDMR & - & 90 $-$ 1100                       &  124 $\pm$ 62  & -  & 138 $\pm$ 95 & -\\
GDMR [Z$_{\odot}$] & & 120 $-$ 150 & - & 129 $\pm$ 9 & -  & 130 $\pm$ 10\\
$\rm [CO/H_{2}]$ &- & 1.0$\times$10$^{-7}$ $-$ 2.0$\times$10$^{-4}$&  (7 $\pm$ 5)$\times$10$^{-5}$ & -  & (7$\pm$ 5)$\times$10$^{-5}$ & - \\
$\rm [CO/H_{2}]$ [Z$_{\odot}$] & & 1.0$\times$10$^{-4}$ $-$ 2.0$\times$10$^{-4}$& - & (14 $\pm$ 3)$\times$10$^{-5}$ & - & (14 $\pm$ 3)$\times$10$^{-5}$ \\
$\rm [CI/H_{2}] $ &- & 1.0$\times$10$^{-7}$ $-$ 1.0$\times$10$^{-4}$ & (3 $\pm$ 2)$\times$10$^{-5}$ &  (5 $\pm$ 2)$\times$10$^{-5}$    & (3 $\pm$ 2)$\times$10$^{-5}$ & (4 $\pm$ 2)$\times$10$^{-5}$ \\
$\rm \beta_{[CI]}$ & - & -5 $-$ 0            &  -0.8 $\pm$ 1.1 & -0.7 $\pm$ 0.9  & -2.5 $\pm$ 1.5 & -2.3 $\pm$ 1.6\\
\hline
Estimated \\
within the model\\
\hline
T$_{\rm dust}$ & K &                        &  52 $\pm$ 5 & 55 $\pm$ 5  & 48 $\pm$ 4 & 51 $\pm$ 4\\
M$_{\rm gas}$ & M$_{\odot}$ &               &  (5.4 $\pm$ 3.4)$\times$10$^{11}$ &  (4.5 $\pm$ 1.8)$\times$10$^{11}$    & (3.1 $\pm$ 2.7)$\times$10$^{10}$ & (2.6 $\pm$ 0.7)$\times$10$^{10}$\\
M$_{\rm dust}$ & M$_{\odot}$ &              &   (4.3 $\pm$ 3.5)$\times$10$^{9}$ & (3.5 $\pm$ 1.4)$\times$10$^{9}$   &  (2.2 $\pm$ 2.5)$\times$10$^{8}$ & (2.0 $\pm$ 0.5)$\times$10$^{8}$\\
\hline
Derived \\
from the model\\
\hline
$\rm L^{\prime}_{CO}$ & $\rm K \ km\ s^{-1} pc^{2}$ &  & (8 $\pm$ 1)$\times$10$^{10}$ & (9 $\pm$ 1)$\times$10$^{10}$ & (9 $\pm$ 2)$\times$10$^{9}$ & (11 $\pm$ 2)$\times$10$^{9}$\\
$\rm \alpha_{CO}$ & $\rm M_{\odot}/K \ km\ s^{-1} pc^{2} $ &  & 7.1 $\pm$ 5.3 &  5.3 $\pm$ 2.7  & 3.9 $\pm$ 4.4 & 2.5 $\pm$ 0.8\\
L$_{\rm FIR}$ & L$_{\odot}$ &  & (19 $\pm$ 1)$\times$10$^{12}$ & (19 $\pm$ 1)$\times$10$^{12}$ & (31 $\pm$ 1)$\times$10$^{11}$ & (31 $\pm$ 2)$\times$10$^{11}$\\
SFR & M$_{\odot}$ yr$^{-1}$ &  &  5046  $\pm$  944 & 5043  $\pm$  949 & 701 $\pm$ 151 &  708 $\pm$ 151\\
t$_{\rm dep}$ & Myr &                           &  107 $\pm$ 70 & 90 $\pm$ 40 & 44 $\pm$ 40 & 36 $\pm$ 12  \\

\hline\hline
\multicolumn{7}{p{\textwidth}}{\hspace{+0.2in}{NOTE. - The input and derived parameters of the \textit{Turbulence} model with units and the explored range. The model outputs are the intrinsic source properties as we use the magnification corrected photometry and line flux densities for modeling. There are 12 input parameters in the model which are explored within the given range.
We present two \textit{Turbulence} models: the first one is modeled by assuming less than solar metallicity and varying the GDMR and the CO abundance. In the second model, to get better constrains on the parameters, we assume solar metallicity in the galaxies ([Z$_{\odot}$]) and fix GDMR and CO abundance accordingly. 
The best-fit parameters from the \textit{Turbulence} model in SPT0311-58 West and East are shown with and without assuming solar metallicity. 
}}
\end{tabular}
\end{adjustwidth}
\end{table*}

\subsubsection{Model Outputs}
\label{sec:modelout}
\begin{figure*}[h!t] 
\hspace*{0.0cm} 
\includegraphics[trim={0.0cm 0 0.0cm 0.0},clip,width=1.0\textwidth]{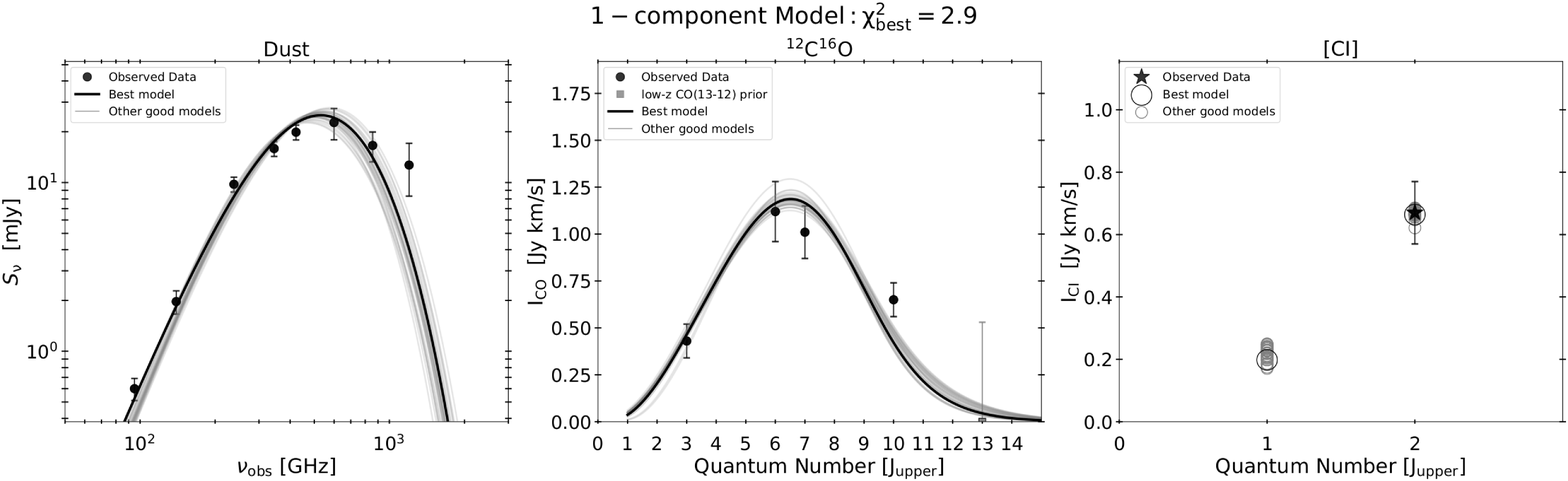}\\
\hspace*{0.0cm} 
\includegraphics[trim={0.0cm 0 0.0cm 0.0},clip,width=1.0\textwidth]{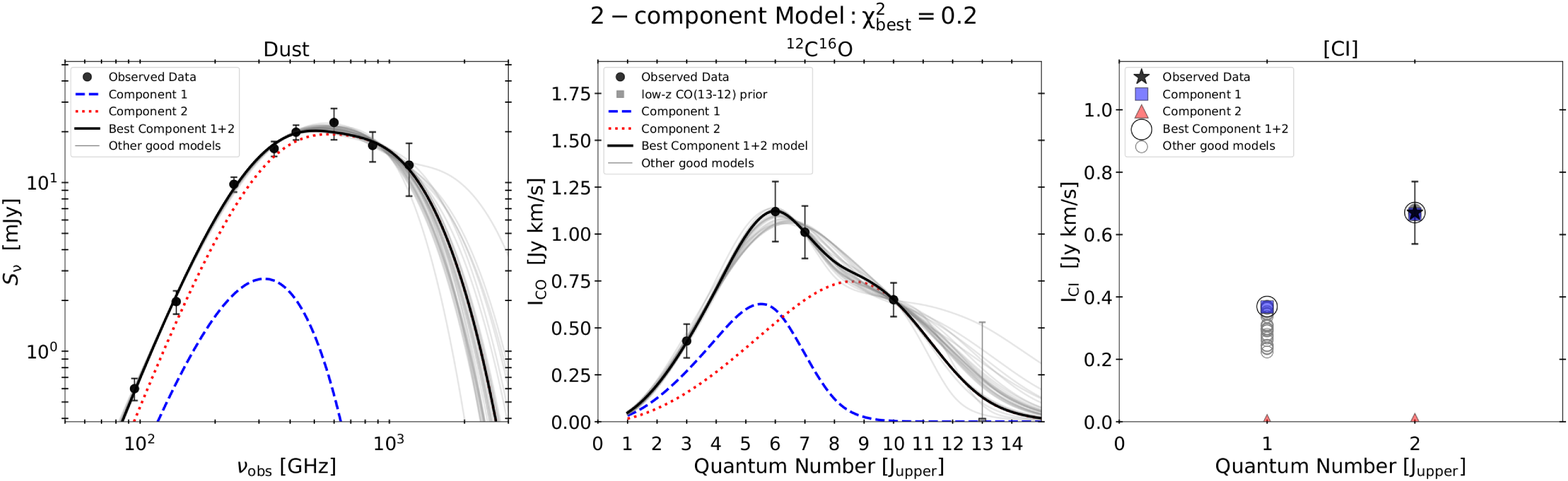}\\
\hspace*{0.0cm}
\includegraphics[trim={0.0cm 0 0.0cm 0.0},clip,width=1.0\textwidth]{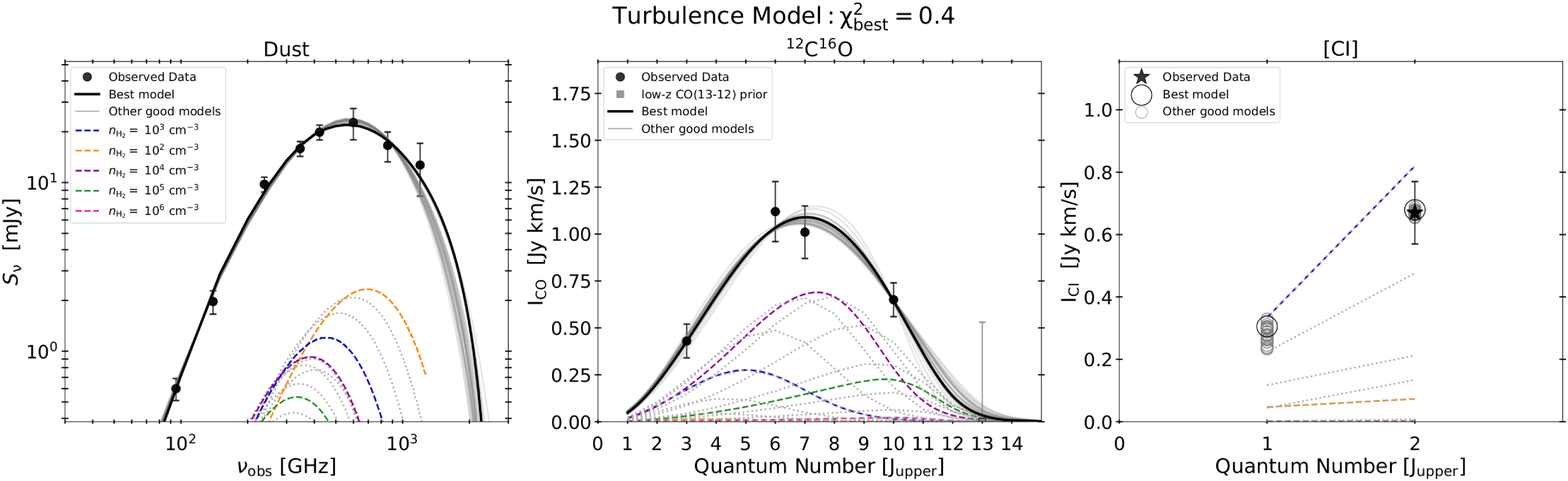}\\
\caption{\textbf{Top:} Dust continuum, CO, and \ci\ SED in W from the \textit{1-component} model. The observed data is shown as black points with error bars. The best fit SED obtained from the model is shown as the thick black line and a the 30 other good models in grey.  The best-fit model in \ci\ is shown as a black circle. \textbf{Middle:} The SED from the \textit{2-component} model in W. The contribution from component 1 is shown in blue and component 2 in red. \textbf{Bottom:} The SED from the \textit{Turbulence} model. We present the contribution to the final SED from the ISM at 5 different gas number densities shown in colored dashed lines. The remaining densities (from the 50 samples) are shown as dashed grey SEDs.
}
\label{westsed}
\end{figure*}

\begin{figure*}[h!t]
\hspace*{0.0cm} 
\includegraphics[trim={0.0cm 0 0.0cm 0.0},clip,width=1.0\textwidth]{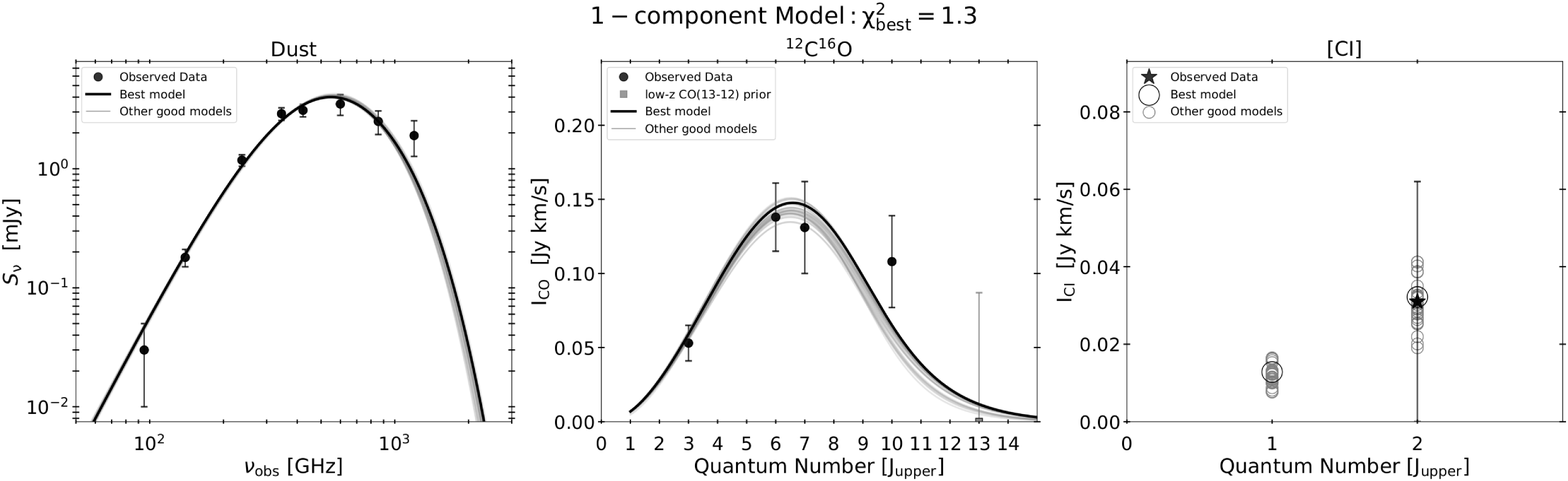}\\
\hspace*{0.0cm} 
\includegraphics[trim={0.0cm 0 0.0cm 0.0},clip,width=1.0\textwidth]{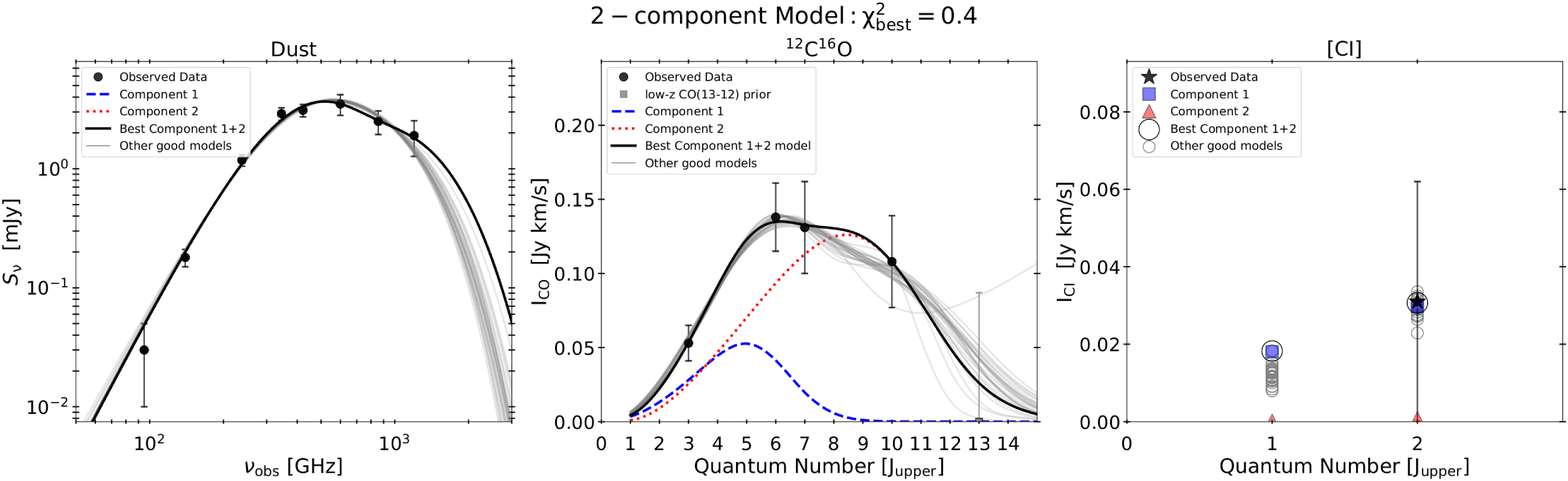}\\
\hspace*{0.0cm}
\includegraphics[trim={0.0cm 0 0.0cm 0.0},clip,width=1.0\textwidth]{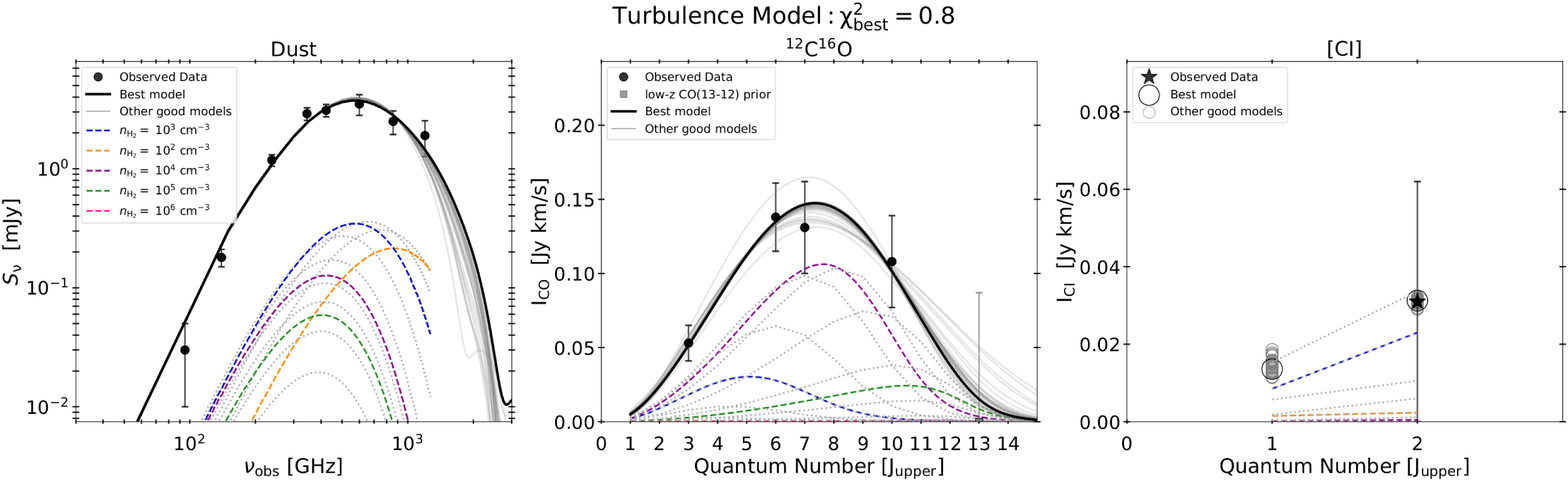}\\
\caption{\textbf{Top:}  Dust continuum, CO, and \ci\ SED in E from the \textit{1-component} model. \textbf{Middle:} The SED from the \textit{2-component} model. \textbf{Bottom:} The SED from the \textit{Turbulence} model.
}
\label{eastsed}
\end{figure*}

The best-fit parameters from the \textit{Turbulence} model in SPT0311-58 W and E assuming less than solar metallicity and solar metallicity are shown in Table \ref{turb_parameters}. The best-fit value for each parameter is taken as the mean of outputs from all the 10$^{7}$ models weighted by $\chi^{2}$, and the error on the parameters is the $\chi^{2}$ weighted standard deviation. In the Appendix, we show the $\chi^{2}$ weighted parameter values from all the \textit{Turbulence} models assuming less than solar metallicity conditions (Figures \ref{westcorner} and \ref{eastcorner}). The SEDs estimated using the best-fit parameters are shown in Figures \ref{westsed} and \ref{eastsed} where the top panel corresponds to the dust, CO, and \ci\ SEDs obtained from the \textit{1-component} model, the middle panel is from the \textit{2-component} model, and the bottom panel is from the \textit{Turbulence} model. In all the models, we show the best-fit SED as the thick black line and the 30 other good $\chi^{2}$ models in grey.  In the \textit{2-component} model, we show the relative contribution of component 1 in blue and component 2 in red. As mentioned in Section \ref{2comp-turb}, in the \textit{Turbulence} model, the density PDF is sampled at 50 densities, which correspond to a source solid angle and the final SED is the sum of SEDs obtained at each of the 50 models. In the \textit{Turbulence} model panel, we show the relative contribution of 5 densities to the final SED as different dashed colored lines and other densities as dashed grey lines. In this analysis, we focus our discussion on the \textit{Turbulence} model.

 The best-fit mean gas densities are log(n$_{\rm H_{2}}$/cm$^{-3}$ ) = (5.3 $\pm$ 1.3) in W and (4.7 $\pm$ 2.4) in E from the \textit{Turbulence} model. In the \textit{Turbulence} model panel of Figures \ref{westsed} and \ref{eastsed}, we see that the larger gas densities (log(n$_{\rm H_{2}}$) $>$ 5) emitted from smaller regions of the galaxy do not contribute significantly to the overall dust and CO emission. In the \textit{2-component} fit, component 1 has a lower gas density and is emitted from a larger area than component 2 in both W and E (see Table \ref{ncomp_parameters} in the Appendix). This indicates that component 1, associated with low-J CO excitations, is primarily tracing diffuse regions of the ISM, while component 2 traces dense and compact gas where high-J CO is excited. The mean of the gas density from the two components is log(n$_{\rm H_{2}}$/cm$^{-3}$ ) = (4.7 $\pm$ 1.2) in W and (4.6 $\pm$ 1.3) in E. These values are consistent with the \textit{Turbulence} model within the uncertainty. The kinetic temperature from the \textit{Turbulence} model is (116 $\pm$ 42) K in W and (166 $\pm$ 65) K in E. This is larger than the dust temperatures, (52 $\pm$ 5) K in W and (48 $\pm$ 4) K in E, since the ratio of T$_{\rm K}$/T$_{\rm dust}$ from the model is greater than 1.0. This could imply that in addition to photoelectric heating, there are other heating mechanisms such as X-rays, cosmic rays, and mechanical energy input from processes such as AGN outflows, mergers, or stellar feedback. However, it has to be noted that, while T$_{\rm K}$/T$_{\rm dust}$ is a function of visual extinction \citep{tielens85}, we consider a constant value at all extinctions.
 
 The best-fit GDMR from the \textit{Turbulence} model is (124 $\pm$ 62) in W and (138 $\pm$ 95) in E, which is similar to the enriched Milky Way value. The best-fit CO abundance is (7 $\pm$ 5) $\times$ 10$^{-5}$ in W and E, consistent with the canonical value of 8 $\times$ 10$^{-5}$ adopted in literature \citep{frerking82}.
 The best-fit $\rm [CI/H_{2}]$ is (3.4 $\pm$ 1.8) $\times$ 10$^{-5}$ in W and (3.4 $\pm$ 2.3) $\times$ 10$^{-5}$ in E. These values are consistent with the carbon abundance in dense star-forming environments where a value of $\sim$5 $\times$ 10$^{-5}$ is reported in the center of local starburst galaxy M82 \citep{white94}. In high redshift (z $>$ 2) samples of sub-millimeter and quasar host galaxies, \citet{walter11} derive a carbon abundance of (8.4 $\pm$ 3.5) $\times$ 10$^{-5}$. Such high values are possible at high redshift where CO molecules are dissociated due to cosmic rays or interstellar radiation, increasing the \ci\ abundance \citep{bisbas05}. In addition to cosmic rays, other factors are also found to affect the \ci\ abundance such as gas density, temperature, and metallicity. In hydrodynamical simulations, the carbon abundance is observed to decrease with increasing metallicity \citep{glover16}. In the \textit{Turbulence} model, the \ci\ abundance decreases with increasing $\rm n_{H_{2}}$ using the negative power law index $\rm \beta_{[CI]}$. This relationship is also manifested in the \textit{2-component} model where the \ci\ abundance is lower in the denser component 2 than the diffuse component 1 by a factor of $\gtrsim$ 3. However, it has to be noted that, due to the lack of \cia\ observations, we cannot reliably confirm the relationship between \ci\ abundance and gas density. The \ci\ emission is mainly dominated by the diffuse component as seen in the \textit{2-component} model (component 1 shown in blue) in Figures \ref{westsed} and \ref{eastsed}.


\subsubsection{CO and \ci\ Excitations}\label{sec:cociexcitation}
\begin{figure}[h!]
\hspace*{-1.0cm}
\includegraphics[trim={0.0cm 1.2cm 0.0cm 4.0},clip,width=0.58\textwidth]{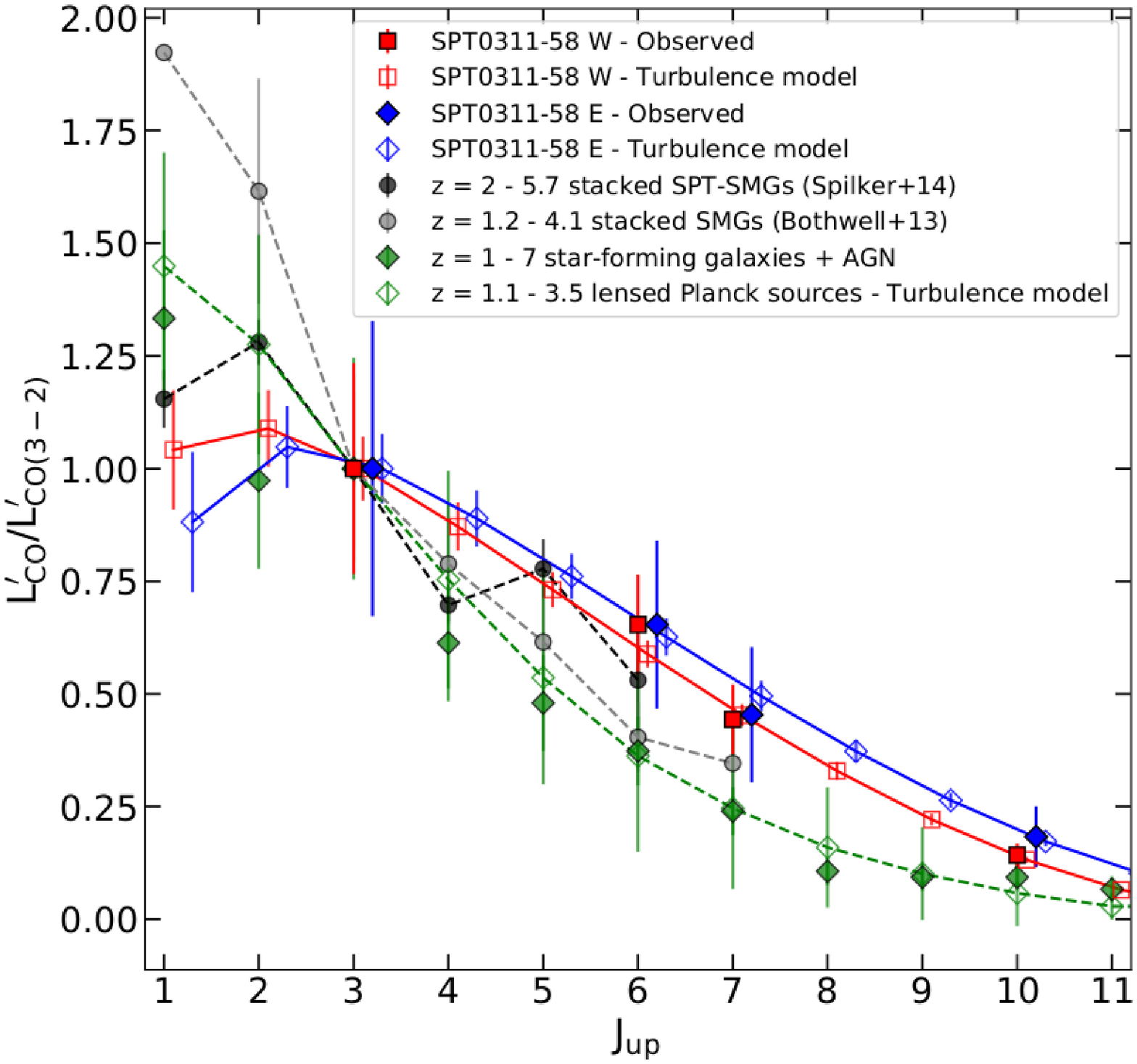}\\
\caption{Brightness temperature ratio at each CO excitation transition. The \textit{Turbulence} model SPT0311-58 values are the mean of the 30 SEDs shown in Figure \ref{westsed} and Figure \ref{eastsed}. 
The gravitationally lensed SPT-SMG sample is from \citet{spilker14} and the unlensed SMGs from z = 1.2 $-$ 4.1 are from \citet{bothwell13}. A combined sample of starburst galaxies and AGN from z = 1 $-$ 7 is taken from \citet{harrington20} and \citet{kirkpatrick19}. The average \textit{Turbulence} model output from the 24 Planck sources \citep{harrington20} are also shown.
}
\label{fig:lvgcoratio}
\end{figure}

From the radiative transfer modeling, we can estimate the intrinsic brightness temperature ($\rm T_{b}$) ratios for all the CO transitions from J=1-15 in terms of the line luminosities as:
\begin{equation} \label{eqn:rj}
    \rm r_{J_{i},J_{j}} = L^{\prime}_{CO(J_{i}->J_{i}-1)}/L^{\prime}_{CO(J_{j}->J_{j}-1)}
\end{equation}
where $\rm J_{j}$ is usually 1. 
In Figure \ref{fig:lvgcoratio} we show the brightness temperature ratios in the two SPT0311-58 galaxies by normalizing to $\rm J_{j} = 3$, which is the lowest observed CO transition. We compare the observed and model CO excitations in two galaxies with the stacked $\rm T_{b}$ from 22 gravitationally lensed SPT-SMGs from \citet{spilker14} and the median values obtained from 32 SMGs from z = 1.2 $-$ 4.1 \citep{bothwell13}. We also include a heterogeneous sample of star-forming galaxies and AGN from \citet{kirkpatrick19} combined with 24 lensed Planck selected sources from \citet{harrington20}, which gives a range of galaxies from z = 1 $-$ 7. The average \textit{Turbulence} model output of the 24 Planck sources from \citet{harrington20} is also shown. 
The \textit{Turbulence} model reproduces the observed values in SPT0311-58 W and E. Both the SPT0311-58 galaxies have a brightness temperature profile similar to the high-redshift SMG samples and a subset of sources from \citet{harrington20} also show similar excitations. 


\begin{table*}
\caption{Brightness temperature ratios from the \textit{Turbulence} model}
\label{tab:rj}
\hspace{-2.3cm}
\begin{tabular}{c c c c c c c c c c}
\hline\hline
Source & $\rm r_{1,3}$ &  $\rm r_{2,3}$&  $\rm r_{4,3}$   &  $\rm r_{5,3}$ &   $\rm r_{6,3}$ &  $\rm r_{7,3}$ &  $\rm r_{8,3}$ &  $\rm r_{9,3}$ &  $\rm r_{10,3}$\\
\hline
W & 1.04 $\pm$ 0.13 & 1.09 $\pm$ 0.08 & 0.87 $\pm$ 0.05 & 0.73 $\pm$. 0.04 & 0.59 $\pm$ 0.03 & 0.45 $\pm$ 0.02 & 0.33 $\pm$ 0.02 & 0.22 $\pm$ 0.01  & 0.13 $\pm$ 0.01 \\
E & 0.88 $\pm$ 0.16 & 1.05 $\pm$ 0.09 & 0.89 $\pm$ 0.06 & 0.76 $\pm$ 0.05 & 0.63 $\pm$ 0.04 & 0.50 $\pm$ 0.03  & 0.37 $\pm$ 0.03 & 0.26 $\pm$ 0.02 & 0.17 $\pm$ 0.01 \\
\hline\hline
\multicolumn{10}{p{\textwidth}}{{NOTE. - The brightness temperature ratios are obtained from Equation \ref{eqn:rj} using the mean of 30 best SEDs from the \textit{Turbulence} LVG modeling.
}}
\end{tabular}
\end{table*}

The non-LTE \textit{2-component} best-fit model outputs the excitation temperature ($\rm T_{ex}$) of \ci\ in \mbox{component 1} and \mbox{component 2}. We calculate the total $\rm T_{ex}$ of \cia\ by performing a flux density weighted sum of $\rm T_{ex}$ in \mbox{component 1} and \mbox{component 2}. 
This gives \mbox{$\rm T_{ex}$ $\sim$ 30 K} and 32 K in W and E, respectively. These values are similar to the typical $\rm T_{ex}$ of 30 K adopted in the literature \citep{walter11}. However, due to a lack of \cia\ observations, we cannot get an accurate estimate of the excitation temperatures from modeling alone.

\begin{figure}
\hspace*{-0.6cm}
\includegraphics[trim={0.0cm 0 0.0cm 0.0},clip,width=0.55\textwidth]{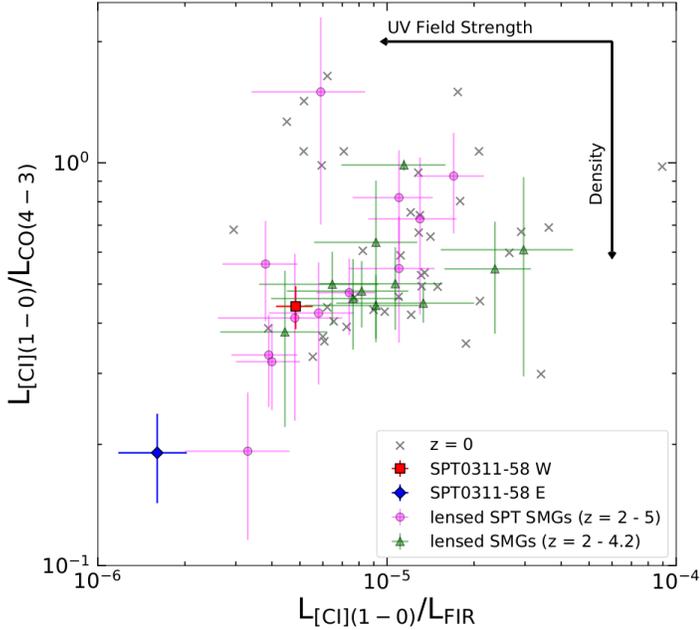}\\
\caption{$\rm L_{CI(1-0)}/L_{CO(4-3)}$ representing total gas to dense gas ratio against $\rm L_{CI(1-0)}/L_{FIR}$. The values obtained from the \textit{Turbulence} model in SPT0311-58 is compared to the observations in local and  high-redshift galaxies. 
The SPT-SMG data is from \citet{bothwell17} and the other lensed SMGs are from \citet{alaghbandzadeh13}. Local galaxies from \textit{Herschel}/SPIRE are from \citet{kamenetzky16}. CO(4$-$3) and \cia\ line luminosities in SPT0311-58 W and E are obtained from the \textit{Turbulence} model, where we model the observed dust, CO, and \cib\ transitions. In SPT0311-58 W and E, we show the mean line luminosities and the uncertainty obtained from running the \textit{Turbulence} model 30 times. 
The direction of increase in the UV field strength and the density are shown at the top right corner of the plot. The model values in W and E agree with the observations in the SPT-SMG sample.}
\label{fig:cico43}
\end{figure}

In Figure \ref{fig:cico43}, we compare the \textit{Turbulence} model \cia\ and $\rm CO(4-3)$ in \mbox{SPT0311-58} W and E to the sources from literature. \cia\ is observed to be a good tracer of $\rm CO(1-0)$ and hence, a good tracer of the bulk of the $\rm H_{2}$ gas mass \citep[e.g.,][]{jiao17,jiao19}. $\rm CO(4-3)$ traces denser $\rm H_{2}$ gas participating in the star formation. \citet{alaghbandzadeh13} observed that the ratio $\rm L_{[CI](1-0)}/L_{CO(4-3)}$ decreases with increasing \lfir. In Figure \ref{fig:cico43} we compare the ratio of \cia/$\rm CO(4-3)$ in SPT0311-58 with the lensed SMGs \citep{alaghbandzadeh13, bothwell17} and the local star-forming galaxies from \citet{kamenetzky16}. The direction in which the UV field strength and the density of the gas increases is also shown in Figure \ref{fig:cico43} \citep[e.g.,][]{kaufman99,valentino20}. SPT0311-58 W and E have ratios similar to the $\sim$40 lensed SPT-SMGs sample detailed in \citet{bothwell17}. W is similar to the mean value of SPT-SMGs while E has a lower value, which might indicate that E has higher UV radiation and is a more compact starburst than W.

\subsubsection{Gas Mass and SFR}\label{sec:gasmasslvg}
The gas mass in the LVG model is estimated using Equation \ref{gasmasseqn}. From the \textit{Turbulence} model, we get \mbox{$\rm M_{gas} = (5.4 \pm 3.4) \times 10^{11}\ M_{\odot}$} in W and \mbox{($\rm 3.1 \pm 2.7) \times 10^{10}\ M_{\odot}$} in E.  
Using the GDMR from the \textit{Turbulence} model, we derive \mbox{$\rm M_{dust} = (4.3 \pm 3.5) \times 10^{9}\ M_{\odot}$} in W and \mbox{$\rm (2.2 \pm 2.5) \times 10^{8}\ M_{\odot}$} in E. By assuming solar metallicity in SPT0311-58, we get \mbox{$\rm M_{gas} = (4.5 \pm 1.8) \times 10^{11}\ M_{\odot}$} in W and \mbox{($\rm 2.6 \pm 0.7) \times 10^{10}\ M_{\odot}$} in E from the \textit{Turbulence} model. Because the GDMR and CO abundance are fixed for solar metallicity, the gas mass is better constrained than in the case where we do not constrain the parameters.

\begin{figure*}
\includegraphics[trim={0.0cm 0.3cm 0.0cm 0.0},clip,width=1.0\textwidth]{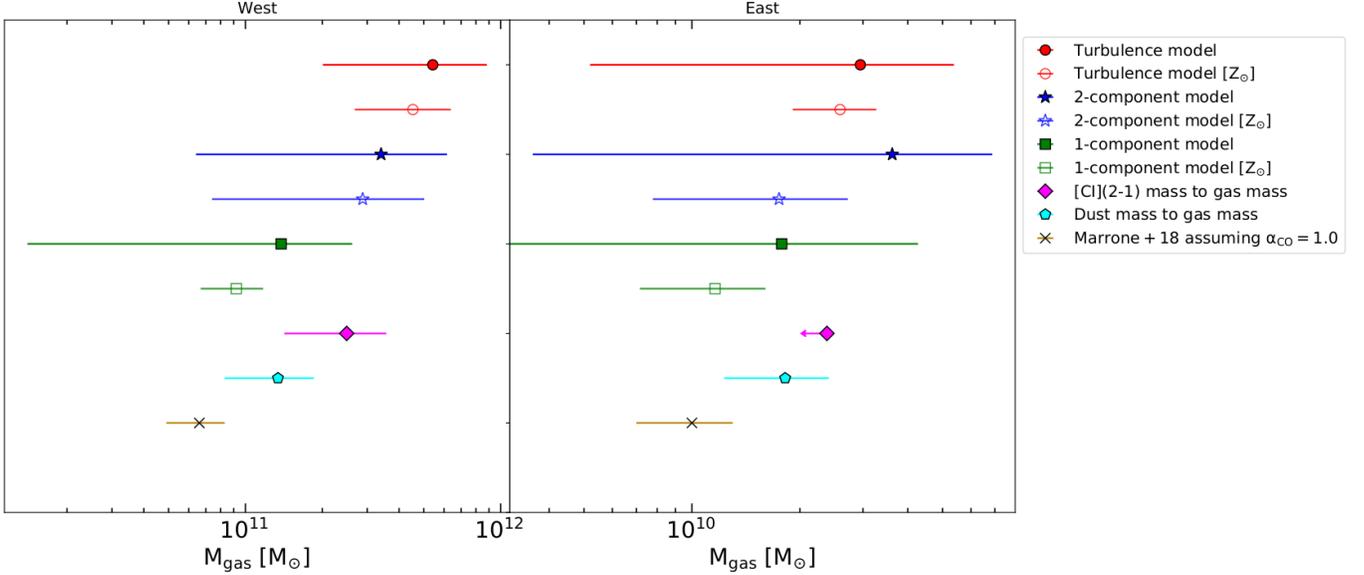}\\
\caption{Comparing the gas mass calculated from different methods in SPT0311-58 W and E. Data points from Top to Bottom: 1) The \textit{Turbulence} model output from Equation \ref{gasmasseqn} and not constraining the GDMR and CO abundance. 2) The \textit{Turbulence} model output by assuming solar metallicity ([Z$_{\odot}$]) and fixing GDMR and CO abundance. 3) Using the same equation in the \textit{2-component} model and adding the gas mass from component 1 and 2. 4) Gas mass from the \textit{2-component} model assuming solar metallicity. 5) Gas mass from the \textit{1-component} model. 6) Gas mass from the \textit{1-component} model assuming solar metallicity. 7) Gas mass from the observed $\rm \ L^{\prime}_{CI(2-1)}$ (Equation \ref{eqn:gasmassci21}) and assuming a typical excitation temperature and \ci\ abundance. 8) Gas mass estimated from dust mass by assuming GDMR = 100. 9) Gas mass from \citet{marrone18} where $\rm CO(1-0)$ is converted from $\rm CO(3-2)$ and assuming a conversion factor $\rm \alpha_{CO} = 1.0$ \mbox{$\rm M_{\odot}/K\ km\ s^{-1}\ pc^{2}$}.
}
\label{fig:gasmass}
\end{figure*}

We compare the gas mass calculated using different methods in Figure \ref{fig:gasmass}. We compare the gas mass from the LVG models with the calculations from \cib, discussed in Section \ref{sec:cigasmass}. We also calculate the gas mass from dust mass (Section \ref{sec:sedfitting}) by assuming GDMR = 100 \citep{sandstrom13}, which includes contribution from Helium. We also show the gas masses reported in \citet{marrone18} where they are obtained by scaling $\rm CO(3-2)$ to $\rm CO(1-0)$ and converting to gas mass by assuming $\rm \alpha_{CO} = 1.0$ \mbox{$\rm M_{\odot}/K\ km\ s^{-1}\ pc^{2}$}. The total gas mass from the \textit{2-component} models is consistent with the \textit{Turbulence} models within uncertainties. The gas mass estimated from \cib\ agrees with the LVG models to within the uncertainties. The gas mass from \cib\ in E is the upper limit due to non-detection of the line. 
SPT0311-58 W and E gas mass estimates reported in \citet{marrone18} and the estimates from dust mass include assumptions about the CO scaling, gas mass conversion factor and the GDMR which could result in discrepancy with the LVG models. 


All the methods used to estimate the gas mass yield a canonical value of CO-to-$\rm H_{2}$ conversion factor, \mbox{$\rm \alpha_{CO} > 0.8$ \mbox{$\rm M_{\odot}/K\ km\ s^{-1}\ pc^{2}$}} as discussed in Section \ref{sec:conversionfactor}.
The intrinsic SFR from the \textit{Turbulence} model is calculated using Equation \ref{eqn:sfr}, giving a SFR = $\rm 5046 \pm 944 \ M_{\odot}$ in W and $\rm 701 \pm 151 \ M_{\odot}$ in E. 

\section{Discussion} \label{sec:discussion}
In this section we discuss the differential magnification across the different CO transitions and dust in \mbox{SPT0311-58}. We also compare the spatial extent of CO and dust. We further estimate the CO-to-$\rm H_{2}$ conversion factor and gas depletion time scales in SPT0311-58 and compare them with the other high-redshift SMGs in literature. Towards the end of this section, we briefly discuss some of the heating mechanisms contributing to the CO emission in SPT0311-58.

\subsection{Magnification and Intrinsic Size}\label{sec:codustsize}

\begin{figure}[h!]
\hspace*{-0.5cm}
\includegraphics[trim={0.0cm 3.0cm 0.0cm 3.5cm},clip,width=0.52\textwidth]{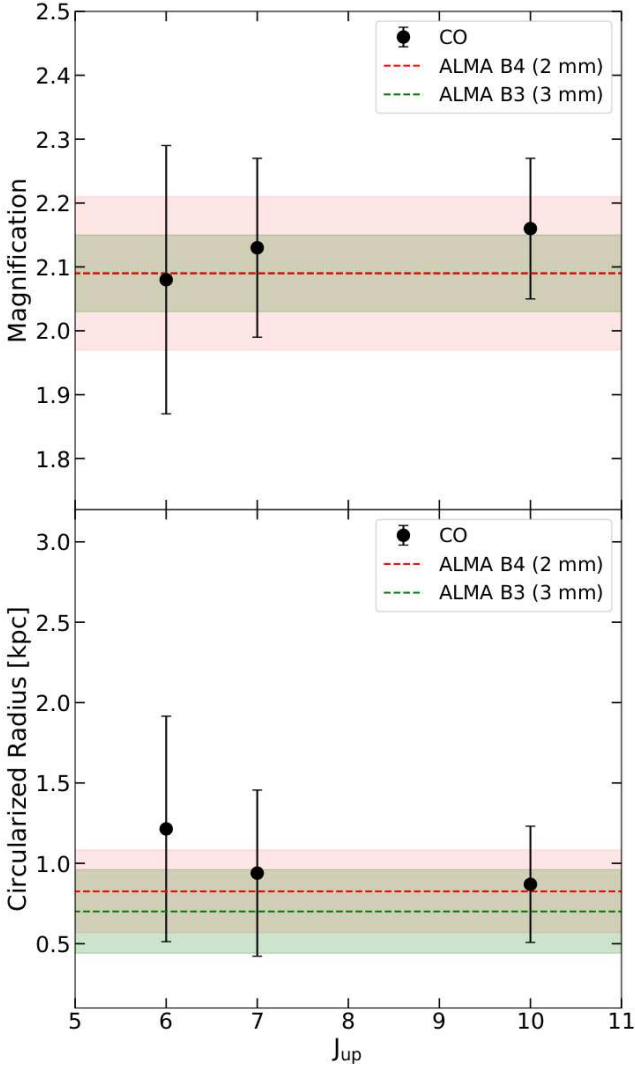}\\
\caption{\textbf{Top:} The magnification of CO and dust emission in SPT0311-58 W. There is no significant differential magnification between the different CO transitions and the dust.
\textbf{Bottom:} CO and dust emission region sizes (circularized radius) in SPT0311-58 W. The higher-J CO emission arises from more compact regions of the galaxy than lower-J CO transitions. CO has been observed to have a marginally larger size than dust. 
}
\label{fig:codustsize}
\end{figure}

In Figure \ref{fig:codustsize} we compare the magnification and intrinsic size of CO and dust continuum regions in SPT0311-58 W obtained from the lens models. The magnification for the dust continuum and CO are given in Table \ref{continuum_prop} and \ref{line_prop} respectively and the lens model parameters are given in Table \ref{tab:cont_parameters} and \ref{tab:co_parameters}. 

The circularized intrinsic radius is calculated from the semimajor axis ($\rm a_{s}$) and semiminor axis ($\rm b_{s}$) as $\rm \sqrt{(a_{s}b_{s})}$.
Differential magnification can occur depending on the position of the source relative to the lensing caustic and the size of the emission region \citep{hezaveh12a,spilker15}. Significant differential magnification between the CO lines, particularly between those tracing different physical conditions of the galaxy, can affect the physical properties derived from the CO SLED \citep{dong19}. From the top panel in Figure \ref{fig:codustsize}, we see that the magnification is consistent between the CO transitions and the dust continuum, i.e., we do not observe differential magnification between these components in SPT0311-58 W. From the bottom panel of the figure, we see that the CO emission area is marginally decreasing with increasing excitation level. This trend has been observed in sources such as M82 \citep{weiss05b} and SPT0346-52 \citep{apostolovski19} where low-J CO transitions such as, $\rm CO(1-0)$ and  $\rm CO(2-1)$ are tracing the diffuse regions of the galaxy while the high-J CO is emitted from compact regions. We also observe that the CO emission regions are larger than the dust emission, although not significantly. 
Our result is consistent with previous literature where CO is observed to have a larger radial extent than dust \citep[e.g.,][]{spilker15,dong19,apostolovski19}. In \citet{calistrorivera18} and \citet{dong19}, the authors discuss several possibilities for such a trend, for example, a temperature gradient across the galaxy giving rise to a lower intensity of continuum emission in the outskirts or spatially varying gas-to-dust ratio. The compact size of the dust continuum in comparison to the cold gas due to a temperature gradient across the galaxy is also discussed in \citet{cochrane19}. In SPT0346-52 \citep{apostolovski19}, it is observed that the radial extent of  dust at \mbox{3 mm} is larger than that at \mbox{2 mm}. In SPT0311-58, the size of the emission region of dust at 2 mm (ALMA B4) is slightly higher than at 3 mm (ALMA B3) but this difference is not statistically significant (bottom panel in Figure \ref{fig:codustsize}). 

\subsection{Gas Mass Conversion Factor}\label{sec:conversionfactor}
\begin{figure*}
\hspace*{-0.4cm}
\includegraphics[trim={0.0cm 0.0cm 0.0cm 0.0},clip,width=1.05\textwidth]{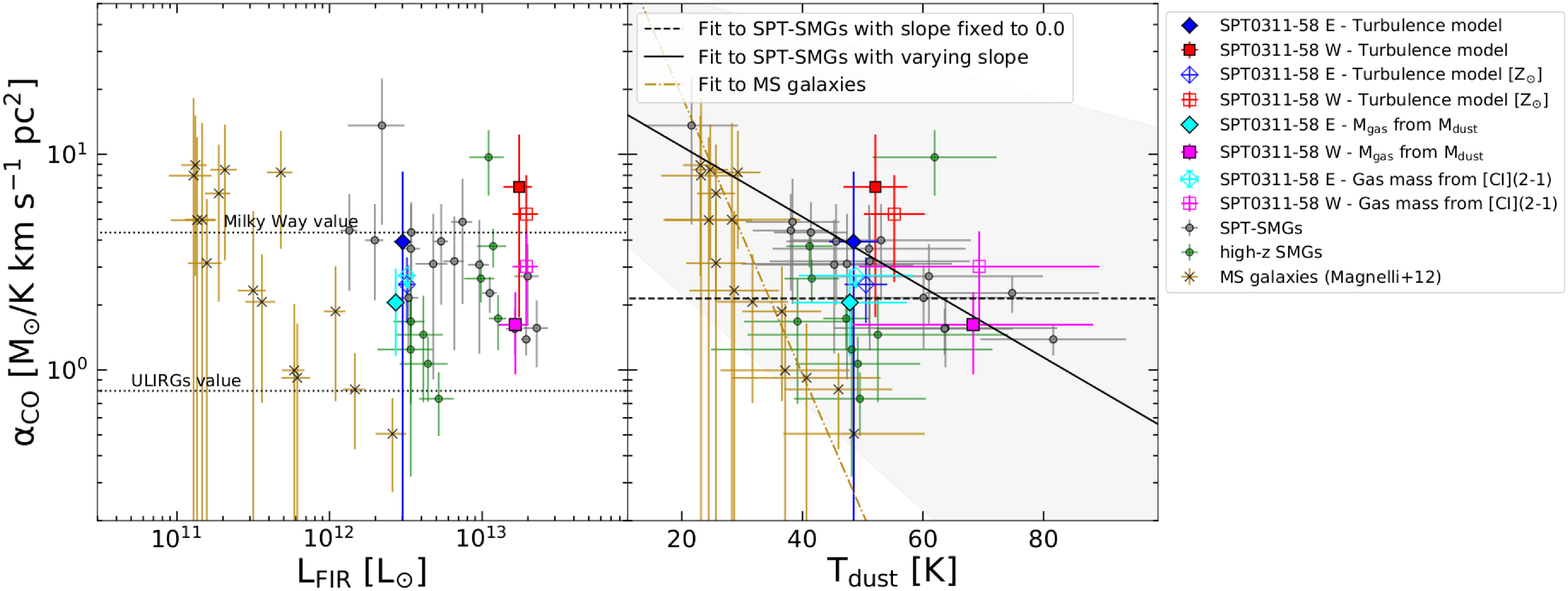}\\
\caption{\textbf{Left:} The main sequence galaxies (MS) are taken from \citep{magnelli12}, the high-redshift SMGs from \citet{carilli10,walter12,fu12,ivison13,fu13,alaghbandzadeh13} and the SPT-SMGs from \citet{aravena16} and \citet{reuter20}. The $\rm \alpha_{CO}$ in the literature sample is estimated using $\rm M_{gas}$ from $\rm M_{dust}$ assuming gas-to-dust mass ratio of 100. The $\rm \alpha_{CO}$ values in SPT0311-58 W and E obtained using $\rm M_{gas}$ from the \textit{Turbulence} model assuming less than solar metallicity are shown as solid red and blue data points. The \textit{Turbulence} model $\rm \alpha_{CO}$ values under solar metallicity assumption are shown as open red and blue data points. The $\rm \alpha_{CO}$ using $\rm M_{gas}$ from $\rm M_{dust}$ are shown as solid magenta and cyan data points in W and E, respectively. The $\rm \alpha_{CO}$ using $\rm M_{gas}$ from \cib\ are shown as open magenta and cyan data points. \textbf{Right:} The correlation between $\rm \alpha_{CO}$ and $\rm T_{dust}$ is statistically significant in the main sequence galaxies and in the SPT-SMGs. 
}
\label{fig:xcolir}
\end{figure*}

Traditionally, the gas mass is estimated by assuming a CO-to-$\rm H_{2}$ conversion factor ($\rm \alpha_{CO}$) where M$_{\rm gas}$ =  $\rm \alpha_{CO} \ L^{\prime}_{CO(1-0)}$. In the literature, $\rm \alpha_{CO} = 0.8$ \mbox{$\rm M_{\odot}/K\ km\ s^{-1}\ pc^{2}$} is typically adopted for ULIRGs \citep[e.g.,][]{downes98,carilli13}. We hereby mention $\rm \alpha_{CO}$ values without units for convenience. There is a large uncertainty in $\rm \alpha_{CO}$ values ranging from \mbox{0.4 $-$ 6} \citep[e.g.,][]{ivison11,papadopoulos12b,carilli13,mashian13}. In \mbox{SPT0311-58}, we calculate $\rm \alpha_{CO}$ using three different estimates of gas mass as shown in Figure \ref{fig:xcolir}. The $\rm CO(1-0)$ luminosity is obtained from the \textit{Turbulence} model where $\rm L^{\prime}_{CO(1-0)} = (8.3 \pm 1.3) \times 10^{10}\ K\ km\ s^{-1}\ pc^{2}$ in W and $\rm (8.8 \pm 2.6) \times 10^{9}\ K\ km\ s^{-1}\ pc^{2}$ in E. In the first method, we derive the conversion factor based on the intrinsic gas mass (Equation \ref{gasmasseqn}) from the \textit{Turbulence} model. Assuming less than solar metallicity in SPT0311-58, we obtain $\rm \alpha_{CO} = 7.1 \pm 5.3$ in W and $\rm 3.9 \pm 4.4$ in E. Assuming solar metallicity GDMR and CO abundance, we obtain $\rm \alpha_{CO} = 5.3 \pm 2.7$ in W and $\rm 2.5 \pm 0.8$ in E. Since we fix two parameters under solar metallicity assumptions, the $\rm \alpha_{CO}$ values are better constrained. 
This is similar to $\rm \alpha_{CO} = 4.8 \pm 2.9$ reported in \citet{strandet17} derived from unresolved observations of SPT0311-58 using the \textit{2-component} model, assuming solar metallicity, and fixing the CO abundance. 
In the second method, we estimate the intrinsic gas mass from the dust mass (Section \ref{sec:sedfitting}) by assuming  GDMR=100.
We obtain $\rm \alpha_{CO} = 1.6 \pm 0.7$ in W and $\rm 2.1 \pm 0.9$ in E. In the third method, we use gas mass from \cib\ as detailed in Section \ref{sec:cigasmass}, obtaining $\rm \alpha_{CO} = 3.0 \pm 1.4$ in W and an upper limit of 2.8 in E. 

In Figure \ref{fig:xcolir}, we compare $\rm \alpha_{CO}$ in SPT0311-58 W and E with the literature sample. In all the literature sources, we estimate the conversion factor using the second method, where we calculate dust mass following
the method detailed in Section \ref{sec:sedfitting}. This is done to be consistent and to reduce uncertainties from different assumptions in dust mass calculations, GDMR values, and gas mass calculated from dynamical mass estimates. In the left and right panels of Figure \ref{fig:xcolir}, we plot $\rm \alpha_{CO}$ as a function of \lfir\ and dust temperature ($\rm T_{dust}$), respectively. In the literature sample, both the \lfir\ and $\rm T_{dust}$ are obtained from the SED fitting procedure as detailed in \ref{sec:sedfitting}. The main sequence galaxy (MS) sample is taken from \citet{magnelli12}, the high-redshift SMGs from \citet{carilli10,walter12,fu12,ivison13,fu13,alaghbandzadeh13}, and the SPT-SMG sample is from \citet{aravena16} and \citet{reuter20}. In \citet{alaghbandzadeh13}, CO(1$-$0) is not observed directly and is derived from CO(4$-$3) using conversion from \citet{bothwell13}. In the SPT sources where CO(1$-$0) is not observed, it is derived from CO(2$-$1) by assuming a line brightness ratio of 0.9 \citep{aravena16}.
The errors are large in sources which have photometry data available at three or fewer wavelengths.
In the figure, we also show the conversion factors typically adopted in literature: 4.6 for the Milky Way and 0.8 for high-redshift ULIRGs \citep{downes98}. 

In SPT0311-58, the values from the \textit{Turbulence} model are closer to the Milky Way value. The conversion factor estimated from the dust mass in the SPT sample is larger than the ULIRGs value of 0.8. This discrepancy in $\rm \alpha_{CO}$ is discussed in \citet{ivison11} where they find that a range of $\rm \alpha_{CO} \sim 5-10$ and $0.4-1$ can both explain the gas excitations in SMGs. Constraining $\rm \alpha_{CO}$ in ULIRGs based solely on low-J CO, which traces an extended low density and warm diffuse gas, results in a lower $\rm \alpha_{CO}$ as it does not account for all the gas mass. A higher density and lower $\rm T_{kin}$ turbulent gas in ULIRGs can dominate most of the gas mass and can increase $\rm \alpha_{CO}$ to galactic values \citep{ivison11,papadopoulos12b,scoville12}. From our \textit{Turbulence} model, we find that the dense gas in both E and W ($\rm log(n_{\rm H_{2}}/cm^{-3}) = 4-5$) has a significant contribution to the overall gas emission which might give galactic values of $\rm \alpha_{CO}$ in SPT0311-58.

In the right panel of Figure \ref{fig:xcolir}, we explore the correlation of $\rm \alpha_{CO}$ and $\rm T_{dust}$. We fit a linear function to the main sequence galaxies and the SPT-SMGs using MCMC and find a statistically significant correlation between $\rm \alpha_{CO}$ and $\rm T_{dust}$. We use an F-test to determine that the model with a negative slope is statistically better than the model with no slope or zero correlation. We obtain a p$-$value of $<$ 0.05 and we reject the null hypothesis that the complex model i.e. the fit with negative slope does not provide more information than the fit with slope fixed to zero. However, we observe no significant correlation if we include all the SMGs in the model. The strong correlation between $\rm \alpha_{CO}$ and $\rm T_{dust}$ in the main sequence galaxies has been discussed in \citet{magnelli12}. We find a similar correlation in the SPT-SMG sample where SPT0311-58 W and E (from the dust mass method) follow a similar trend as the other SPT-SMGs. 
It has to be noted that, while $\rm T_{dust}$ and dust mass are estimated from the same photometry, the correlation between $\rm M_{dust}$ and $\rm T_{dust}$ is not as steep as the $\rm \alpha_{CO}$ and $\rm T_{dust}$ correlation. This empirical relation between $\rm \alpha_{CO}$ and $\rm T_{dust}$ can be used to select the appropriate $\rm \alpha_{CO}$ value for gas mass calculations in normal and starburst galaxies. One caveat is that the dust SED fitting, dust mass calculation and the gas mass from dust mass have to be estimated using the same assumptions in all the sources to understand this correlation. 


From the high-redshift SMG sample together with SPT-SMGs, we estimate a mean $\rm \alpha_{CO} = 3.2 \pm 2.7$.

\subsection{Depletion Time}\label{sec:depletiontime}
\begin{figure}[h!t]
\hspace*{-0.5cm}
\includegraphics[trim={0.0cm 0 0.0cm 0.0cm},clip,width=0.55\textwidth]{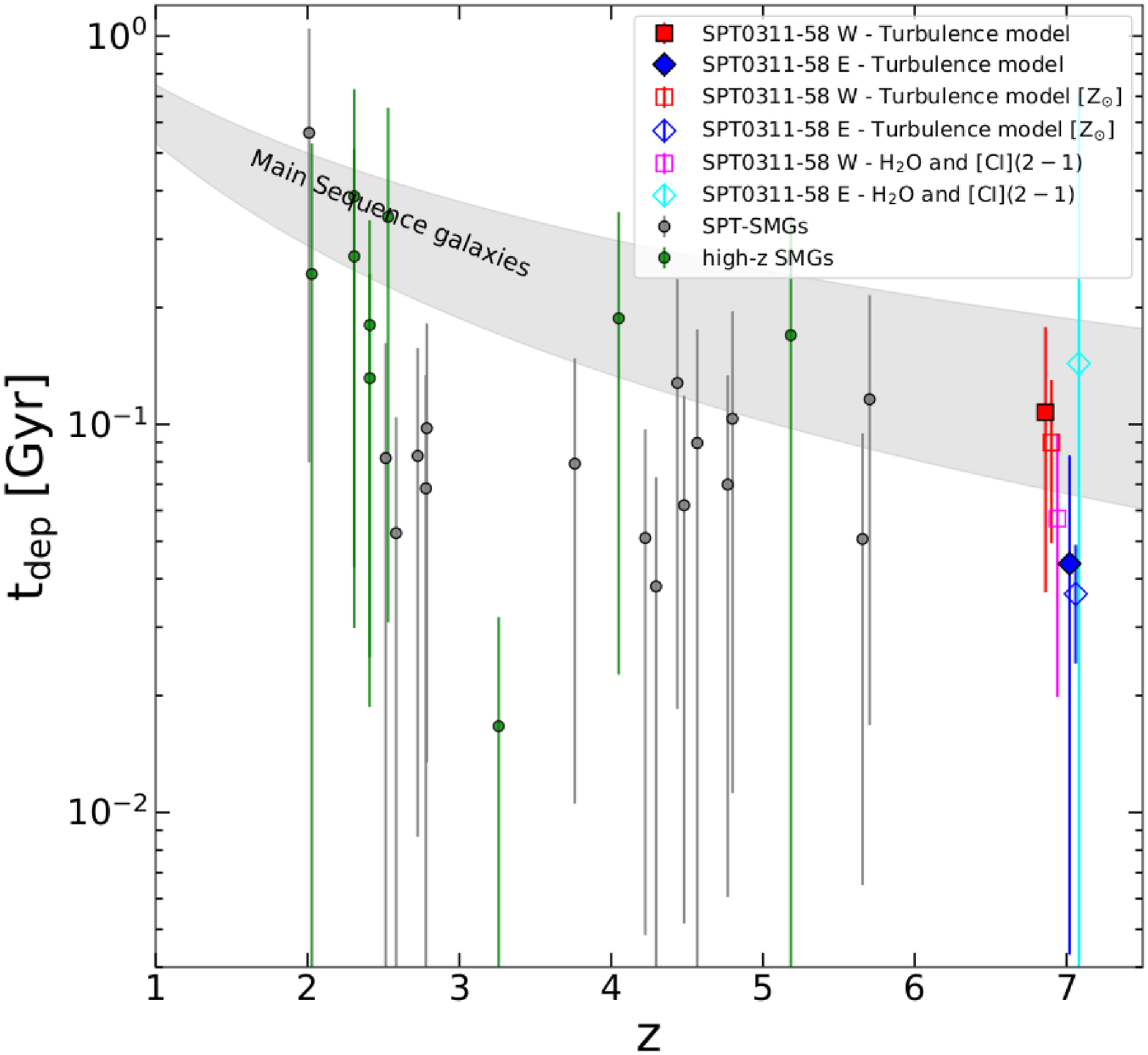}
\caption{Depletion time as a function of redshift. The values in SPT0311-58 are estimated using two methods as outlined in Section \ref{sec:depletiontime}. The depletion time from the \textit{Turbulence} model, assuming less than solar metallicity, is shown as solid red and blue data points in W and E, respectively. Assuming solar metallicity ($\rm Z_{\odot}$), the depletion time is better constrained and shown as open red and blue data. The depletion time estimated using SFR from \watera\ and gas mass from \cib\ are shown as magenta and cyan in W and E.
The high-redshift gravitationally lensed SPT-SMGs are from \citet{aravena16} and \citet{reuter20} and the other high-redshift SMGs are taken from \citet{carilli10,walter12,fu12,ivison13,fu13,alaghbandzadeh13}. The main sequence galaxies are described in \citet{saintonge13} as $\rm t_{dep} = 1.5(1+z)^{\alpha}$ where $\alpha$ is from -1.5 \citep{dave12} to -1.0 \citep{magnelli13}, which is shown as the grey shaded region. There is no clear evidence of evolution of depletion time with redshift in the SMG sample above \mbox{z $>$ 3}.
}
\label{fig:depletiontime}
\end{figure}

The gas depletion time scale ($\rm t_{dep}$) is defined as $\rm M_{gas}/SFR$. We calculate and compare the depletion time scales in SPT0311-58 W and E with the literature sources in Figure \ref{fig:depletiontime}. We estimate $\rm t_{dep}$ in \mbox{SPT0311-58} using two methods. In the first method, the gas mass and SFR are from the \textit{Turbulence} model (Section \ref{sec:gasmasslvg}). We estimate $\rm t_{dep} = 107 \pm 70 \ Myr$ and $\rm 44 \pm 40 \ Myr$ in W and E, respectively, assuming less than solar metallicity. Under the assumption of solar metallicity in SPT0311-58, we estimate $\rm t_{dep} = 90 \pm 40 \ Myr$ and $\rm 36 \pm 12 \ Myr$ in W and E, respectively. 
The second method is using the observed lines where gas mass is calculated from the observed \cib\ (Section \ref{sec:cigasmass}) and SFR from \watera\ (Section \ref{sec:lh2olfir}). We obtain $\rm t_{dep} = (57 \pm 37) \ Myr$ in W and (143 $\pm$ 570) Myr in E. The non-detection of \cib\ and \watera\ in E gives large uncertainty on the depletion time estimate. The literature high-redshift SMGs are taken from \citet{carilli10,walter12,fu12,ivison13,fu13,alaghbandzadeh13} and the SPT-SMGs from \citet{aravena16} and \citet{reuter20}. The gas mass in the literature sources is calculated from the \coe\ luminosity by assuming the mean $\rm \alpha_{CO} = 3.2 \pm 2.7$ from the high-redshift SMG sample and the SPT-SMG sample (Figure \ref{fig:xcolir}). The SFR is calculated using Equation \ref{eqn:sfr} where the \lir\ is obtained from the modified blackbody SED fit described in Section \ref{sec:sedfitting}. The main sequence galaxies are described in \citet{saintonge13} as $\rm t_{dep} = 1.5(1+z)^{\alpha}$ where $\alpha$ is from -1.5 \citep{dave12} to -1.0 \citep{magnelli13}, which is shown as the grey shaded region.

From Figure \ref{fig:depletiontime}, we observe that the depletion time in the SMGs between z=2$-$3 follow a trend similar to the main sequence galaxies where $\rm t_{dep}$ decreases with increasing redshift \citep[e.g.,][]{saintonge13,tacconi13}. However, this evolution does not seem to exist in the SMG sample at z $>$ 3. 
In SPT0311-58 W and E, the depletion time obtained from both the methods are within the typical range for high-redshift SMGs.


\subsection{Energy Budget}\label{sec:energybudget}
In this section, we discuss the heating and cooling budget of the neutral gas in SPT0311-58 with a focus on the dense molecular gas traced by CO. Some of the important cooling lines in the neutral gas regions include \cii\ (158 $\mu$m), \ci\ (369 $\mu$m, 609 $\mu$m), \oi\ (63 $\mu$m), \sii\ (35 $\mu$m), and CO. The total neutral gas cooling budget is the sum of luminosities of the lines. We combine the observations of \cii\ from \citet{marrone18} and the total CO and \ci\ luminosities from the LVG modeling to estimate the cooling power. However, since we do not have observations of the two important coolants \oi\ and \sii, which are found to contribute to more than \mbox{50\%} of the total cooling budget \citep{rosenberg15}, we cannot provide the complete picture of the neutral gas. The total power of CO ($\rm \sum_{J=1}^{J=15} L_{CO_{J}}$) and neutral carbon ($\rm \sum_{J=1}^{J=2} L_{[CI]_{J}}$) from the \textit{Turbulence} model and \cii\ from \citet{marrone18} are given in Table \ref{tab:cooling}. 
We also show the contribution of CO and \ci\ to the total cooling by assuming that \cii\ contributes to $\sim30 \%$ to the total cooling \citet{rosenberg15}.
In dense molecular gas regions (high visual extinction), CO transitions are the dominant cooling lines \citep{tielens85}, where molecular collisions play an important role over photoelectric heating. From the \textit{Turbulence} model, we derive a gas density $\rm > 10^{4}\ cm^{-3}$ and a gas kinetic temperature $\rm \sim 100 - 170\ K$ in \mbox{SPT0311-58}, which is the dense molecular gas regime. Over the depletion time of the molecular gas, i.e. $\sim100$ Myr and 40 Myr in W and E, we estimate the total energy output from CO emission to be $\rm \sim 5 \times 10^{58}$ ergs and \mbox{$\rm \sim3 \times 10^{57}$ ergs}, respectively. 

\begin{table*}
\centering
\caption{Intrinsic cooling power and percent of the total cooling fraction}
\label{tab:cooling}
\makebox[0pt][c]{\parbox{1.0\textwidth}{%
\begin{minipage}[b]{0.8\textwidth}
\begin{tabular}{c @{\hskip 0.1in} c@{\hskip 0.2in} c @{\hskip 0.5in} c@{\hskip 0.2in} c@{\hskip 0.5in} c@{\hskip 0.1in} c}
\hline\hline
Cooling channel &  \multicolumn{2}{c}{Power}  &  \multicolumn{2}{c}{Fraction relative to \cii} & \multicolumn{2}{c}{Total cooling \%} \\
 & \multicolumn{2}{c}{[$\rm ergs\ s^{-1}$]} &   & & &  \\
 & W & E & W & E & W & E \\
\hline
CO &  $\sim 1\times \rm10^{43}$ & $\sim2 \times \rm 10^{42}$ &  0.36 &  0.12 & $\sim$ 10.7\% &  $\sim$ 3.7\% \\
\ci &  $\sim 2\times \rm10^{42}$ & $\sim8 \times \rm 10^{40}$ &  0.04 &  0.004 & $\sim$ 1.2\% &  $\sim$ 0.1\% \\
\cii & $\sim 4\times \rm 10^{43}$ &  $\sim2 \times \rm10^{43}$ & -  & -  & -  & -  \\
\hline\hline
\multicolumn{7}{p{\textwidth}}{{NOTE. - The intrinsic cooling power of CO is the sum $\rm \sum_{J=1}^{J=15} L_{CO_{J}}$ and \ci\ is $\rm \sum_{J=1}^{J=2} L_{[CI]_{J}}$ from the \textit{Turbulence} model. The \cii\ intrinsic power is obtained from \citet{marrone18}.
The percent cooling fraction for CO and \ci\ is obtained by considering the fraction of the cooling channel luminosity relative to the \cii\ luminosity and assuming that \cii\ contributes to $\sim$30$\%$ of the total cooling \citep{rosenberg15}.
}}
\end{tabular}
\end{minipage}}}
\end{table*}

The \textit{Turbulence} LVG model has a free parameter ($\rm T_{kin}$/$\rm T_{dust}$) to account for heating from sources such as X-rays, cosmic-rays, and mechanical heating, in addition to photoelectric heating by modeling $\rm T_{kin}$ and $\rm T_{dust}$ simultaneously. $\rm T_{kin}/T_{dust} = 2.0 \pm 0.9$ and $3.4 \pm 1.3$ in SPT0311-58 W and E, respectively, which suggests that there are other heating mechanisms in addition to photoelectric heating (traced by $\rm T_{dust}$).
A high value of $\rm T_{kin}/T_{dust}$ could also occur in photo-dissociation regions at low extinctions or low density \citep{tielens85}. However, since both galaxies have a high gas density where high-J CO is excited, we can consider heating from other processes. One caveat is that we do not consider the change in $\rm T_{kin}$/$\rm T_{dust}$ with density in the LVG model. Following \citet{harrington20}, we calculate the contribution of turbulent kinetic energy to the CO line emission in the dense molecular gas. The turbulent energy can be calculated from the turbulence line width and the gas mass (Table \ref{turb_parameters}) as $\rm E_{turb} = 0.5\ M_{gas} \times dv_{turb}^{2}$. We get a turbulence energy of $\rm \sim 1 \times 10^{59}\ ergs$ in W and $\rm \sim 1 \times 10^{57}\ ergs$ in E, similar to the total CO cooling energy. 
This shows that turbulence energy is sufficient to heat the molecular gas in SPT0311-58.
There are several possible sources of turbulence-driven mechanical heating such as the presence of AGN outflows, stellar winds or supernovae explosions.  

We also estimate heating from star formation such as stellar feedback and supernovae explosions based on the calculations from \citet{harrington20} and references therein. The SPT sources have a negligible contribution from AGN to the total infrared luminosity, even in the galaxy with one of the highest star-formation rate densities \citep{ma16}. Hence, the presence of an AGN is probably not a significant source of molecular gas heating in SPT0311-58. It is also argued in \citet{harrington20} that the X-ray luminosity from non-AGN sources is not a major source of heating. We estimate heating from stellar outflows and supernovae. Stellar feedback in galaxies is primarily contributed by massive stars, such as O-type, which evolve into core collapse supernovae, neutron stars, or blackholes. Using an O-type stellar wind luminosity of $\sim 10^{49-51}$ ergs in a lifetime of 5 Myr \citep{leitherer99,smith14,ramachandran19} and assuming O-type stars comprise 0.2\% of the total number of stars formed (Kroupa initial mass function), we can calculate the total energy output over the depletion time of the galaxies. We estimate a total stellar feedback energy of $\sim 10^{58-60}$ ergs in W and $\sim 7 \times 10^{56-58}$ ergs in E. Stars with masses in the range of $\rm \sim 10-40\ M_{\odot}$ collapse as supernovae \citep{heger03}, which emit an energy of $\sim 10^{51}$ ergs. Assuming the stars with these stellar masses comprise $\sim 7\%$ of the total stellar mass, we estimate a total energy output from supernovae explosions of $\sim 3 \times 10^{60}$ ergs in W and $\sim 2 \times 10^{59}$ ergs in E, over the depletion time scale. This is an upper limit as the energy input from supernovae explosions is not a continuous process. About 1\% or less of the supernovae energy goes into turbulent energy \citep[e.g.][]{iffrig15,martizzi16}, which is consistent with the CO cooling energy.

The energy estimates show that the mechanical heating from star formation (stellar outflows and supernovae explosions), some of which is converted into turbulent energy through a turbulent energy cascade from galaxy scales to smaller physical scales, may contribute significantly to the CO cooling budget over the molecular gas depletion timescale.

\section{Summary and Conclusion} \label{sec:conlusion}
SPT0311-58 is comprised of two intensly star-forming galaxies, West (W) and East (E), at a redshift of 6.9, in the Epoch of Reionization. We characterize the physical properties of the galaxies using new observations of \coa, \cob, \coc, \cib, and \watera\ transitions. We perform lensing reconstruction by assuming a S$\rm \acute{e}$rsic source profile using \texttt{visilens} \citep{spilker16}. We show that there is no significant differential magnification between dust and CO. We run non-LTE LVG radiative transfer models, which consider dust, CO, and \ci\ simultaneously: \textit{1-component}, \textit{2-component}, and \textit{Turbulence} models. In the \textit{1-component} and \textit{2-component} models, the ISM is modeled by one and two gas densities, respectively. The \textit{Turbulence} model is more sophisticated where the gas density is sampled from a log-normal PDF in turbulent gas. The main results and conclusion are given below: 

\begin{itemize}
\itemsep0.2em 
\item{We detect \watera\ in SPT0311-58 W, which is the most distant detection of water in a galaxy without any evidence for active galactic nuclei in the literature. The $\rm L_{H_{2}O}/L_{FIR}$ ratio in SPT0311-58 is consistent with other high-redshift galaxies. 
From the \lwater-\lfir\ correlation, we estimate a SFR of $\rm 4356 \pm 2143\ M_{\odot}yr^{-1}$ in W and an upper limit of $\rm 385\ M_{\odot}yr^{-1}$ in E. The SFR calculated from \lir\ is $\rm 5046 \pm 944\ M_{\odot}yr^{-1}$ in W and $\rm 701 \pm 151\ M_{\odot}yr^{-1}$ in E. Both the calculations give consistent values within errors in W. These measurements of \water\ are broadly consistent with the possibility that the cascade transition from FIR absorption pathways may trace total FIR luminosity and thus star formation.}

\item{The CO SLED and brightness temperature ratios in SPT0311-58 W and E are consistent with the other high-redshift starburst galaxies. We explore $\rm L_{[CII]}/L_{FIR}$ versus $\rm L_{CO(10-9)}/L_{CO(6-5)}$ as an indicator of the presence of heating mechanisms in addition to photoelectric heating. We observe that $\rm L_{[CII]}/L_{FIR}$ decreases with increasing $\rm L_{CO(10-9)}/L_{CO(6-5)}$. The ratio of $\rm L_{CO(10-9)}/L_{CO(6-5)}$ in SPT0311-58 is comparable to low-redshift AGN and Class II and III galaxies \citep{rosenberg15}, which suggests that there are additional heating mechanisms, but we cannot confirm the presence of AGN. 
}
\item{By comparing the radial extent of dust and CO transitions, we observe that the CO is emitted from a larger area than dust, although the result is not significant. We also observe that the CO emission region is marginally decreasing with increasing excitation level.}
\item{The mean density derived from the \textit{Turbulence} LVG model is $\rm log(n_{H_{2}}/ cm^{-3}) = (5.3 \pm 1.3)$ in W and $\rm (4.7 \pm 2.4)$ in E. The gas mass from the \textit{Turbulence} model is $\rm 5.4 \pm 3.4\ \times10^{11}\ M_{\odot}$ in W and $\rm 3.1 \pm 2.7\ \times10^{10}\ M_{\odot}$ in E. From the predicted LVG model $\rm CO(1-0)$, we estimate a gas conversion factor $\rm \alpha_{CO} = 7.1 \pm 5.3$ and $\rm 3.9 \pm 4.4$ in W and E, respectively. This is consistent with the $\rm \alpha_{CO}$ in the high-redshift SMGs within the uncertainties. From the high-redshift SMG sample together with SPT-SMGs, we estimate a mean $\rm \alpha_{CO} = 3.2 \pm 2.7$.
} 
\item{From the \textit{Turbulence} model, we estimate depletion time scale of $\rm 107 \pm 70\ Myr$ in W and $\rm 44 \pm 40\ Myr$ in E. We observe that there is no evidence for the evolution of depletion time with redshift amongst the SMG sample at z $>$ 3. The gas depletion time in SPT0311-58 W and E is within the range of the other high-redshift starburst galaxies.}

\item{The ratio of  $\rm T_{kin}/T_{dust}$ is $>$ 1 in both galaxies, which indicates that there are additional heating mechanisms such as X-rays, stellar outflows, and supernovae in addition to photoelectric heating in dense molecular gas. The mechanical heating from stellar outflows and supernovae explosions, some of which is converted into turbulent energy, may contribute significantly to the total CO cooling over the depletion timescale. }
\end{itemize}
SPT0311-58 is one of the most well characterized galaxies in the EoR. We observed the brightest source at a high spatial resolution of $\sim$2-3 kpc and could detect the CO lines with a peak signal-to-noise ratio greater than 4. These resolved observations at z$\sim$7 highlight the power of ALMA. SPT0311-58 is the highest redshift source from the SPT-SZ survey and it is expected that more than 100 sources at z $>$ 7 can be found in the SPT-3G survey \citep[e.g.,][]{benson14,guns21}.


\section{Acknowledgements} 
The SPT is supported by the NSF through grant OPP-1852617. D.P.M., J.D.V., and S.J. acknowledge support from the US NSF under grants AST-1715213 and AST-1716127. S.J. acknowledges support from the US NSF NRAO under grant SOSPA7-006. J.D.V. acknowledges support from an A. P. Sloan Foundation Fellowship. M.A. and J.D.V. acknowledge support from the Center for AstroPhysical Surveys at the National Center for Supercomputing Applications in Urbana, IL.
M.A. has been supported by the grant CONICYT+PCI+REDES 190194. T.R.G. acknowledges the Cosmic Dawn Center of Excellence funded by the Danish National Research Foundation under grant No. 140. The Flatiron Institute is supported by the Simons Foundation. This paper makes use of the following ALMA data: ADS/JAO.ALMA \#2017.1.01168, ADS/JAO.ALMA \#2016.1.01293, and ADS/JAO.ALMA \#2015.1.00504.S. ALMA is a partnership of ESO (representing its member states), NSF (USA) and NINS (Japan), together with NRC (Canada), MOST and ASIAA (Taiwan), and KASI (Republic of Korea), in cooperation with the Republic of Chile. The Joint ALMA Observatory is operated by ESO, AUI/NRAO, and NAOJ. The National Radio Astronomy Observatory is a facility of the National Science Foundation operated under cooperative agreement by Associated Universities, Inc.

\software{CASA (v5.1.1; \citealt{mcmullin07}), emcee (\citealt{foremanmackey13}), visilens (\citealt{spilker16})}

\bibliography{./spt_smg}
\bibliographystyle{aasjournal}


\section{Appendix}
\subsection{Far-Infrared Photometry}\label{app:photometry}

\begin{table*}[h]
\centering
\caption{Intrinsic far-infrared photometry}
\label{tab:photometry}
\makebox[0pt][c]{\parbox{0.5\textwidth}{%
\begin{minipage}[b]{0.4\textwidth}
\begin{tabular}{c c c}
\hline\hline

Wavelength &    East  &      West  \\

[$\mu$m] & [mJy]   &  [mJy]  \\
\hline
250 &  1.9 $\pm$ 0.6 & 12.7 $\pm$ 4.4 \\
350 &  2.5 $\pm$ 0.6 & 16.6 $\pm$ 3.3 \\
500 &  3.5 $\pm$ 0.7 & 22.7 $\pm$ 4.8 \\
710 &  3.1 $\pm$ 0.4 & 19.9 $\pm$ 2.0 \\
869 &  2.9 $\pm$ 0.4 & 15.9 $\pm$ 1.6 \\
1260 &  1.18 $\pm$ 0.03 & 9.8 $\pm$ 1.0 \\
2140 &  0.18 $\pm$ 0.03 & 2.0 $\pm$ 0.3 \\
3150 &  0.03 $\pm$ 0.02 & 0.6 $\pm$ 0.1 \\

\hline\hline

\multicolumn{3}{p{\textwidth}}{\hspace{+0.2in}{NOTE. - The continuum photometry given in the table is corrected for magnification. The photometry from 250 $\mu$m to 1260 $\mu$m are taken from Extended Data Table 3 in \citet{marrone18}. The continuum flux densities at \mbox{2140 $\mu$m} and \mbox{3150 $\mu$m} (140 GHz and 95 GHz respectively) are given in Table \ref{tab:cont_parameters}. A 15\% calibration error is added in quadrature to the statistical error to account for the uncertainty from absolute flux calibration and lens modeling. 
}}

\end{tabular}
\end{minipage}}}
\end{table*}

\subsection{Lens Models}\label{app:lensmodels}
\begin{figure*}[h]
\includegraphics[trim={3.0cm 0 0.0cm 0.0cm},clip,width=1.1\textwidth]{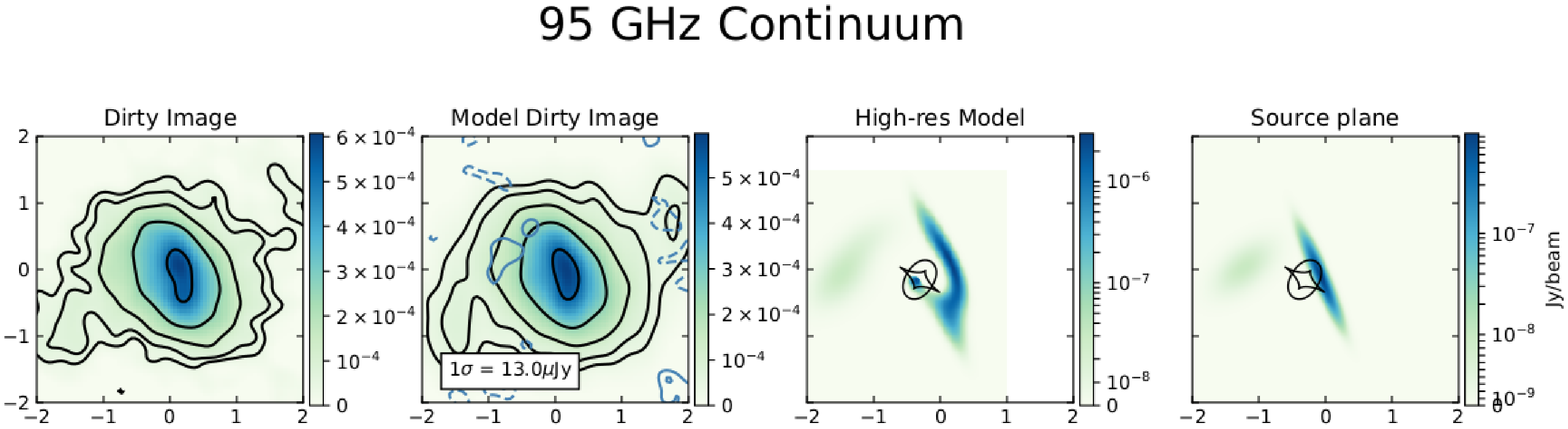}
\hspace{-2.0cm}
\includegraphics[trim={3.0cm 0 0.0cm 0.2cm},clip,width=1.1\textwidth]{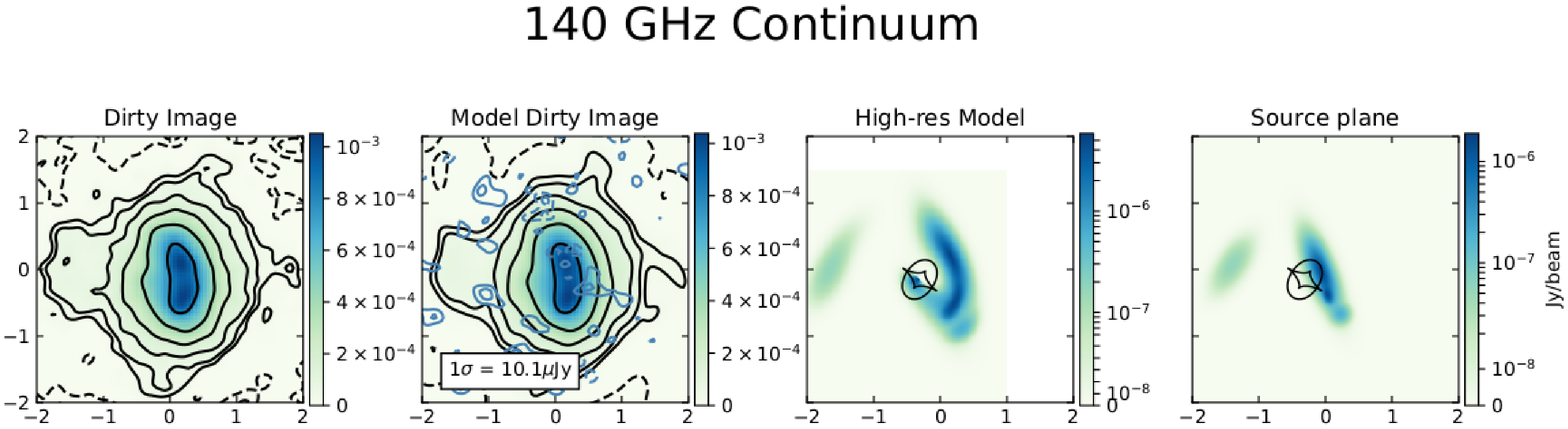}
\caption{Lens models in continuum at 95 GHz (top panel) and 140 GHz (bottom panel). The first panel is from the observations and the second panel is the high-resolution model convolved with the telescope beam. The contours are at $\pm$[3, 5, 10, 20, 40, 80] $\times \sigma$ where $\sigma$ is the RMS noise in the map. The residual \mbox{(observed - model)} contours are shown in blue at $\pm$[2,3,4,5] $\times \sigma$. The third panel is the high-resolution source plane obtained from the model with the lensing caustic shown in black. The last panel is the source plane reconstruction. The third and fourth panels are shown in logarithmic scale to emphasize the features.}
\label{fig:contlensmodel}
\end{figure*}

\begin{figure*}[]
\centering
\includegraphics[trim={3.0cm 0 0.0cm 0.0cm},clip,width=1.1\textwidth]{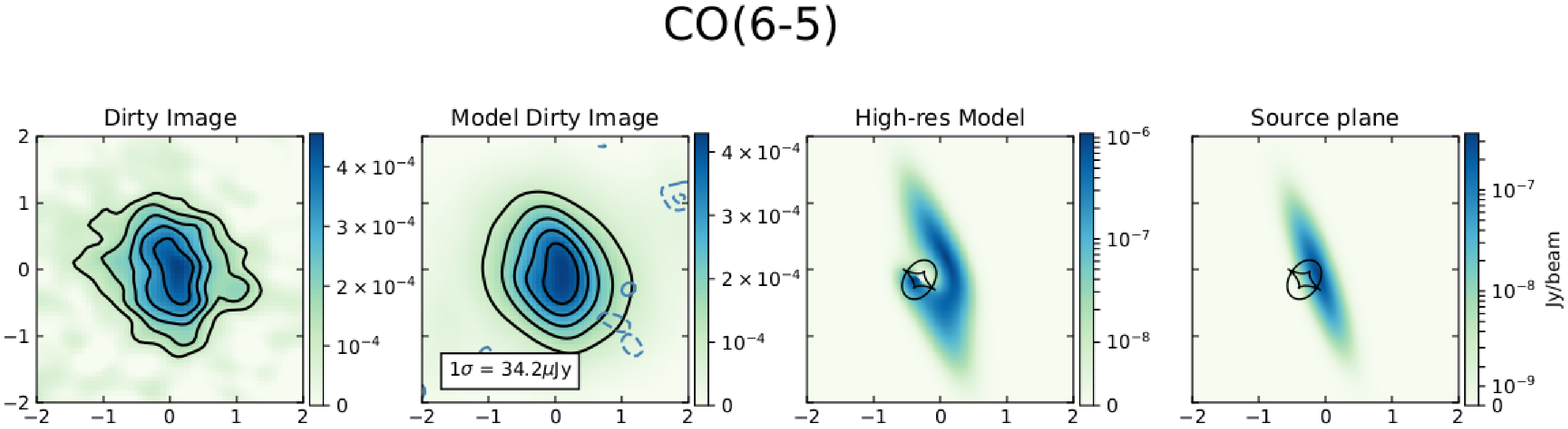}
\hspace{-2.0cm}
\includegraphics[trim={3.0cm 0 0.0cm 0.2cm},clip,width=1.1\textwidth]{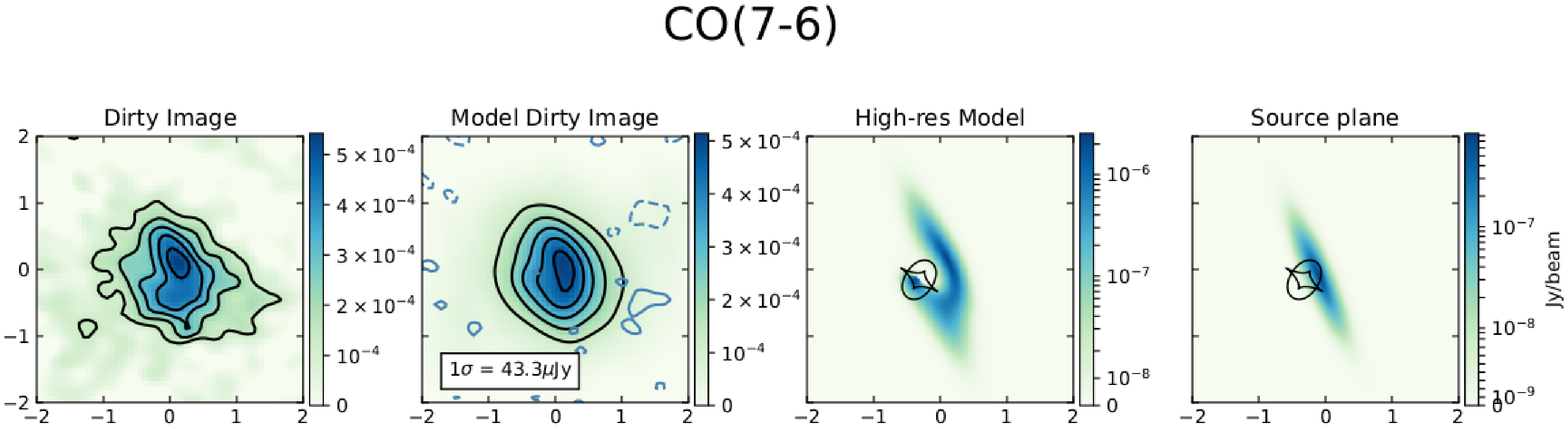}
\hspace{-2.0cm}
\includegraphics[trim={3.0cm 0 0.0cm 0.2cm},clip,width=1.1\textwidth]{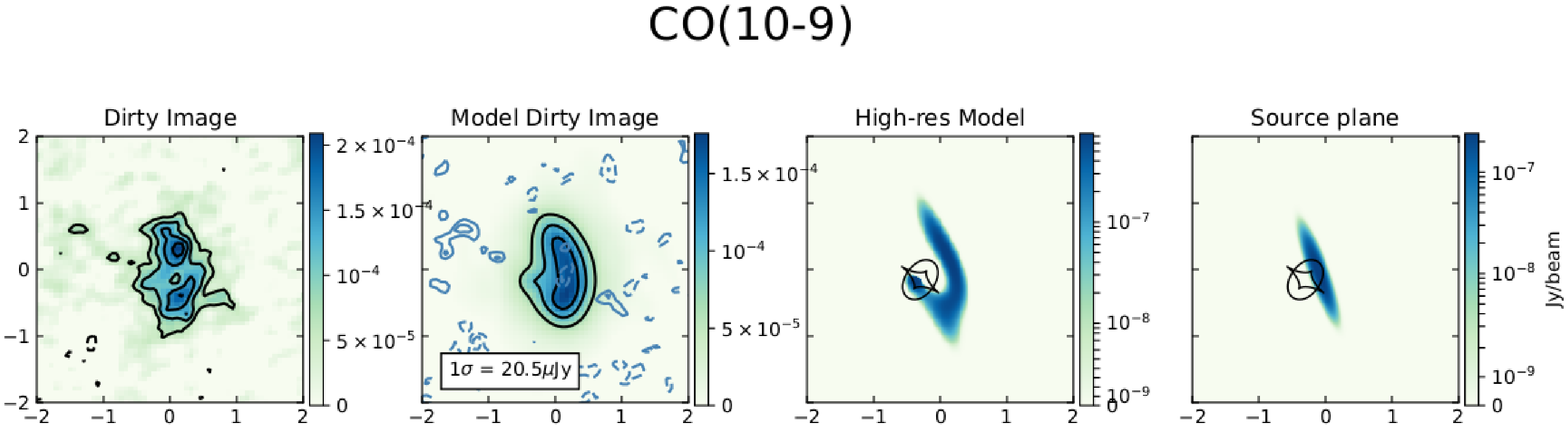}
\caption{Lens models in the velocity integrated single channel \coa\ (top), \cob\ (middle), and \coc\ (bottom). The contours in the dirty images are at $\pm$[3, 5, 7, 9, 11] $\times \sigma$. The residual description is the same as Figure \ref{fig:contlensmodel}.}
\label{fig:linelensmodel}
\end{figure*}

We present the source plane reconstruction of the continuum and spectral lines in W in Figures \ref{fig:contlensmodel} and \ref{fig:linelensmodel}. The first two panels show the observations and the beam-convolved image from the model, respectively. The third panel is the high-resolution image from the lens model with the lensing caustic shown. The final panel shows the source plane reconstructed image.


\begin{table*}[h]
\centering
\caption{Lens parameters}
\label{tab:lens_parameters}
\begin{tabular}{c*{5}{>{\raggedleft\arraybackslash}p{10.0em}}}
\hline\hline
$\rm x_{L}$ &    $\rm y_{L}$ &   $\rm e_{L}$ &   $\rm M_{L}$ &   $\rm \theta_{L}$\\

[ \arcsec\ ] & [ \arcsec\ ] &   & [$\rm 10^{11}\ M_{\odot}$] &  [degree] \\
\hline
 -0.31 $\pm$ 0.02 &  -0.15 $\pm$ 0.01 &   0.68 $\pm$ 0.07 &  0.30 $\pm$ 0.02 & 56.09 $\pm$ 3.99\\
\hline\hline
\multicolumn{5}{p{\textwidth}}{\hspace{+0.2in}{NOTE. - $\rm x_{L}$ and $\rm y_{L}$ is the position of the lens relative to the phase center. $\rm e_{L}$ is the ellipticity of the lens. $\rm M_{L}$ is the mass of the lens. $\rm \theta_{L}$ is the position angle of the major axis counter-clockwise from East. These parameters are consistent with the lens parameters from \citet{marrone18}}.}
\end{tabular}
\end{table*}

\begin{table*}
\centering
\caption{Continuum source parameters in W}
\label{tab:cont_parameters}
\begin{tabular}{c*{8}{>{\raggedleft\arraybackslash}p{6.2em}}}
\hline\hline
$\rm \nu_{obs}$ & $\rm x_{S}$ &    $\rm y_{S}$ &  $\rm S_{cont}$  & $\rm a_{S}$   & $\rm n_{S}$ & $\rm b_{S}/a_{S}$ & $\rm \phi_{S}$ \\
$\rm [GHz]$ &  [ \arcsec\ ] &   [ \arcsec\ ] & [mJy]  &  [ \arcsec\ ] &  & & [degree] \\
\hline
95 & 0.24 $\pm$ 0.01 &  0.04 $\pm$ 0.01 &  0.60 $\pm$ 0.02 &  0.33 $\pm$ 0.01 & 0.67 $\pm$ 0.17 & 0.15 $\pm$ 0.02 & 113 $\pm$ 1 \\
140 & 0.24 $\pm$ 0.01 &  0.11 $\pm$ 0.01 &  1.46 $\pm$ 0.06 &  0.29 $\pm$ 0.01 & 0.84 $\pm$ 0.09 & 0.32 $\pm$ 0.03 & 111 $\pm$ 2 \\
 & 0.34 $\pm$ 0.02 & -0.24 $\pm$ 0.02 &  0.19 $\pm$ 0.07 &  0.08 $\pm$ 0.01 & 0.54 $\pm$ 0.17 & 0.58 $\pm$ 0.09 & -72 $\pm$ 10 \\
 & 0.54 $\pm$ 0.04 & -0.53 $\pm$ 0.04 &  0.06 $\pm$ 0.02 &  0.08 $\pm$ 0.02 &  &  &  \\
\hline\hline
\multicolumn{8}{p{\textwidth}}{\hspace{+0.2in}{NOTE. - The best model includes one source at 95 GHz and 3 sources at 140 GHz. $\rm x_{S}$ and $\rm y_{S}$ is the position relative to the lens. $\rm S_{cont}$ is the  continuum flux density of the source. $\rm a_{S}$ is the half light radius of the major axis of the S$\rm \acute{e}$rsic profile or radius of the Gaussian. $\rm n_{S}$ is the S$\rm \acute{e}$rsic index (fixed to 0.5 for a Gaussian source). $\rm b_{S}/a_{S}$ is the axis ratio. $\rm \phi_{S}$ is the position angle counter-clockwise from East.
}}
\end{tabular}
\end{table*}

\begin{table*}
\centering
\caption{Single channel CO source parameters in W}
\label{tab:co_parameters}
\begin{tabular}{c*{8}{>{\raggedleft\arraybackslash}p{6.2em}}}
\hline\hline
Line & $\rm x_{S}$ &    $\rm y_{S}$ &  $\rm S_{line}$  & $\rm a_{S}$   & $\rm n_{S}$ & $\rm b_{S}/a_{S}$ & $\rm \phi_{S}$ \\
 &  [ \arcsec\ ] &   [ \arcsec\ ] & [mJy]  &  [ \arcsec\ ] &  & & [degree] \\
\hline
\coa\ & 0.20 $\pm$ 0.02 &  0.08 $\pm$ 0.04 &  0.52 $\pm$ 0.05 &  0.48 $\pm$ 0.06 & 0.84 $\pm$ 0.47 & 0.22 $\pm$ 0.07 & 109 $\pm$ 4 \\
\cob\ & 0.21 $\pm$ 0.02 &  0.06 $\pm$ 0.03 &  0.61 $\pm$ 0.06 &  0.38 $\pm$ 0.05 & 0.96 $\pm$ 0.53 & 0.21 $\pm$ 0.06 & 113 $\pm$ 4 \\
\coc\ & 0.19 $\pm$ 0.01 &  0.10 $\pm$ 0.02 &  0.35 $\pm$ 0.03 &  0.38 $\pm$ 0.03 & 0.32 $\pm$ 0.23 & 0.18 $\pm$ 0.03 & 110 $\pm$ 3 \\

\hline\hline
\multicolumn{8}{p{\textwidth}}{\hspace{+0.2in}{NOTE. - $\rm x_{S}$ and $\rm y_{S}$ is the position relative to the lens. $\rm S_{line}$ is the line flux density of the source. $\rm a_{S}$ is the half light radius of the major axis. $\rm n_{S}$ is the S$\rm \acute{e}$rsic index. $\rm b_{S}/a_{S}$ is the axis ratio. $\rm \phi_{S}$ is the position angle counter-clockwise from East.}}
\end{tabular}
\end{table*}

\newpage
\subsection{LVG Models}\label{app:lvgmodels}

The best-fit \textit{1-component} and \textit{2-component} model parameters are shown in Table \ref{ncomp_parameters}.
\begin{table*}
\caption{1 and 2-component model parameters}
\label{ncomp_parameters}
\begin{adjustwidth}{-2.0cm}{}
\begin{tabular}{c @{\hskip 0.1in}c@{\hskip 0.1in} c @{\hskip 0.2in}c@{\hskip 0.1in} c@{\hskip 0.2in} c@{\hskip 0.1in} c}
\hline\hline

Parameter &   \multicolumn{2}{c}{\textit{1-component} model}  &  \multicolumn{4}{c}{\textit{2-component} model}\\
 & W & E & \multicolumn{2}{c}{W} & \multicolumn{2}{c}{E}  \\
 & & & component 1 & component 2 & component 1 & component 2\\
\hline
Model input\\
\hline
log(n$_{\rm H_{2}}$)  & 4.0 $\pm$ 1.6 & 3.9 $\pm$ 1.1 &
3.9 $\pm$ 0.4 & 5.6 $\pm$ 1.1  & 3.6 $\pm$ 0.4 & 5.5 $\pm$ 1.2\\ 
T$_{\rm K}$ & 222 $\pm$ 79 & 225 $\pm$ 94 & 
88 $\pm$ 65 & 168 $\pm$ 93 & 100 $\pm$ 80 & 142 $\pm$ 76 \\
T$_{\rm K}$/T$_{\rm dust}$ & 4.0 $\pm$ 1.3 & 3.9 $\pm$ 1.6 &
2.7 $\pm$ 1.4 & 1.5 $\pm$ 0.5 & 2.5 $\pm$ 1.5 & 1.7 $\pm$ 0.8 \\
$\rm \beta_{T_{dust}}$ & 1.8 $\pm$ 0.1 & 1.9 $\pm$ 0.1 &
1.9 $\pm$ 0.1 & 1.9 $\pm$ 0.1 &  2.0 $\pm$ 0.1 & 2.0 $\pm$ 0.1\\
R$_{\rm eff}$ & 1676 $\pm$ 646 & 702 $\pm$ 844 &
2458 $\pm$ 920 & 651 $\pm$ 380 & 1028 $\pm$ 918 & 356 $\pm$ 778 \\
$\rm \kappa_{vir}$ & 1.4 $\pm$ 0.6 & 1.4 $\pm$ 0.5 &
1.9 $\pm$ 0.6 & 1.7 $\pm$ 0.4 & 2.0 $\pm$ 0.6 & 1.8 $\pm$ 0.4 \\ 
dv$_{\rm turb}$ & 161 $\pm$ 40 & 156 $\pm$ 38 &
134 $\pm$ 42 & 146 $\pm$ 32 & 112 $\pm$ 42 & 116 $\pm$ 44 \\
GDMR & 110 $\pm$ 56 & 149 $\pm$ 131 & 
155 $\pm$ 97 & 155 $\pm$ 97 & 245 $\pm$ 148 & 245 $\pm$ 148 \\
$\rm [CO/H_{2}]$ & (6 $\pm$ 5)$\times$10$^{-5}$ & (10$\pm$ 7)$\times$10$^{-5}$ & (8 $\pm$ 5)$\times$10$^{-5}$ & (7 $\pm$ 6)$\times$10$^{-5}$ & (12 $\pm$ 5)$\times$10$^{-5}$ & (7 $\pm$ 6)$\times$10$^{-5}$ \\
$\rm [CI/H_{2}]$ & (6 $\pm$ 3)$\times$10$^{-5}$ & (3$\pm$ 2)$\times$10$^{-5}$ & (6 $\pm$ 2)$\times$10$^{-5}$ & (2 $\pm$ 2)$\times$10$^{-5}$ & (2 $\pm$ 1)$\times$10$^{-5}$ & (0.6 $\pm$ 0.6)$\times$10$^{-5}$ \\
\hline
Estimated \\
within the model\\
\hline
T$_{\rm dust}$ & 54 $\pm$ 4 & 58 $\pm$ 4 & 
31 $\pm$ 9 & 106 $\pm$ 37 & 38 $\pm$ 16 & 80 $\pm$ 16 \\
M$_{\rm gas}$ & (1.4 $\pm$ 1.2)$\times$10$^{11}$ & (1.8 $\pm$ 2.5)$\times$10$^{10}$ & 
(2.4 $\pm$ 2.4)$\times$10$^{11}$ & (1.1 $\pm$ 2.0)$\times$10$^{11}$ &
(2.0 $\pm$ 2.3)$\times$10$^{10}$ & (1.6 $\pm$ 2.2)$\times$10$^{10}$ \\
M$_{\rm dust}$ & (1.2 $\pm$ 1.3)$\times$10$^{9}$ & (1.2 $\pm$ 2.0)$\times$10$^{8}$ & 
(1.5 $\pm$ 1.8)$\times$10$^{9}$ & (0.7 $\pm$ 1.3)$\times$10$^{9}$ &
(0.8 $\pm$ 1.0)$\times$10$^{8}$ & (0.7 $\pm$ 1.0)$\times$10$^{8}$ \\
\hline
Derived \\
from the model\\
\hline
$\rm L^{\prime}_{CO}$ & (6 $\pm$ 3)$\times$10$^{10}$ & (9 $\pm$ 4)$\times$10$^{9}$ & 
(5 $\pm$ 1)$\times$10$^{10}$ & (1 $\pm$ 0.6)$\times$10$^{10}$ &
(2 $\pm$ 1)$\times$10$^{9}$ & (8 $\pm$ 3)$\times$10$^{8}$ \\
L$_{\rm FIR}$ & (19 $\pm$ 2)$\times$10$^{12}$ & (31 $\pm$ 2)$\times$10$^{11}$ & (2 $\pm$ 2)$\times$10$^{12}$ & (16 $\pm$ 2)$\times$10$^{12}$ &
(6 $\pm$ 6)$\times$10$^{11}$ & (25 $\pm$ 5)$\times$10$^{11}$\\

\hline\hline
\multicolumn{7}{p{\textwidth}}{\hspace{+0.2in}{NOTE. -  The input and derived parameters of the \textit{1-component} and \textit{2-component} models assuming less than solar metallicity. The units and the explored range are same as the \textit{Turbulence} model given in Table \ref{turb_parameters}. The model outputs are the intrinsic source properties as we use the magnification corrected photometry and line flux densities for modeling. 
}}
\end{tabular}
\end{adjustwidth}
\end{table*}

\begin{figure*}[h!t] 
\hspace*{0.0cm}
\includegraphics[trim={0.0cm 0 0.0cm 0.0},clip,width=1.05\textwidth]{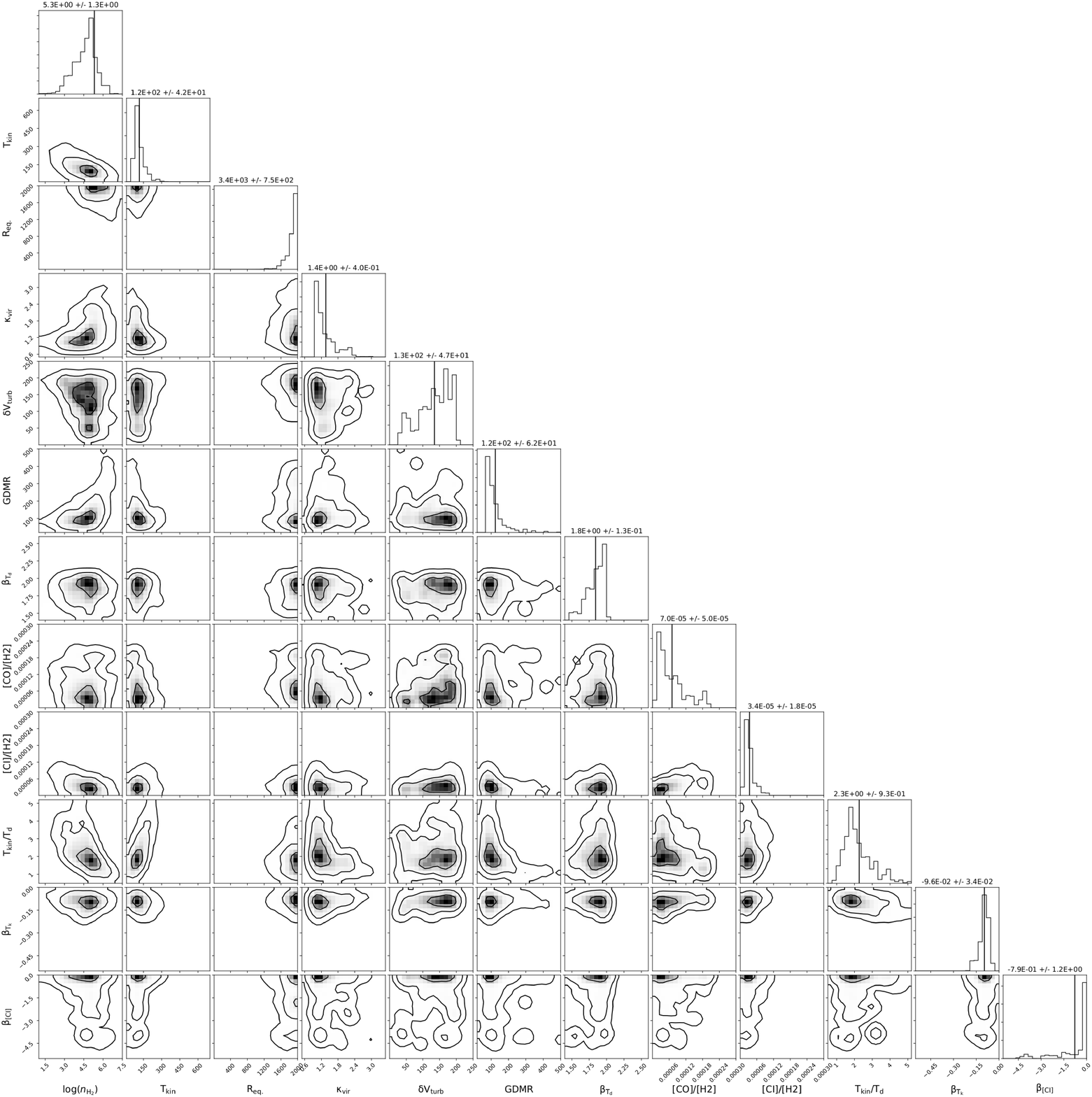}
\caption{\textit{Turbulence} model output in W. The contours are at 1,2 and 3$\sigma$ confidence levels. The mean value of each parameter is shown as a black vertical line on the histogram. Note that $\rm \beta_{[CI]}$ is not constrained because of lack of observations of \cia.}
\label{westcorner}
\end{figure*}

\begin{figure*}[h!t] 
\hspace*{0.0cm}
\includegraphics[trim={0.0cm 0 0.0cm 0.0},clip,width=1.05\textwidth]{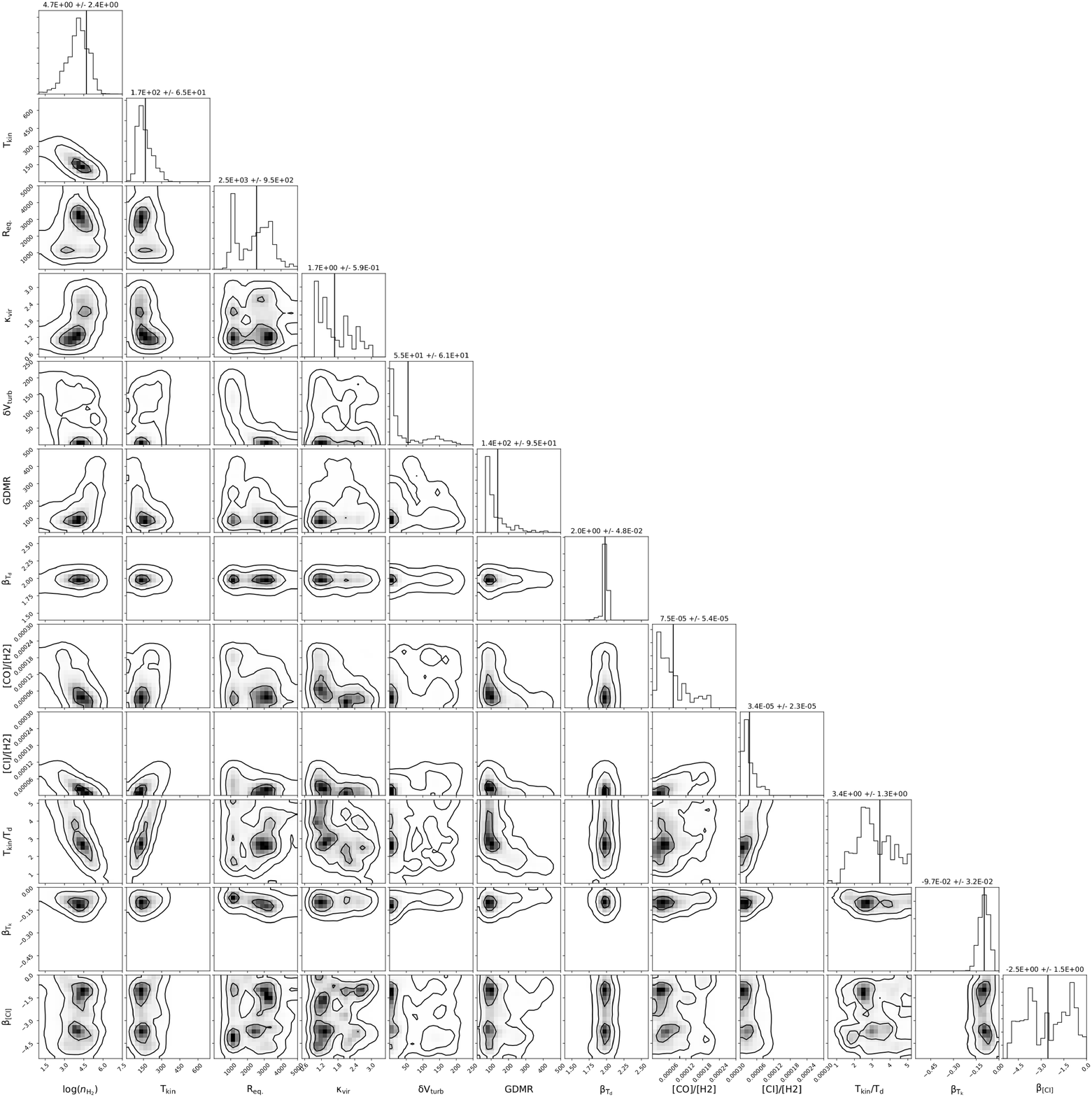}
\caption{\textit{Turbulence} model output in E. The contours are at 1,2 and 3$\sigma$ confidence levels. The mean value of each parameter is shown as a black vertical line on the histogram. Note that $\rm \beta_{[CI]}$ is not constrained because of lack of observations of \cia\ and \cib.}
\label{eastcorner}
\end{figure*}

We also present the output parameters from $\sim$10$^{7}$ \textit{Turbulence} models in W and E in Figures \ref{westcorner} and \ref{eastcorner}, respectively. The mean value is indicated in the parameter histograms.

\end{document}